%% file: CALS_2020.tex
\documentclass[acmsmall, review=false, nonacm=true, printfolios=true]{acmart}
\usepackage{amsmath}

\AtBeginDocument{%
  \providecommand\BibTeX{{%
    \normalfont B\kern-0.5em{\scshape i\kern-0.25em b}\kern-0.8em\TeX}}}

\setcopyright{acmcopyright}
\copyrightyear{2018}
\acmYear{2018}
\acmDOI{10.1145/1122445.1122456}

\acmConference[Woodstock '18]{Woodstock '18: ACM Symposium on Neural
  Gaze Detection}{June 03--05, 2018}{Woodstock, NY}
\acmBooktitle{Woodstock '18: ACM Symposium on Neural Gaze Detection,
  June 03--05, 2018, Woodstock, NY}
\acmPrice{15.00}
\acmISBN{978-1-4503-XXXX-X/18/06}

\input{lib/CALS_2020_include.tex}

\begin{document}

\title{Concurrent Alternating Least Squares for multiple simultaneous Canonical Polyadic Decompositions}
\renewcommand{\shorttitle}{Concurrent ALS for multiple simultaneous CP decompositions}

\author{Christos Psarras}
\orcid{0000-0001-6057-7491}
\affiliation{%
  \institution{RWTH Aachen University}
  \streetaddress{Schinkelstr. 2}
  \city{Aachen}
  \state{North Rhine-Westphalia}
  \postcode{52062}
  \country{Germany}}
\email{psarras@aices.rwth-aachen.de}

\author{Lars Karlsson}
\affiliation{%
  \institution{Ume{\aa} Universitet}
  \streetaddress{Plan 4, MIT-huset}
  \city{Ume{\aa}}
  \country{Sweden}}
\email{larsk@cs.umu.se}

\author{Rasmus Bro}
\orcid{0000-0002-7641-4854}
\affiliation{%
    \institution{University of Copenhagen}
    \streetaddress{Rolighedsvej 26}
    \city{Copenhagen}
    \state{}
    \postcode{1958 Frederiksberg C}
    \country{Denmark}}
\email{rb@food.ku.dk}

\author{Paolo Bientinesi}
\affiliation{%
	\institution{Ume{\aa} Universitet}
	\streetaddress{Plan 4, MIT-huset}
	\city{Ume{\aa}}
	\country{Sweden}}
\email{larsk@cs.umu.se}

\renewcommand{\shortauthors}{Psarras, Karlsson, Bro and Bientinesi}

\begin{abstract}
  Tensor decompositions, such as CANDECOMP/PARAFAC (CP), are widely used in a variety of applications, such as chemometrics, signal processing, and machine learning.
  A broadly used method for computing such decompositions relies on the Alternating Least Squares (ALS) algorithm.
  When the number of components is small, regardless of its implementation, ALS exhibits low arithmetic intensity, which severely hinders its performance and makes GPU offloading ineffective.
  We observe that, in practice, experts often have to compute multiple decompositions of the same tensor, each with a small number of components (typically fewer than 20), to ultimately find the best ones to use for the application at hand.
  In this paper, we illustrate how multiple decompositions of the same tensor can be fused together at the algorithmic level to increase the arithmetic intensity.
  Therefore, it becomes possible to make efficient use of GPUs for further speedups;
  at the same time the technique is compatible with many enhancements typically used in ALS, such as line search, extrapolation, and non-negativity constraints.
  We introduce the Concurrent ALS algorithm and library, which offers an interface to MATLAB, and a mechanism to effectively deal with the issue that decompositions complete at different times.
  Experimental results on artificial and real datasets demonstrate a shorter time to completion due to increased arithmetic intensity.
\end{abstract}

\ccsdesc[500]{Mathematics of computing~Mathematical software performance}
\ccsdesc[500]{Software and its engineering~Software performance}

\keywords{Tensor, decomposition, high-performance}


\maketitle

\input{sections/01-introduction.tex}
\input{sections/02-related.tex}
\input{sections/03-algorithm.tex}
\input{sections/04-convergence.tex}
\input{sections/05-features.tex}
\input{sections/06-results.tex}
\input{sections/07-conclusion.tex}



%
%

\begin{acks}
  Financial support from the Deutsche Forschungsgemeinschaft (German Research Foundation) through grant IRTG 2379 is gratefully acknowledged.
\end{acks}

\bibliographystyle{ACM-Reference-Format}
\bibliography{CALS_2020.bib}


\end{document}

%% file: lib/CALS_2020_include.tex
\usepackage[ruled,vlined,linesnumbered]{algorithm2e}

\SetKwComment{tcp}{$\triangleright$}{}
\SetKwInput{KwInput}{Input}
\SetKwInput{KwOutput}{Output}

\usepackage{lipsum}
\usepackage{tikz}
\usepackage{adjustbox}   
\usepackage{subcaption}  
\usepackage{layouts}  
\usepackage{multirow}

\newcommand{\changed}[1]{#1}

\def\decompname{CP}

\newcommand{\mathcalbf}[1]{\pmb{\mathcal{#1}}}

\newcommand{\vect}[1]{\ensuremath{\mathbf{#1}}}
\newcommand{\mat}[1]{\ensuremath{\mathbf{#1}}}
\newcommand{\set}[1]{\ensuremath{\mathcal{#1}}}
\newcommand{\tensor}[1]{\ensuremath{\mathcalbf{#1}}}
\newcommand{\unfolding}[2]{\ensuremath{\mat{#1}_{(#2)}}}

\newcommand{\inr}[1]{\ensuremath{\in \mathbb{R}^{#1}}}

\newcommand{\factor}[2]{\ensuremath{\mathbf{U}_{#2}^{(#1)}}}
\newcommand{\krp}[2]{\ensuremath{\mathbf{M}_{#2}^{(#1)}}}
\newcommand{\hadam}[2]{\ensuremath{\mathbf{H}_{#2}^{(#1)}}}

\newcommand{\multifactor}[2]{\ensuremath{\mathbf{\overline{U}}_{#2}^{(#1)}}}
\newcommand{\multikrp}[2]{\ensuremath{\mathbf{\overline{M}}_{#2}^{(#1)}}}

\newcommand{\range}[2]{#1, $\ldots$ , #2}

%% file: sections/01-introduction.tex

\section{Introduction}

The Canonical Polyadic Decomposition (CPD or CP), also known as PARAllel FACtor analysis (PARAFAC), is a tensor decomposition or "tensor model" that has found applications in several domains, including chemometrics~\cite{andersen:2003}, signal processing and machine learning~\cite{sidiropoulos:2016}.
A target tensor is (often approximately) decomposed into a sum of rank-1 tensors, commonly referred to as "components".
A \decompname{} decomposition can be computed by, for example, the Alternating Least Squares (CP-ALS) algorithm~\cite{harshman:1970,carroll:1970}.
This is an iterative algorithm that searches for a local minimum to the sum of squared residuals starting from some given starting point with a given number of components.
Since only a local minimum is found, the result of the decomposition is highly susceptible to both the chosen number of components and the starting point.
For this reason, application experts often resort to performing tens or hundreds of independent decompositions of a tensor, varying the number of components and/or the starting point, to determine the quality of each generated solution.

\changed{Particularly in the field of fluorescence of dissolved matter \cite{Murphy:2013}, the process of determining the rank of a target dataset (tensor) involves splitting large datasets into smaller chunks and for each one of them following a ``trial-and-error'' approach, where multiple CP models with the same or different number of components are fitted to the sub-sampled tensors to determine their rank. 
Alternatives to this method, such as performing Higher Order Singular Value Decomposition (HOSVD) to the initial tensor and fitting CP models to the resulting, compressed, tensor G, are not used in practice in this field.
The primary reasons being that missing data, non-negativity and artifacts such as Rayleigh and Raman scattering make the use of HOSVD compression problematic.}

Thus far, research has focused on improving methods that compute a single decomposition.
In this paper, we show that it is possible to combine several decompositions at an algorithmic level to make more efficient use of the hardware.
Even though the proposed technique does not decrease the time to complete a single decomposition, the total time to complete the whole set of decompositions can nevertheless be greatly reduced.

For dense tensors, the computational cost of CP-ALS is typically dominated by the so-called MTTKRP operation~\cite{Hayashi:2018} (see also Section~\ref{sec:als-concurrent-als} ahead).
The processor is a bottleneck for MTTKRP only when the number of components is large. 
Otherwise, the performance will be limited by the transfer of data back and forth between main memory and processor;
we say that MTTKRP is "memory-bound" when the components are few, since the \emph{arithmetic intensity}~\cite{Williams:2009} (the number of floating point operations divided by the number of memory accesses) is proportional to the number of components.
Low arithmetic intensity leads to poor utilization of the CPU and also reduces the benefit from multi-threaded execution and GPU offloading.
Since the issue is inherent to MTTKRP (and by extension to CP-ALS and many of its alternatives), we need to look beyond the narrow problem of computing a single decomposition in order to make progress.

\vspace{\textfloatsep}
\noindent\begin{minipage}{\textwidth}
             \centering
             \begin{minipage}[t]{.49\textwidth}
               \begin{algorithm}[H]
                 \DontPrintSemicolon
                 \KwInput{\tensor{T}: The tensor to decompose. \newline
                   \set{S}: Set of $K$ starting points.}
                 \KwOutput{\set{P}: Set of $K$ \decompname{} models fitted to the tensor \tensor{T}.}
                 $\set{P} \gets \emptyset$\;
                 \ForEach{$\tensor A \in \set{S}$}{
                   $\tensor B \gets$ \decompname{}-ALS($\tensor T$, $\tensor A$)\;
                   $\set{P} \gets \set{P} \cup \{ \tensor B \}$\;
                 }
                 \Return{$\set{P}$}\;
                   \caption{Common usage scenario of \decompname{}-ALS.}
                   \label{alg:app-als}
               \end{algorithm}
             \end{minipage}
             \begin{minipage}[t]{.49\textwidth}
               \begin{algorithm}[H]
                 \DontPrintSemicolon
                 \KwInput{\tensor{T}: The tensor to decompose. \newline
                 \set{S}: Set of $K$ starting points.}
                 \KwOutput{\set{P}: Set of $K$ \decompname{} models fitted to the tensor \tensor{T}.}
                 $\set P \leftarrow$ CALS($\tensor{T}$, $\set S$)\;
                 \Return{$\set{P}$}\;
                   \caption{Common usage scenario of CALS.}
                   \label{alg:app-cals}
               \end{algorithm}
             \end{minipage}
\end{minipage}
\vspace{\textfloatsep}

Taking into account that a typical workflow of an application expert involves computing a set of \decompname{} decompositions (with varying numbers of components and starting points), instead of optimizing only the CP-ALS procedure, we consider the workflow as a whole (see Algorithm~\ref{alg:app-als}).
We introduce the \emph{Concurrent ALS} (CALS) algorithm\footnote{Since our algorithm is used to compute the \decompname{} decomposition, its proper name is CP-CALS. However, to avoid repetition and make it easier for the reader to differentiate between CP-CALS and CP-ALS, throughout the paper we refer to CP-CALS as just CALS.}, which extends the standard CP-ALS algorithm such that it concurrently computes \emph{a set of} \decompname{} decompositions.
We provide CALS as a C++ library, with an interface to MATLAB and GPU offloading via CUDA.
Within CALS, we combine multiple invocations of \decompname{}-ALS at an algorithmic level such that the arithmetic intensity increases --- without numerically affecting any of the invocations.
Crucially, CALS remains compatible with many enhancements typically used in CP-ALS; to demonstrate, we included line search and non-negativity constraints in our library implementation.
A challenge is that different instances of \decompname{}-ALS require a varying number of iterations before they converge.
Therefore, we include a mechanism to dynamically insert and remove instances with minimal impact on performance.

\paragraph{Contributions}
These are the highlights:

\begin{itemize}
  
\item CALS achieves a higher arithmetic intensity than a sequence of CP-ALS invocations, without numerically affecting the computation, and therefore completes a set of decompositions faster.
  
\item We showcase how CALS makes offloading to a GPU worthwhile by increasing the granularity of the central MTTKRP operation, which further increases the speed.

\item We demonstrate that CALS is compatible with enhancements of CP-ALS that preserve the central MTTKRP operation by incorporating line search and non-negativity constraints to the CALS library.

\item To help application experts take full advantage of the CALS features within their existing source code, we also provide an interface to MATLAB.

\end{itemize}

\paragraph{Organization}
The rest of the paper is organized as follows.
In Section~\ref{sec:related}, we provide an overview of related research.
In Section~\ref{sec:als-concurrent-als}, we review the standard \decompname{}-ALS algorithm and introduce the basics of CALS.
We describe how CALS handles the issue of uneven convergence in Section~\ref{sec:handling-convergence}.
We present several features of CALS in Section~\ref{sec:features}.
In Section~\ref{sec:experiments}, we show experimental results that support the claim that CALS reduces the time to completion by increasing the arithmetic intensity.
We also include preliminary experiments on a real dataset from fluorescence spectroscopy to demonstrate its practicality.

\paragraph{Notation}
For vectors and matrices, we use bold lowercase and uppercase roman letters, respectively, e.g., \vect{v} and \mat{U}.
For tensors, we follow the notation in~\cite{Kolda:2009};
specifically, we use bold calligraphic fonts, e.g., \tensor{T}.
The order (number of indices or modes) of a tensor is denoted by uppercase roman letters, e.g., $N$.
For each mode $n$, a tensor \tensor{T} can be unfolded (matricized) into a matrix, denoted by \unfolding{T}{n}, where the columns are the mode-$n$ fibers of \tensor{T}, i.e., the vectors obtained by fixing all indices except for mode~$n$.
Sets are denoted by non-bold calligraphic fonts, e.g., $\mathcal S$.
Given two matrices $\mat A$ and $\mat B$, with the same number of columns, the Khatri-Rao product, denoted by $\mat A \odot \mat B$, is the column-wise Kronecker product of $\mat A$ and $\mat B$.


%% file: sections/02-related.tex
\section{Related Work}
\label{sec:related}

Several variations of the \decompname{} decomposition have been developed to meet the needs of applications.
Examples of constraints on the factor matrices include non-negativity~\cite{bro:1997}, orthogonality~\cite{2012_sorensen}, and coherence~\cite{2018_farias}.
In so called dictionary-based variants, the columns of a factor matrix are constrained to a given set~\cite{2018_cohen}.
Other variations include weighting~\cite{1997_paatero}, missing values~\cite{2005_tomasi}, and alternative objective functions such as the Kullback-Leibler divergence~\cite{2015_hansen}.

Whether the given tensor is dense, sparse, or presented in factored form (e.g., Tucker or \decompname{}) has a big impact on data structures and algorithms~\cite{2007_bader}.
Other classes of methods besides ALS have been proposed, e.g., methods based on eigendecompositions~\cite{Sanchez:1986} and methods based on gradient-based (all-at-once) optimization~\cite{2011_acar}.

Several modifications to \decompname{}-ALS have been proposed.
Line search~\cite{2008_rajih} and extrapolation~\cite{2019_ang} procedures accelerated convergence and appear to help avoid local minima.
Pairwise perturbation is a recently proposed acceleration technique that uses error-controlled approximations in order to reduce the arithmetic cost of generating the \decompname{}-ALS subproblems~\cite{2019_ma}.
For very large tensors, randomization can reduce the time and space complexity of a \decompname{}-ALS iteration.
Some randomization methods use random sampling of the tensor~\cite{2016_vervliet} while others sample the Khatri-Rao product~\cite{2018_battaglino}. 
Compression-based techniques replace the tensor with an approximation of lower rank, thereby shrinking the effective size of the tensor decomposition problem~\cite{1998c_bro}.
A \decompname{} decomposition of the compressed tensor is computed and then inflated to form a decomposition of the large tensor.

In the context of large dense tensors and small \changed{number of components}, the time complexity of the MTTKRP dominates the cost of \decompname{}-ALS.
Efficient (parallel) algorithms for the MTTKRP have therefore been a target for research.
Naive permute-and-multiply algorithms, which explicitly permute data to generate the unfolding prior to a matrix multiplication, are easy to implement but suffer
from large overheads due to the repeated permutations of data.
The overhead can be reduced, but not eliminated, with a high-performance tensor transposition library~\cite{springer:2017}.
A leap forward was presented in~\cite{2013_phan}, where the authors showed how an MTTKRP can be done without any permutations at all.
They also removed redundant computations across the sequence of related MTTKRPs that appear in a \decompname{}-ALS iteration.
An alternative approach to avoiding permutation is to recognize that the mode-$n$ unfoldings have a natural block structure. 
This can be exploited to create a one-step algorithm without permutations~\cite{Hayashi:2018}.
\changed{Despite the large number of software targeting tensor computations~\cite{psarras:2021} (and operations similar to MTTKRP),
    the efforts have largely been fragmented among different fields of application (e.g., machine learning, computational chemistry) and languages (e.g., Matlab, Python, C++).
    Ultimately, despite the large common ground of operations shared among these fields, a universal specification similar to BLAS and
    LAPACK has not emerged yet, often leading to the independent and redundant (and often not intercompatible) development of would-be kernels such as MTTKRP.}

\changed{Several projects have been developed that support CP-ALS, including Cyclops~\cite{Solomonik:2013}, PLANC~\cite{Ramakrishnan:2016}, Partensor~\cite{Lourakis:2018}, SPLATT~\cite{Smith:2015}, Genten~\cite{Phipps:2019}, to name just a few---a more comprehensive list of tensor-related software can be found in~\cite{psarras:2021}.}
The technique we propose in this paper complements rather than competes with many of these related efforts.
What we propose can be applied alongside other enhancements and even to other methods besides \decompname{}-ALS.
\changed{Indeed, the method described can be applied to any CP algorithm that relies on MTTKRP and where the target tensor, which appears in the MTTKRP, does not change for different instances of the problem (e.g. different starting points or number of components).}
This is important because there is good reason to believe that it is the accumulative effect of many disparate techniques that will lead to the best performance. 


%% file: sections/03-algorithm.tex

\section{ALS and Concurrent ALS (CALS)}
\label{sec:als-concurrent-als}

We begin by reviewing the regular \decompname{}-ALS algorithm in Section~\ref{sec:als-concurrent-als:als} and introduce the CALS algorithm in Section~\ref{sec:als-concurrent-als:cals}.

\subsection{ALS}
\label{sec:als-concurrent-als:als}

\begin{algorithm}[htbp]
    \SetAlgoLined
    \DontPrintSemicolon
    \KwInput{
        \tensor{T}\inr{I_1, \ldots{\,} , I_N}: The tensor to decompose. \newline
        \range{\factor{1}{}}{\factor{N}{}}: The factor matrices of the starting point \changed{($R$ components)}.
      }
    \KwOutput{\range{\factor{1}{}}{\factor{N}{}}: The computed \decompname{} decomposition of $\tensor T$.}
    \Repeat{convergence detected or maximum number of iterations reached}
    {
        \For{\range{$n = 1, 2$}{$N$}}{
            $\krp{n}{} \leftarrow \mat{T}_{(n)} (\odot_{i\neq n} \factor{i}{})$ \tcp*[r]{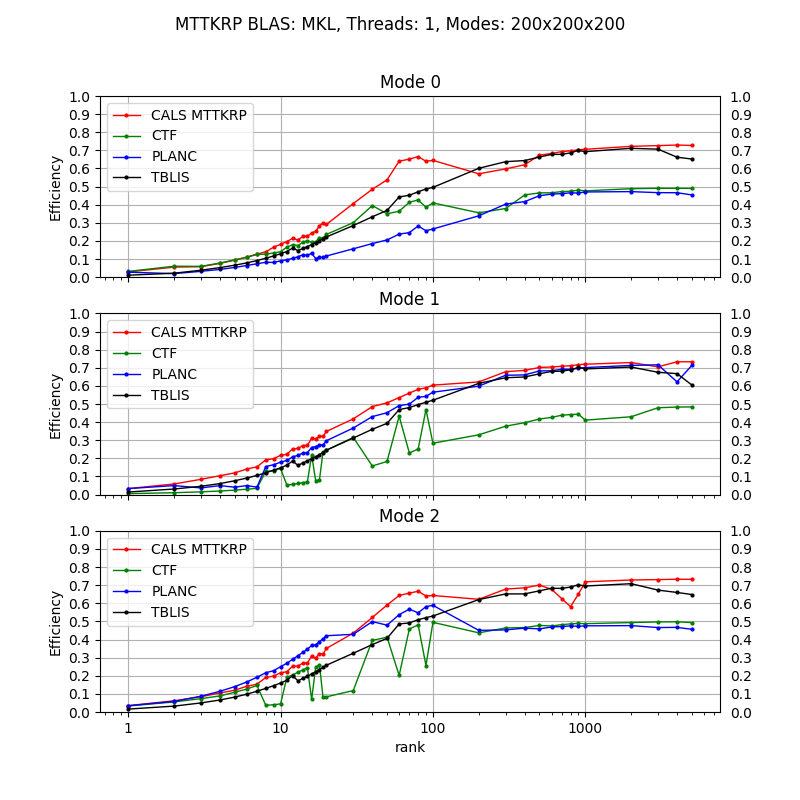} \label{alg:als:mttkrp}
            $\hadam{n}{} \leftarrow \ast_{i\neq n}({\factor{i}{}}^{T}\factor{i}{})$ \tcp*[r]{Hadamard product of Gramians} \label{alg:als:hadamard}
            $\factor{n}{} \leftarrow \krp{n}{} {\hadam{n}{}}^{\dagger}$ \tcp*[r]{${\hadam{n}{}}^{\dagger}$: pseudoinverse of \hadam{n}{}} \label{alg:als:update}
        }
        \changed{$e \leftarrow ||\tensor{X}||^{2} - (\oplus (\ast({\factor{i}{}}^{T}\factor{i}{}))) - 2 (\oplus (\factor{N}{} \ast \krp{N}{}))$} \tcp*[r]{\changed{Error calculation}} \label{alg:als:error}
    }
    \caption{\decompname{}-ALS: Alternating least squares method for \decompname{} decomposition.}
    \label{alg:als}
\end{algorithm}

Algorithm~\ref{alg:als} shows the standard alternating least squares method for \decompname{} decomposition (\decompname{}-ALS).
Given a starting point, the factor matrices are repeatedly updated one-by-one in sequence until either convergence is detected \changed{(fit falls below a tolerance threshold)} or some maximum number of iterations has been reached.
When updating the factor matrix for mode-$n$, the gradient of the objective function with respect to \factor{n}{} is set to zero and the resulting (linear) least squares problem is solved exactly via the normal equations. 

Computationally, the most expensive step is the Matricized Tensor Times Khatri-Rao Product (MTTKRP) in line~\ref{alg:als:mttkrp}.
Conceptually, the mode-$n$ unfolding \unfolding{T}{n} is multiplied with the Khatri-Rao Product (KRP) of all factor matrices except \factor{n}{}.
This involves $2 R \prod_{i} I_{i}$ flops (not counting the KRP, which accounts for a lower order term) and $\prod_{i} I_{i}$ accesses to tensor elements.
Thus, the arithmetic intensity of the MTTKRP is $2R$ flops per tensor element access.
In line~\ref{alg:als:hadamard}, the Gramians of each factor matrix (${\factor{i}{}}^{T}\factor{i}{}$) are multiplied together, to form $\hadam{n}{}$, using the Hadamard product.
Since $\hadam{n}{}$ is of size $R \times R$, the cost of solving the linear system in line~\ref{alg:als:update} is $\mathcal{O}(R^{2}I_{n}+R^{3})$, which, for small $R$, is much less than the cost of MTTKRP.
\changed{In line~\ref{alg:als:error}, the error is calculated after an iteration of the algorithm using the fast error calculation formula, described in \cite{Phan:2013}.
    This operation has a computational cost of $\mathcal{O}(\prod_{i} I_{i} + R^{2} + I_{N}R)$, even though the term $\prod_{i} I_{i}$ needs only be computed once at the start of the algorithm.}

\begin{figure}
    \input{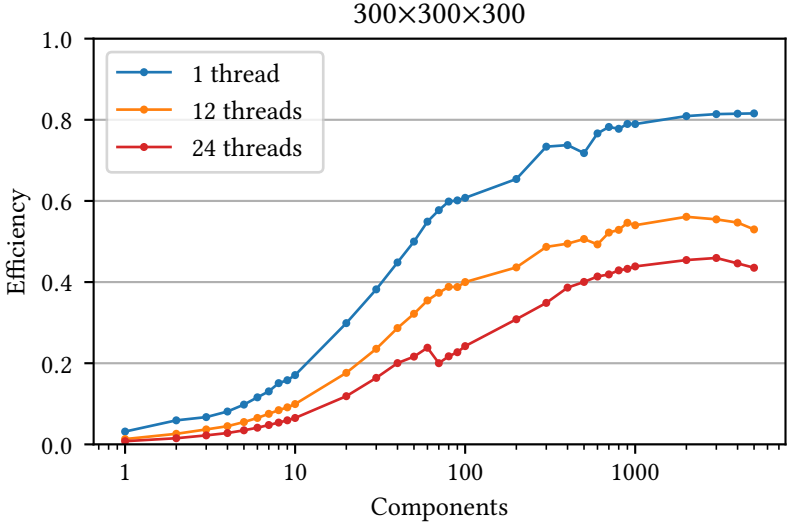}
    \caption{Efficiency of MTTKRP on a $300 \times 300 \times 300$ tensor for increasing \changed{number of components}.
    }
    \label{fig:MTTKRP_rank_sweep}
\end{figure}

Figure~\ref{fig:MTTKRP_rank_sweep} illustrates the practical effect of the $2R$ arithmetic intensity of MTTKRP on its computational efficiency.
Efficiency measures the performance of an algorithm relative to the Theoretical Peak Performance (TPP) of the machine it runs on, and is given by the formula:
\begin{displaymath}
  \textsc{efficiency} = \frac{\textsc{performance}}{\textsc{tpp}} = \frac{\textsc{total \#flops} / \textsc{time}}{\textsc{tpp}}
\end{displaymath}
\changed{TPP is defined as the peak number of double precision floating point operations per second (flops/sec). It is given by multiplying the frequency of the CPU (freq), with the number of threads used (nt), the number of doubles that can fit in a Fused Multiply Add (FMA) vector register (nd) and the number of said vector registers on the CPU (nv).
    
    \begin{equation}
    TPP = 2 * freq * nt * nd * nv
    \label{eq:TPP}
    \end{equation}}
For \changed{components} up to $20$, the efficiency is $< 30\%$ for single-threaded and $< 20\%$ for multi-threaded execution.
An efficiency over $50\%$ is only reached for \changed{components} over $60$ and $500$ with $1$ and $12$ threads, respectively, and never reached (at least for \changed{components} up to $5000$) with $24$ threads.
According to Figure~\ref{fig:MTTKRP_rank_sweep}, the efficiency, for a tensor of size $300 \times 300 \times 300$, increases with $R$ until eventually reaching a plateau, which in this case occurs around $R = 1000$.
Hence, the computational resources tend to be better utilized when the decomposition has many components rather than few.
Given that MTTKRP accounts for the bulk of the cost of \decompname{}-ALS, the efficiency profiles of MTTKRP and \decompname{}-ALS are similar.
The particular implementations used to maximize efficiency are explained in Section~\ref{sec:features:MTTKRP}.

\subsection{Concurrent ALS (CALS)}
\label{sec:als-concurrent-als:cals}

A straightforward way to run $K$ instances of \decompname{}-ALS on the same tensor is to run them one-by-one in sequence as in Algorithm~\ref{alg:app-als} or, in the case of multi-threading, to run multiple instances in parallel.
As explained above, the arithmetic intensity when fitting model~$i$ \changed{with $R_{i}$ components} will be only $2R_{i}$.
The gist of CALS is to reorganize the computations involved in $K$ instances of \decompname{}-ALS into a form which achieves higher arithmetic intensity (and hence higher efficiency), namely $2\sum_{i=1}^{K} R_{i}$.
In particular, if $K = 100$ and $R_{i} = 10$ for all $i$, then from Figure~\ref{fig:MTTKRP_rank_sweep} we expect about $17\%$ MTTKRP efficiency for a sequence of single-threaded \decompname{}-ALS instances.
With CALS we will soon see that we expect an MTTKRP efficiency closer to that observed for $R = \sum_{i=1}^{K} R_{i} = 1000$ or about $78\%$ --- a speedup of $4.5 \times$.

The key idea of CALS is to fuse $K$ small and independent MTTKRPs with low arithmetic intensity into one large MTTKRP with higher arithmetic intensity.
To see how this can be done, first note that (for appropriately sized matrices) independent KRPs can be fused into a larger KRP:
\begin{equation}
  \label{eq:MTTKRP}
  \begin{bmatrix}
    \mat A_{1} \odot \mat B_{1} & \mat A_{2} \odot \mat B_{2}
  \end{bmatrix}
  =
  \begin{bmatrix}
    \mat A_{1} & \mat A_{2}
  \end{bmatrix}
  \odot
  \begin{bmatrix}
    \mat B_{1} & \mat B_{2}
  \end{bmatrix}.
\end{equation}
This allows the fusion of two (or more) independent MTTKRPs:
\begin{displaymath}
  \begin{split}
    \begin{bmatrix}
        \textsc{mttkrp}(\mat X,\mat A_1, \mat B_1) & \textsc{mttkrp}(\mat X,\mat A_2, \mat B_2)
    \end{bmatrix}
    & =
    \begin{bmatrix}
      \mat X ( \mat A_1 \odot \mat B_1 ) & \mat X ( \mat A_2 \odot \mat B_2 )
    \end{bmatrix} \\
    &= \mat X
    \begin{bmatrix}
      \mat A_1 \odot \mat B_1 & \mat A_2 \odot \mat B_2
    \end{bmatrix} \\
    &= \mat X
    \left(
    \begin{bmatrix}
      \mat A_1 & \mat A_2
    \end{bmatrix}
    \odot
    \begin{bmatrix}
      \mat B_1 & \mat B_2
    \end{bmatrix}
    \right) \\
    & =
    \textsc{mttkrp}(\mat X,
    \begin{bmatrix}
        \mat A_1 & \mat A_2
    \end{bmatrix}
    ,
    \begin{bmatrix}
        \mat B_1 & \mat B_2
    \end{bmatrix}).
  \end{split}
\end{displaymath}
Starting from two independent MTTKRPs, the common matrix factor $\mat X$ is extracted and the KRPs are fused together using (\ref{eq:MTTKRP}).
The result is a single, larger MTTKRP. 

Generalized to a tensor \tensor{T} of order $N$, the mode-$n$ MTTKRPs for two independent \decompname{}-ALS instances (on the same tensor) can be fused:
\begin{equation}
  \label{eq:multimatrix}
  \begin{bmatrix}
    \mat{T}_{(n)} (\odot_{i\neq n} \factor{i}{1}) & \mat{T}_{(n)} (\odot_{i\neq n} \factor{i}{2})
  \end{bmatrix}
  =
  \mat{T}_{(n)}
  \left(
    \odot_{i \neq n}
    \begin{bmatrix}
      \factor{i}{1} & \factor{i}{2}
    \end{bmatrix}
  \right).  
\end{equation}
This naturally extends to any number of \decompname{}-ALS instances.

Seemingly, to apply the fused MTTKRP in (\ref{eq:multimatrix}), it suffices to concatenate factor matrices and extract submatrices from the result corresponding to the small MTTKRPs.
This could potentially be very expensive. 
One should instead work directly on the concatenated form of the factor matrices.
Let $\mat A_{1}, \mat A_{2}, \ldots, \mat A_{K}$ be matrices with the same number of rows.
Then we refer to the horizontal concatenation
$\overline{\mat A} =
\begin{bmatrix}
  \mat A_{1} & \mat A_{2} & \ldots & \mat A_{K}
\end{bmatrix}$
as a \emph{multi-matrix}.
The $k$-th constituent matrix of $\overline{\mat A}$ is denoted by $\overline{\mat A}_{|k}$, so $\overline{\mat A}_{|k} \equiv \mat A_{k}$. 

\begin{figure}[htbp]
    \includegraphics[scale=.8]{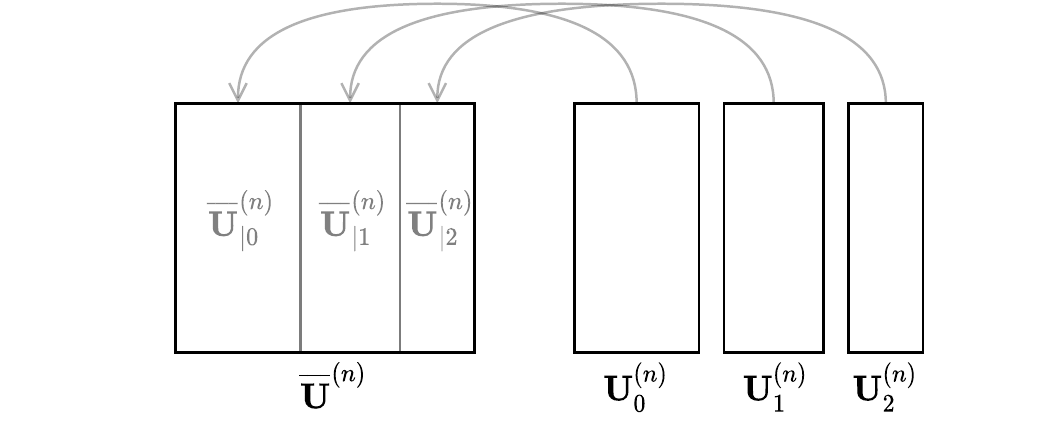}
    \caption{Horizontally concatenating factor matrices into a multi-matrix.}
    \label{fig:multimatrix}
\end{figure}

During initialization, CALS creates $N$ multi-matrices $\multifactor{i}{}$, one for each mode $i = 1, 2, \ldots, N$ by concatenating the initial factor matrices from each starting point\footnote{
 In practice, each multi-matrix is allocated to a fixed size, optionally specified by the user.
 Then the multi-matrices are filled with horizontally concatenated factor matrices from each starting point, until they are full.
 When models converge, as the algorithm progresses, space is freed up, and new starting points can take their place.
 More details about this mechanism are provided in Section~\ref{sec:handling-convergence}.}, as shown in Figure~\ref{fig:multimatrix}.
When a single factor matrix is needed, say to compute a Gramian, then it is accessed as a submatrix of the corresponding multi-matrix.
The fused MTTKRP is readily computed using the multi-matrices as input.
The outputs of the small MTTKRPs are the constituent matrices of the multi-matrix output of the fused MTTKRP.
In summary, after initialization neither concatenation nor extraction is necessary.
 
\begin{algorithm}[htbp]
  \SetAlgoLined
  \DontPrintSemicolon
  \KwInput{\tensor{T}\inr{I_1, \ldots{\,} , I_N}: The tensor to decompose. \newline
        \range{\factor{1}{1}}{\factor{N}{K}}: The factor matrices of the $K$ initial points (\changed{number of components} $R_{i}$ for $i = 1, 2, \ldots K$).}
  \KwOutput{\range{\factor{1}{1}}{\factor{N}{K}}: The $K$ computed \decompname{} decompositions of \tensor{T}.}
  \For(\tcp*[f]{Initialize one factor multi-matrix for each mode.}){\range{$n = 1, 2$}{$N$}}{
    \For{\range{$k = 1, 2$}{$K$}}{
        $\multifactor{n}{|k} \leftarrow \factor{n}{k}$ \label{alg:cals:multifactor}\;
     }
   }
  \Repeat(\tcp*[f]{Concurrently run $K$ instances of \decompname{}-ALS.}){convergence detected for all instances or maximum number of iterations reached}{
    \For{\range{$n = 1, 2$}{$N$}}{
      $\multikrp{n}{} \leftarrow \mat{T}_{(n)}(\odot_{i\neq n} \multifactor{i}{})$ \label{alg:cals:mttkrp}\;
      \For{\range{$k = 1, 2$}{$K$}}{
        $\hadam{n}{k} \leftarrow \ast_{i\neq n}({\multifactor{i}{|k}}^{T}\multifactor{i}{|k}) \label{alg:cals:hadamard}$\;
        $\multifactor{n}{|k} \leftarrow \multikrp{n}{|k}{\hadam{n}{k}}^{\dagger}$ \label{alg:cals:update}\;
       }
     }
   }
   \caption{CALS: Concurrent alternating least squares method for a set of independent \decompname{} decompositions of one tensor. Basic algorithm without handling of uneven convergence.}
   \label{alg:cals}
\end{algorithm}

A simplified version of CALS is described in Algorithm~\ref{alg:cals}.
(In Section~\ref{sec:handling-convergence}, we will extend the algorithm to appropriately handle a large number of instances and convergence at different iterations.)
First, the $N$ factor multi-matrices are initialized by packing the factor matrices from the starting points.
Then the main phase begins.
The $K$ small MTTKRPs are fused into one large MTTKRP in line~\ref{alg:cals:mttkrp}.
The remaining parts of the algorithm are essentially the same as for regular \decompname{}-ALS (Algorithm~\ref{alg:als}) and the instances are treated separately in a loop. 
In particular, lines~\ref{alg:cals:hadamard} and~\ref{alg:cals:update} of Algorithm~\ref{alg:cals} correspond to lines~\ref{alg:als:hadamard} and~\ref{alg:als:update} of Algorithm~\ref{alg:als}.

The efficiency of CALS is determined by the efficiency of MTTKRP (see Figure~\ref{fig:MTTKRP_rank_sweep}) for \changed{a number of components equal to} $\sum_{i = 1}^{K} R_{i}$, regardless of the individual \changed{numbers of components of each model}.
In contrast, for \decompname{}-ALS, the efficiency when computing the $i$th instance is determined by $R_{i}$.


%% file: sections/04-convergence.tex

\section{Handling Convergence}
\label{sec:handling-convergence}

The idea of CALS is to concurrently run multiple instances of \decompname{}-ALS on the same tensor.
The \changed{number of components} of the individual decompositions may or may not be equal.
The starting points can also be different.
The CALS algorithm synchronously advances all instances one iteration at a time.
But the number of iterations required for an instance to converge can vary from a few to thousands.
If convergence is detected for one instance, then that instance should be removed from further processing.
Naturally, one may also want to add new instances during the execution.
In this section, we describe how CALS handles the dynamic insertion and removal of instances and present the full CALS algorithm (Algorithm~\ref{alg:cals_full}).

\begin{algorithm}[htbp]
    \SetAlgoLined
    \DontPrintSemicolon
    \KwInput{\tensor{T}\inr{I_1, \ldots{\,} , I_N}: The tensor to decompose. \newline
    $Q_{\mathrm{in}}$: Input queue with starting points.}
    \KwOutput{$Q_{\mathrm{out}}$: Output queue for computed \decompname{} decompositions.}

    $n_{\rm active} \gets 0$ \tcp*{Initialize.}
    Allocate buffers for $N$ factor multi-matrices $\multifactor{i}{}$ for $i = 1, 2, \ldots, N$\;
    \Repeat(\tcp*[f]{Main loop.}){$Q_{\rm in}$ is empty and $n_{\rm active} = 0$}{
      \While(\tcp*[f]{Fill the factor multi-matrices.}){$Q_{\rm in}$ is not empty}{ \label{alg:cals_full:init_start}
        $\tensor A \leftarrow \textsc{Front}(Q_{\rm in})$ \tcp*{Look at first element of queue.}
        \eIf {$\tensor A$ can be inserted into the factor multi-matrices}{
          $n_{\rm active} \gets n_{\rm active} + 1$\;
          Insert $\tensor A$ into the factor multi-matrices\;
          $\textsc{Dequeue}(Q_{\rm in})$\;
        }{
          Exit the loop\;
        }
        \label{alg:cals_full:init_end}
      }

      \For(\tcp*[f]{Advance all instances one iteration.}){\range{$n = 1, 2$}{$N$}}{ \label{alg:cals_full:iteration_start}

        $\multikrp{n}{} \leftarrow \mat{T}_{(n)}(\odot_{i\neq n} \multifactor{i}{})$ \tcp*{Fused MTTKRP.} \label{alg:cals_full:mttkrp}
        \For(\tcp*[f]{Separate processing of active instances.}){\range{$k = 1, 2$}{$n_{\rm active}$}}{
          $\hadam{n}{k} \leftarrow \ast_{i\neq n}({\multifactor{i}{|k}}^{T}\multifactor{i}{|k})$ \label{alg:cals_full:hadamard}\;
          $\multifactor{n}{|k} \leftarrow \multikrp{n}{|k}{\hadam{n}{k}}^{\dagger}$ \label{alg:cals_full:update}\;
        }
        \label{alg:cals_full:iteration_end}
      }
      \For(\tcp*[f]{Remove converged and stalled instances.}){\range{$k = 1, 2$}{$n_{\rm active}$}}{ \label{alg:cals_full:error_start}
        Let $\tensor P$ denote the $k$th active instance: $\multifactor{1}{|k}, \ldots, \multifactor{N}{|k}$\;
        $E, F \leftarrow$ error and fit of $\tensor P$\;
        \If{$F - F_{prev} < {\rm tol}$ or maximum number of iterations reached}{
          Add $\tensor P$ to $Q_{out}$\;
          Remove the $k$th constituent matrix from each factor multi-matrix\;
          $n_{\rm active} \gets n_{\rm active} - 1$\;
        }
        \label{alg:cals_full:error_end}
      }
    }
\caption{CALS: Concurrent alternating least squares method for a set of independent \decompname{} decompositions of one tensor. Full algorithm.}
\label{alg:cals_full}
\end{algorithm}

In the CALS library, a multi-matrix is stored in column-major format at the beginning of a memory buffer of fixed size.
Constituent matrices are therefore contiguous in memory, which implies that a multi-matrix can grow at the end without moving data around.
One multi-matrix is allocated for each mode for the purpose of storing factor matrices.
The size of the buffer for mode~$n$ is set to $I_{n} R^{*}$, where $R^{*}$ is a user-defined constant specifying the maximum width (number of columns) of a multi-matrix.
($R^{*}$ must be at least as large as the largest \changed{number of components}.)
Ideally, $R^{*}$ is set to the value which maximizes MTTKRP's performance, which, according to Figure~\ref{fig:MTTKRP_rank_sweep}, for a $300 \times 300 \times 300$ tensor, plateaus at approximately $R^{*} = 1000$.
Choosing a larger $R^{*}$ would not affect performance significantly in this case.
This parameter can be chosen to trade off performance versus memory consumption.
The buffer size is enough to concurrently run instances with a \changed{sum of number of components} $\leq R^{*}$.
Any additional instances will have to wait in a queue until space is freed up by converged instances.
\changed{Based on our experience, a value of R* $\approx$ 4000--5000 suffices for a wide range of problems with similarly sized tensors (in terms of volume) to the ones presented in this paper.}


Algorithm~\ref{alg:cals_full} takes a tensor and a queue of starting points as input.
The algorithm starts by allocating buffers for the factor multi-matrices and then enters the main loop, which concludes when the input queue is empty and there are no more active instances.
At the start of each iteration (lines~\ref{alg:cals_full:init_start}--\ref{alg:cals_full:init_end}), the factor multi-matrices are filled with starting points from the input queue until either the buffers are full or the input queue is empty.
Next (lines~\ref{alg:cals_full:iteration_start}--\ref{alg:cals_full:iteration_end}), all active instances are advanced one iteration.
At the end of each iteration (lines~\ref{alg:cals_full:error_start}--\ref{alg:cals_full:error_end}), the error and fit of each instance is computed.
If the difference in fit between two consecutive iterations is lower than some tolerance, then the instance is considered converged and its factor matrices are copied out of the factor multi-matrices and placed in the output queue.
The removal of instances may cause the buffers of the multi-matrices to become fragmented, i.e., they may contain gaps of unused space in between constituent matrices.
To get rid of the fragmentation, a compression routine packs the constituent matrices contiguously starting at the beginning of the respective buffers.



%% file: sections/05-features.tex

\section{Software features}
\label{sec:features}

In this section, we describe the various components of CALS (available on Github\footnote{\url{https://github.com/HPAC/CP-CALS}}), which include MTTKRP, line search, and non-negativity constraints as well as features,
such as GPU offloading and the MATLAB interface.

CALS is written in C++ and depends on the BLAS and LAPACK libraries (supported by, e.g., Intel\textsuperscript{\textregistered} MKL~\cite{mkl}, BLIS~\cite{blis:2015}, and OpenBLAS~\cite{openblas:2012}).
Optional dependencies are CUDA~\cite{cuda} (for GPU offloading) and MATLAB~\cite{matlab}.

\subsection{MTTKRP}
\label{sec:features:MTTKRP}

The MTTKRP operation accounts for the vast majority of the cost of both \decompname{}-ALS and CALS.
There does not exist a highly optimized black-box implementation of MTTKRP that performs well for all modes and sizes.
Recent research (e.g.,~\cite{Hayashi:2018}) has shown that the keys to fast MTTKRP is to
(a) avoid data permutations in memory,
(b) cast the computations in terms of GEMM BLAS operations, and
(c) avoid explicitly computing Khatri-Rao products in certain cases.

To this end, CALS includes a family of algorithmic variants for MTTKRP.
Depending on such things as the size and shape of the tensor, the \changed{number of components} of the decomposition, the mode of the MTTKRP, and the method used for parallelization, the variants vary greatly in terms of performance.
No single variant is best in all cases.
Finding an effective way of determining a good variant for a particular case is discussed in~\cite{Peise:2015} and is beyond the scope of this paper.
For third-order tensors, we hardcoded the choice of the best variant for each case and each algorithm (\decompname{}-ALS or CALS), according to benchmarks.
For higher-order tensors, CALS falls back to computing the Khatri-Rao product explicitly and then performing one, or several, matrix multiplications.

\subsection{Line search}

Line search is a technique that can reduce the total number of iterations to reach convergence in \decompname{}-ALS and to some extent can also help to avoid getting trapped in bad local minima.
For these reasons, line search is used by most high-quality implementations of \decompname{}-ALS (and other methods).
In order to demonstrate that CALS is compatible with line search, we include a basic version (see, e.g.,~\cite{harshman:1970,bro:1998}) as an optional feature.
Let $\tensor P^{(i)}$ and $\tensor P^{(i+1)}$ denote points after the \decompname{}-ALS iterations~$i$ and $i+1$, respectively.
Then after iteration $i+1$, line search linearly extrapolates to a new point $\tensor P^{\mathrm{new}} = \tensor P^{(i)} +
\alpha (\tensor P^{(i+1)} - \tensor P^{(i)})$, where $\alpha > 1$ determines the amount of extrapolation.
If $\tensor P^{\mathrm{new}}$ provides a better fit than $\tensor P^{(i+1)}$, then $\tensor P^{(i+1)}$ is replaced with $\tensor P^{\mathrm{new}}$.
Otherwise, the extrapolated point is rejected.
Regarding the choice of $\alpha$, several proposals have been made.
For example, Bro~\cite{bro:1998} recommends setting $\alpha = \sqrt[3]{i}$ based on experience. 

In CALS, line search is applied independently to each instance as follows.
Let \multifactor{n}{|k, i} and \multifactor{n}{|k, i+1} be the factor matrices for each mode $n$, of an instance $k$, for iterations $i$ and $i+1$ respectively.
The new, extrapolated, factor matrices are given by $\multifactor{n}{|k, \mathrm{new}} = \multifactor{n}{|k, i} + \alpha (\multifactor{n}{|k, i+1} - \multifactor{n}{|k, i})$, for each $n$, where $\alpha > 1$ is a user-defined constant.
Given $\multifactor{n}{|k, \mathrm{new}}$, the errors after iterations $i$ and $i + 1$ are computed and the extrapolated point is used only if it shows an improvement.

\subsection{Non-negativity constraints}

CALS is also compatible with active set-based approaches for enforcing non-negativity constraints to the factor matrices.
To demonstrate this, we included the active set-based method proposed in~\cite{bro:1997} as a feature in CALS.

The non-negativity constraints propagate down to the least squares subproblem that updates a factor matrix (see line~\ref{alg:als:update} in Algorithm~\ref{alg:als}).
The subproblem changes into a \emph{non-negative} least squares problem with multiple right-hand sides.
Mirroring regular \decompname{}-ALS, the method proposed in~\cite{bro:1997} computes the MTTKRP and the Hadamard product of the Gramians only once per inner iteration. 
What changes compared to \decompname{}-ALS is the update step, which is replaced by an iterative search for the right set of active constraints.
Since the size of the matrices involved in this search is of the order of the \changed{number of components} of an individual decomposition, the cost is negligible in our context of small \changed{number of components} and large tensors.

In CALS, the active set-based method is applied to each decomposition independently after the fused MTTKRP in line~\ref{alg:cals:mttkrp} of Algorithm~\ref{alg:cals}.
The final active sets are saved for the next iteration for the purposes of warm starting, as described and justified in~\cite{bro:1997}.

\changed{\subsection{CPU parallelization methods} \label{sec:cpu_parallelization}
Notwithstanding the CALS methodology, there are two other (more obvious) ways of parallelizing multiple, independent instances of CP-ALS on multi-threaded CPUs; one is to run them in sequence, while each instance makes use of multi-threaded BLAS and LAPACK (CP-ALS); another is to launch multiple CP-ALS instances concurrently, by means of OpenMP, while each instance makes use of single threaded BLAS and LAPACK (OMP ALS).

Depending on the size of the target tensor, the numbers of components of the models being fitted and the CPU topology of the target system, different parallelization methods might perform better or worse. For this reason, the CALS library supports all the parallelization methods mentioned above.}

\subsection{GPU offloading}

GPUs offer tremendous computational capabilities, especially for matrix operations.
While the MTTKRP in \decompname{}-ALS can in principle be offloaded to a GPU, the associated overhead likely outweighs the gains due to the low arithmetic intensity.
The larger arithmetic intensity achieved by CALS makes it more suitable for offloading its MTTKRP to a GPU.
To take advantage of this extra capability CALS has to offer, we developed a CUDA interface to MTTKRP.

The elements required to perform MTTKRP on the GPU are the factor multi-matrices and the tensor.
Since the tensor does not change, we transfer it to the GPU during initialization and let it remain there until the end.
We also transfer the multi-matrices to the GPU during initialization, and then, for each iteration, only the multi-matrix being updated is transferred back and forth between the GPU and CPU.
Specifically, after line~\ref{alg:cals:mttkrp} of Algorithm~\ref{alg:cals}, \multikrp{n}{} is brought to the CPU to update \multifactor{n}{|k} for each instance $k$.
Once all instances have had their updates completed, \multifactor{n}{} is sent back to the GPU.
For the algorithmic variants of MTTKRP that require an explicit KRP, the KRP is computed on the CPU and transferred to the GPU.

Section~\ref{sec:experiments:GPU_utilization} showcases the additional speedup gained by from offloading computation to the GPU.

\subsection{MATLAB interface}

Some of the most popular software packages used in applications (e.g., Tensor Toolbox~\cite{TTB_Software,TTB_Dense}, the N-way Toolbox~\cite{nway_toolbox}, and Tensorlab~\cite{tensorlab}) are developed for MATLAB.
To make CALS accessible to MATLAB users, we provide a MATLAB interface via MEX through a function called \texttt{cp\_cals}.
The arguments of this function are similar to the \texttt{cp\_als} function in Tensor Toolbox with some extra optional arguments that enable features such as line search, non-negativity, and GPU offloading.


%% file: sections/06-results.tex

\section{Experiments}
\label{sec:experiments}

In this section, we demonstrate the performance improvements of CALS over \decompname{}-ALS.
First, we isolate the performance impact of the increased arithmetic intensity achieved by CALS.
To this end, we present two synthetic experiments, one targeting speedup and the other efficiency.
Second, we showcase the performance improvements that CALS can offer in real world \changed{applications in the field of fluorescence of dissolved matter}. 
Third, we highlight the effect of GPU offloading \changed{in} CALS.

\changed{
    We observed that in many applications which make use of CP-ALS---including fluorescence of dissolved matter, hyperspectral imaging, neuroimaging, machine learning, and statistics---the vast majority of tensors used are 3D~\cite{bro:online, eeg, imaging, adatm, parcube, partensor, planc, splatt, tensord, tensorly, threeway}.
    Therefore, while the CALS methodology can be applied to higher order tensors, we limit our experimentation to 3D tensors.}

The experiments were performed on a Linux-based system with an Intel\textregistered{} Xeon\textregistered{} Platinum 8160 Processor \changed{(Turbo Boost enabled, Hyper-Threading disabled)}, which contains 24 physical cores split in 2 NUMA regions with 12 cores each.
\changed{All experiments were conducted with double precision arithmetic and} we report results for 1 thread, 12 threads (one NUMA region), and 24 threads (both NUMA regions).
\changed{
%
    
    The CPU used for the experiments contains 2 AVX512 vector registers ($nv = 2$) of length 64 bytes ($nd = 8$ doubles) each, which perform 2 flops/cycle (FMA), totaling $2 * nd * nv = 32$ flops/cycle. Table \ref{tab:TPP} calculates TPP for the thread configurations used, based on Equation~\ref{eq:TPP}.
}
    \begin{table}[h!]
    \centering
    \begin{tabular}{lccc}
    	\toprule
        System & TPP (GFlops/sec) & freq (Ghz) & \texttt{nt} \\ \midrule

    	      \multirow{3}*{CPU} & 112        &    3.5     &      1      \\
    	       & 998        &    2.6     &     12      \\
    	       & 1536       &     2      &     24 \\ \midrule
 
              GPU & 7000 & -& -\\ \bottomrule

    \end{tabular}
    \caption{Theoretical Peak Performance (TPP) for the target system. The values for the CPU were calculated using Eq.~\ref{eq:TPP}, while the value for the GPU was taken from the manufacturer's specifications.}
    \label{tab:TPP}
    \end{table}

The code was compiled using GCC\footnote{GCC version 9} and linked with the Intel\textregistered{} Math Kernel Library version 19.0, which implements a superset of BLAS and LAPACK.
For the CUDA experiments, an Nvidia Tesla V100 was used\footnote{Driver version: 450.51.06, CUDA Version: 11.0}.
For the MATLAB experiments, MATLAB version 2019b was used.
The source code is available online\footnote{\href{https://github.com/HPAC/CP-CALS}{\changed{https://github.com/HPAC/CP-CALS}}}.

\subsection{Arithmetic intensity}

We present experiments that isolate the impact of increased arithmetic intensity.

\subsubsection{Speedup per iteration}

Since CALS achieves a higher arithmetic intensity than \decompname{}-ALS, we expect CALS to complete one iteration on $K$ instances faster than \decompname{}-ALS completes one iteration on $K$ instances.
We measured the size of this speedup with the following experiment.
For a given tensor and \changed{number of components}, we ran both CALS and \decompname{}-ALS on $K = 20$ random starting points.
All instances ran for exactly $50$ iterations and we measured the total time for each method.
We then calculated the average speedup per iteration using the \changed{following formula (CALS and CP-ALS use the same number of threads):}
\begin{displaymath}
  \textsc{speedup per iteration} = \frac{\textsc{time for cp-als}}{\textsc{time for cals}}.
\end{displaymath}
The size of the tensor, the \changed{number of components of the models}, and the multi-threading configuration all play important roles in determining the execution time.
We therefore repeated the experiment for all \changed{components} $R \in \{ 1, 2, \ldots, 20 \}$ on cubic tensors of size \changed{$m \times m \times m$ for all $m \in \{ 100, 200, 300 \}$}.
For each problem configuration, we tested $1$ thread, $24$ threads, and CUDA ($24$ CPU threads plus one GPU).

\begin{figure}
    \input{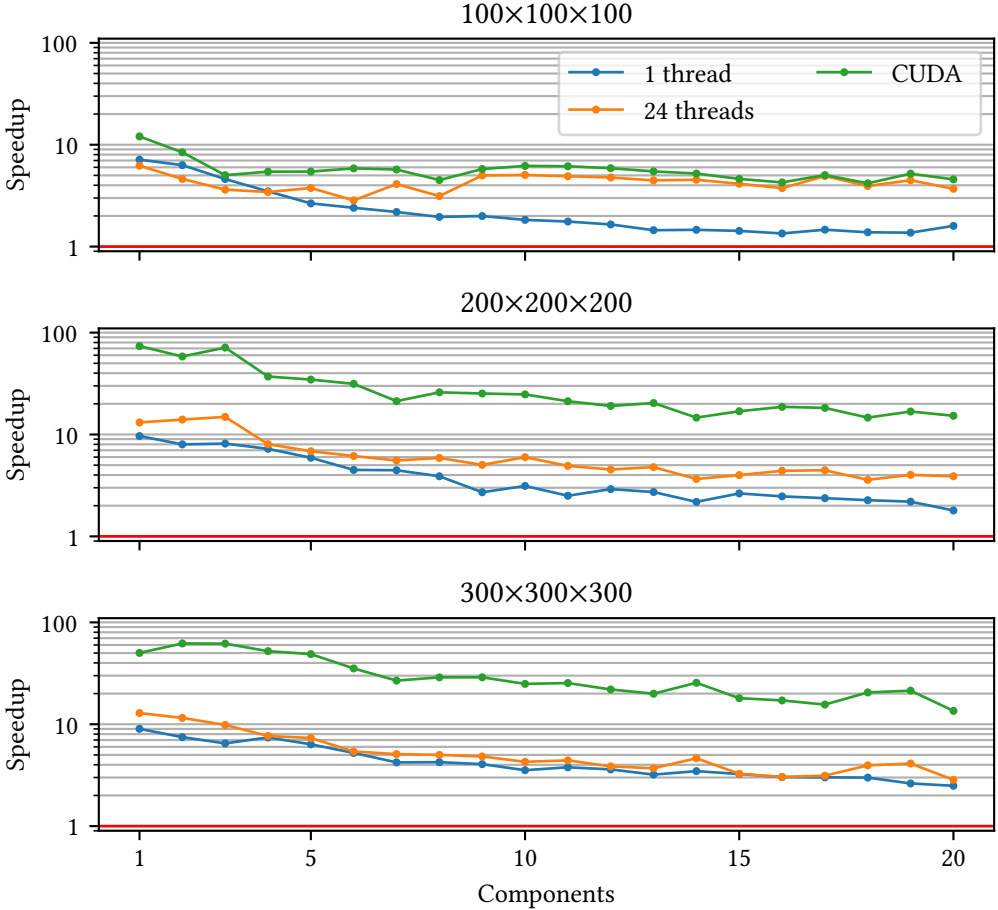}
    \caption{Average speedup per iteration of CALS over \decompname{}-ALS.}
    \label{fig:Speedup_modes}
\end{figure}

The results are shown in Figure~\ref{fig:Speedup_modes}.
CALS is faster than \decompname{}-ALS across all \changed{numbers of components}, and especially so on the smaller \changed{numbers of components} where MTTKRP is particularly inefficient.
The speedup seems to increase with the volume of the tensor.
In particular, GPU offloading leads to significant speedups over multi-threaded \decompname{}-ALS on the larger tensors, reaching up to $60 \times$ for the smallest \changed{numbers of components}.

\subsubsection{Efficiency}
\label{sec:Efficiency}
The efficiency of CALS and \decompname{}-ALS, relative to the machine's theoretical capabilities, was measured with the following experiment.
We used a set of $K = 400$ instances distributed over $1$ through $20$ \changed{components} with $20$ instances for each \changed{number of components}.
Hence, the sum of all \changed{components} is $\sum_{R=1}^{20} 20 R = 4200$. 
All instances ran for exactly $50$ iterations.
We repeated the experiment for cubic tensors of size \changed{$m \times m \times m$ for all $m \in \{ 100, 200, 300 \}$}.
For each problem configuration, we tested $1$ thread, $12$ threads, and $24$ threads.

\begin{figure}
    \input{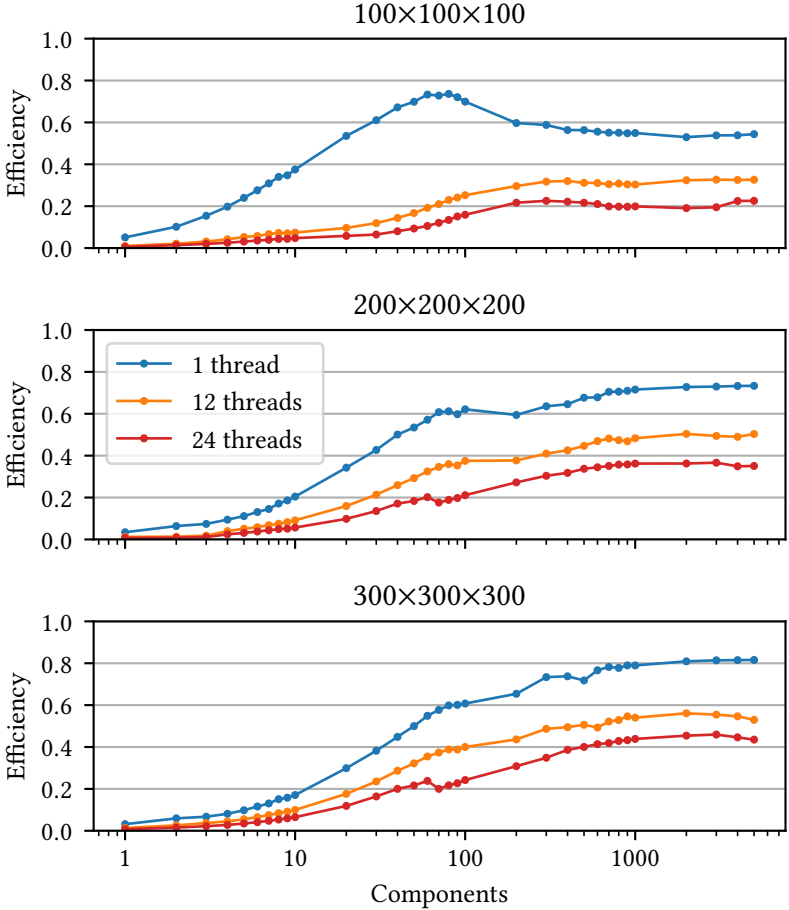}
    \caption{\changed{Performance of the best algorithmic variant of MTTKRP supported by CALS.}}
    \label{fig:MTTKRP_benchmarks}
\end{figure}

The buffer size for CALS ($R^{*}$) was chosen to maximize performance in the BLAS based on the MTTKRP benchmarks reported in Figure~\ref{fig:MTTKRP_benchmarks}.
Specifically, $R^{*} = 4200$ in all cases except for the single-threaded case on the $100 \times 100 \times 100$ tensor, where a suboptimal implementation of the BLAS routine \texttt{dgemm} made $R^{*} = 90$ a better choice.

\begin{figure}
    \input{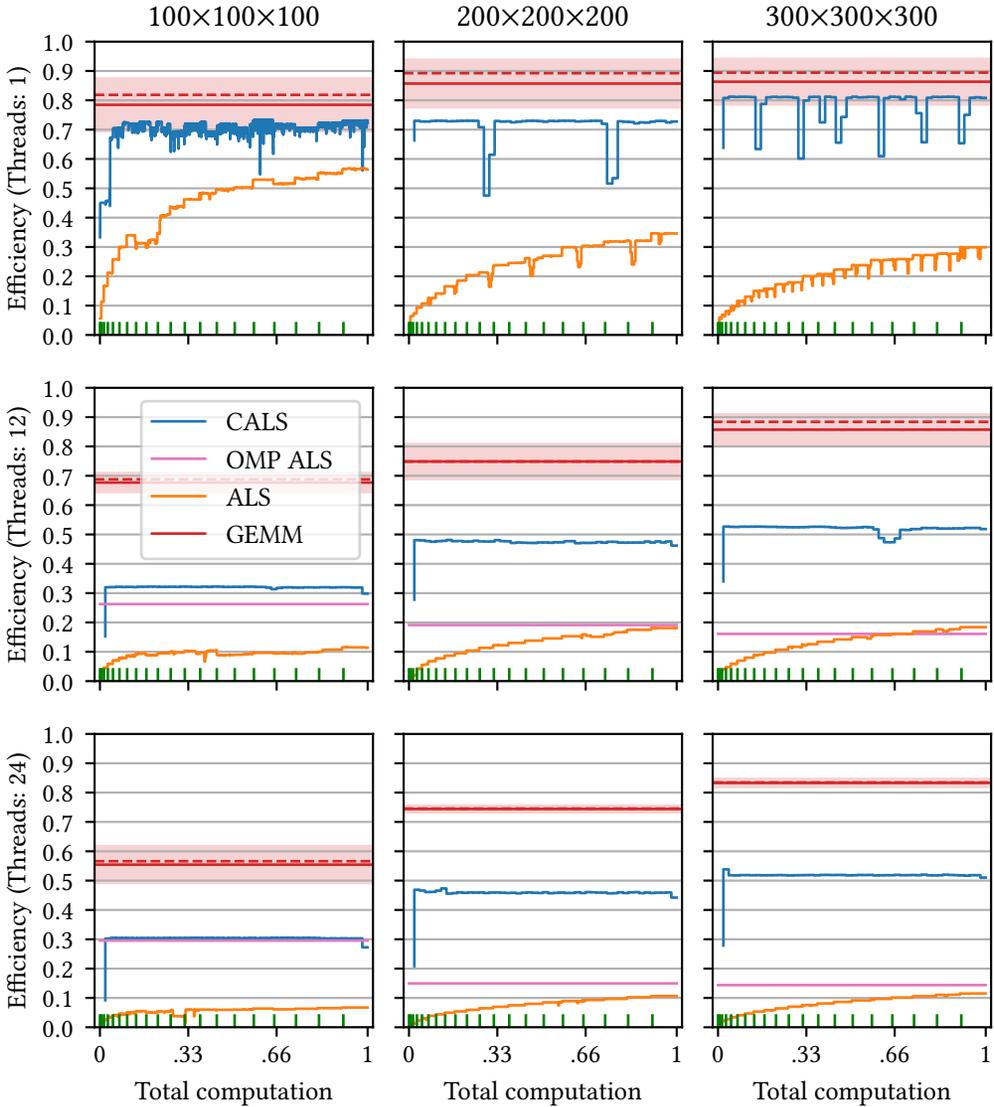}
    \caption{
      The efficiency of CALS, \decompname{}-ALS and OMP ALS during the computation of $400$ decompositions \changed{with number of components} $1$ through $20$ and $20$ decompositions for each \changed{number of components}.
      \decompname{}-ALS processes the \changed{models in ascending order with respect to their number of components}.
      For reference, the \texttt{dgemm} efficiency is shown as a red solid line (mean), dashed line (median), and region (mean $\pm$ one standard deviation); see the text for details.
    }
    \label{fig:arithmetic_intensity}
\end{figure}

The results of the experiment are shown in Figure~\ref{fig:arithmetic_intensity}.
This figure requires a bit of explanation to comprehend. 
We partitioned the total computation into segments: One segment per iteration for CALS and one segment per instance for \decompname{}-ALS.
For each segment, we measured the time and counted the FLOPs.
We then calculated the efficiency for each time segment using the formula
\begin{displaymath}
  \textsc{efficiency} = \frac{\textsc{\#flops} / \textsc{time}}{\textsc{tpp}}.
\end{displaymath}
The figure shows the efficiency versus the fraction of the total computation performed (measured in FLOPs and normalized).
Each time segment corresponds with one piece of the piece-wise constant graphs.
Note that both CALS and \decompname{}-ALS perform the same number of flops.

Since reaching the theoretical peak efficiency of $100\%$ is impossible in practice, we also plot the efficiency of the BLAS routine \texttt{dgemm} as an indicator of practical peak efficiency.
The practical peak is not reproducible, so we characterize it using the mean, median, and standard deviation over many repetitions.
We ran \texttt{dgemm} on square matrices of size \changed{$p \times p$, where $p = \sqrt{m^{3}}$}.
From $100$ samples, we calculated the arithmetic mean, median, and standard deviation and converted times to efficiency using the formula above.
The mean is shown as a solid line, the median as a dashed line, and the mean $\pm$ one standard deviation is shown as a colored region.

\decompname{}-ALS processes the \changed{models in ascending order based on their number of components}, and the transition points are indicated by green tick marks above the horizontal axis.
Observe that the efficiency steadily increases during the computation, which follows the MTTKRP benchmarks reported in Figure~\ref{fig:MTTKRP_benchmarks}.

On the other hand, \changed{by considering the sum of components of all models for the MTTKRP operation, CALS is able to use MTTKRP at the points in Figure~\ref{fig:MTTKRP_benchmarks} where it achieves its largest efficiency.}
The occasional drops in performance for CALS, especially in the single-threaded case, are likely due to CPU frequency throttling.
The locations of these drops are not reproducible, from which we conclude that they are not caused by the algorithm. 

\subsection{\changed{Real Applications}}
\label{sec:experiments:application}

\changed{We demonstrate the effectiveness of CALS on three real datasets (of size $405
  \times 136 \times 19$, $255 \times 286 \times 25$, and $299 \times 301 \times 41$) found in the field of fluorescence of dissolved
  organic matter (two of the datasets are publicly available) \cite{bro:2005, lawaetz2012}. In this field, it is common to set the tolerance between $10^{-6}$ -- $10^{-9}$ (we selected $10^{-6}$), and the maximum number of iterations to the thousands (we selected $1000$). Inspired by the customary workflow of experts in the field, we fit several randomly initialized models, spanning a relatively wide range of components, aiming to get an initial estimate of the rank of the target tensor. In practice, this step of the analysis is repeated several times, experimenting with different data processing methods on the target tensor to overcome noise, outliers and define suitable settings for handling artifacts such as Rayleigh and Raman scattering. For the purposes of this demonstration, we fit a total of $K=400$ models, ranging from 1 to 20 components each (20 randomly initialized models per number of components).

We compare CALS with \decompname{}-ALS and the \texttt{cp\_als} function from Tensor Toolbox.
For multi-threaded execution we experiment with the two kinds of parallelization for CP-ALS described in Section~\ref{sec:cpu_parallelization}, CP-ALS and OMP ALS.}

\begin{figure}
    \centering
    \adjustbox{max width=0.8\textwidth}{
    \begin{subfigure}[b]{\textwidth}
        \input{data/ALS_v_CALS_MKL_real_405-136-19.pgf}
    \end{subfigure}
    }
\adjustbox{max width=0.8\textwidth}{
    \begin{subfigure}[b]{\textwidth}
        \input{data/ALS_v_CALS_MKL_real_255-281-25.pgf}
    \end{subfigure}
}
\adjustbox{max width=0.8\textwidth}{
    \begin{subfigure}[b]{\textwidth}
        \input{data/ALS_v_CALS_MKL_real_299-301-41.pgf}
    \end{subfigure}
}
    \caption{Demonstration of the effectiveness of CALS on real applications. A total of 400 models (with a number of components spanning from 1 to 20) were fitted to each target tensor. The tolerance and maximum number of iterations were set to $10^{-6}$ and $1000$ respectively.}
    \label{fig:EEM_experiments}
\end{figure}
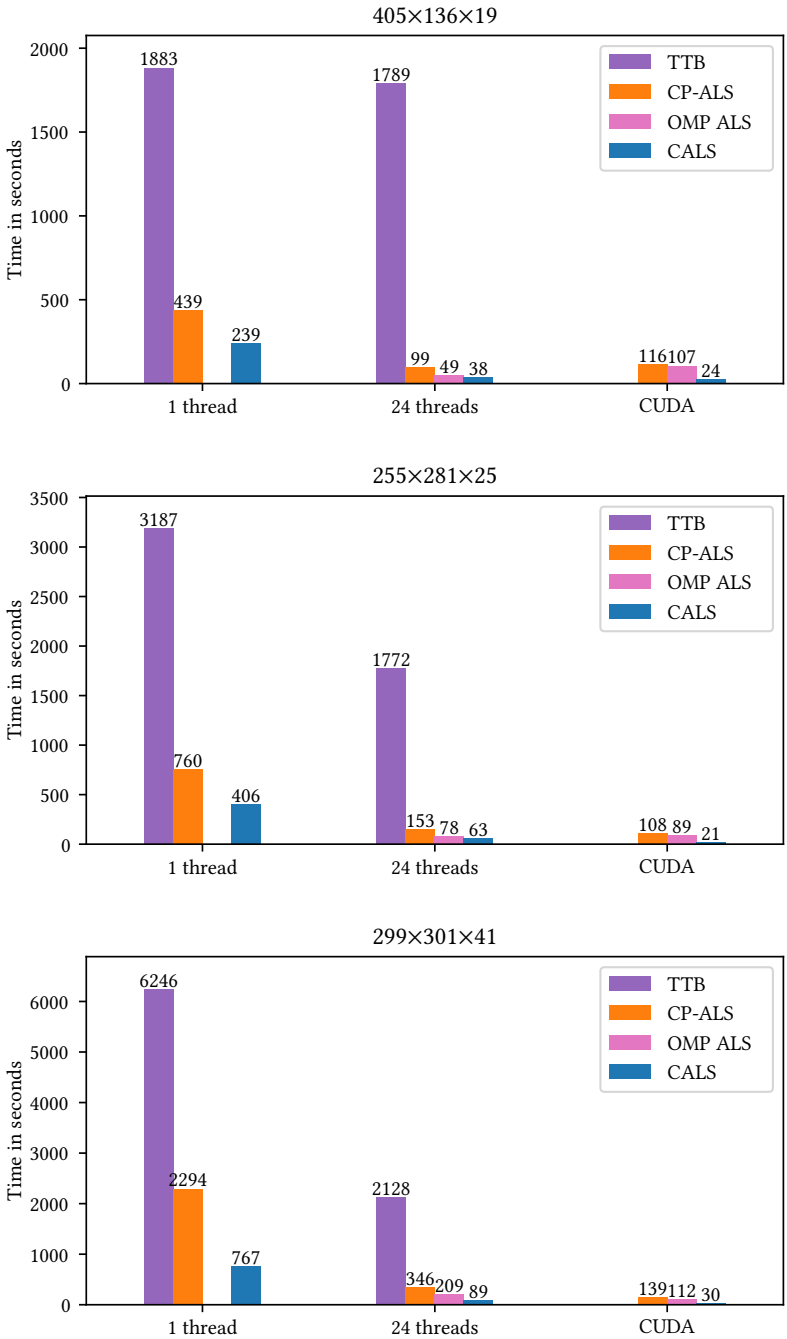

\changed{The results are shown in Figure~\ref{fig:EEM_experiments}.
CALS is faster than Tensor Toolbox and \decompname{}-ALS for single threaded execution, particularly for the larger dataset, achieving $\times 8$ and $\times 3$ speedups respectively.
Note that CALS makes better use of GPU offloading, and for the larger dataset its CUDA implementation reaches a $\times 4.7$ and $\times 3.7$ speedup over \decompname{}-ALS and OMP ALS respectively.}

\subsection{GPU utilization}
\label{sec:experiments:GPU_utilization}

\begin{figure}
    \adjustbox{max width=0.8\textwidth}{
    \input{data/CUDA_v_CALS_MKL.pgf}}
    \caption{Times for CALS, CP-ALS and OMP ALS with GPU offloading (and $24$ threads).}
    \label{fig:CUDA}
\end{figure}
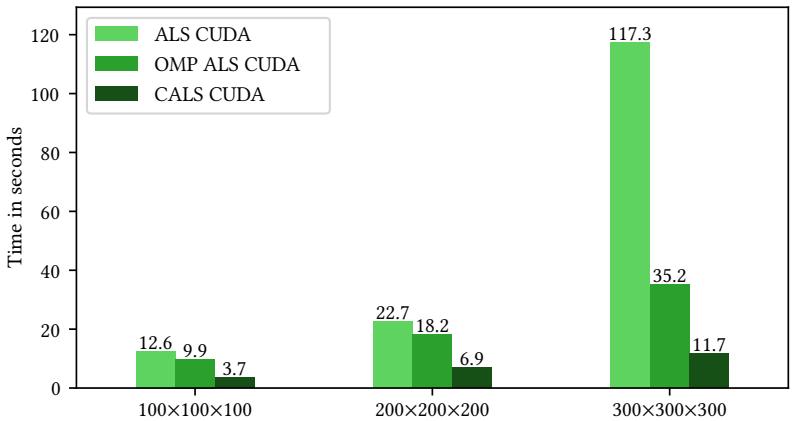

\changed{To demonstrate the effectiveness of utilizing the GPU within CALS, we repeated the experiments in Section~\ref{sec:Efficiency} using 24 threads and the GPU.
The resulting speedups are summarized in Figure~\ref{fig:CUDA}.
Using a GPU is evidently more beneficial for larger tensors.
CALS with GPU offloading achieves speedups of $\times 3.4$ and $\times 2.7$ for tensor size $100 \times 100 \times 100$ compared to \decompname{}-ALS and OMP ALS respectively.
Similarly, the speedups are increased to $\times 9.8$ and $\times 2.9$ for tensor size $300 \times 300 \times 300$.}

%% file: data/ALS_v_CALS_MKL_real_405-136-19.pgf
\begingroup%
\makeatletter%
\begin{pgfpicture}%
\pgfpathrectangle{\pgfpointorigin}{\pgfqpoint{5.478070in}{3.000000in}}%
\pgfusepath{use as bounding box, clip}%
\begin{pgfscope}%
\pgfsetbuttcap%
\pgfsetmiterjoin%
\definecolor{currentfill}{rgb}{1.000000,1.000000,1.000000}%
\pgfsetfillcolor{currentfill}%
\pgfsetlinewidth{0.000000pt}%
\definecolor{currentstroke}{rgb}{1.000000,1.000000,1.000000}%
\pgfsetstrokecolor{currentstroke}%
\pgfsetdash{}{0pt}%
\pgfpathmoveto{\pgfqpoint{0.000000in}{0.000000in}}%
\pgfpathlineto{\pgfqpoint{5.478070in}{0.000000in}}%
\pgfpathlineto{\pgfqpoint{5.478070in}{3.000000in}}%
\pgfpathlineto{\pgfqpoint{0.000000in}{3.000000in}}%
\pgfpathclose%
\pgfusepath{fill}%
\end{pgfscope}%
\begin{pgfscope}%
\pgfsetbuttcap%
\pgfsetmiterjoin%
\definecolor{currentfill}{rgb}{1.000000,1.000000,1.000000}%
\pgfsetfillcolor{currentfill}%
\pgfsetlinewidth{0.000000pt}%
\definecolor{currentstroke}{rgb}{0.000000,0.000000,0.000000}%
\pgfsetstrokecolor{currentstroke}%
\pgfsetstrokeopacity{0.000000}%
\pgfsetdash{}{0pt}%
\pgfpathmoveto{\pgfqpoint{0.703890in}{0.370555in}}%
\pgfpathlineto{\pgfqpoint{5.328070in}{0.370555in}}%
\pgfpathlineto{\pgfqpoint{5.328070in}{2.651000in}}%
\pgfpathlineto{\pgfqpoint{0.703890in}{2.651000in}}%
\pgfpathclose%
\pgfusepath{fill}%
\end{pgfscope}%
\begin{pgfscope}%
\pgfpathrectangle{\pgfqpoint{0.703890in}{0.370555in}}{\pgfqpoint{4.624180in}{2.280445in}}%
\pgfusepath{clip}%
\pgfsetbuttcap%
\pgfsetmiterjoin%
\definecolor{currentfill}{rgb}{0.580392,0.403922,0.741176}%
\pgfsetfillcolor{currentfill}%
\pgfsetlinewidth{0.000000pt}%
\definecolor{currentstroke}{rgb}{0.000000,0.000000,0.000000}%
\pgfsetstrokecolor{currentstroke}%
\pgfsetstrokeopacity{0.000000}%
\pgfsetdash{}{0pt}%
\pgfpathmoveto{\pgfqpoint{1.089238in}{0.370555in}}%
\pgfpathlineto{\pgfqpoint{1.281912in}{0.370555in}}%
\pgfpathlineto{\pgfqpoint{1.281912in}{2.438986in}}%
\pgfpathlineto{\pgfqpoint{1.089238in}{2.438986in}}%
\pgfpathclose%
\pgfusepath{fill}%
\end{pgfscope}%
\begin{pgfscope}%
\pgfpathrectangle{\pgfqpoint{0.703890in}{0.370555in}}{\pgfqpoint{4.624180in}{2.280445in}}%
\pgfusepath{clip}%
\pgfsetbuttcap%
\pgfsetmiterjoin%
\definecolor{currentfill}{rgb}{0.580392,0.403922,0.741176}%
\pgfsetfillcolor{currentfill}%
\pgfsetlinewidth{0.000000pt}%
\definecolor{currentstroke}{rgb}{0.000000,0.000000,0.000000}%
\pgfsetstrokecolor{currentstroke}%
\pgfsetstrokeopacity{0.000000}%
\pgfsetdash{}{0pt}%
\pgfpathmoveto{\pgfqpoint{2.630631in}{0.370555in}}%
\pgfpathlineto{\pgfqpoint{2.823306in}{0.370555in}}%
\pgfpathlineto{\pgfqpoint{2.823306in}{2.336032in}}%
\pgfpathlineto{\pgfqpoint{2.630631in}{2.336032in}}%
\pgfpathclose%
\pgfusepath{fill}%
\end{pgfscope}%
\begin{pgfscope}%
\pgfpathrectangle{\pgfqpoint{0.703890in}{0.370555in}}{\pgfqpoint{4.624180in}{2.280445in}}%
\pgfusepath{clip}%
\pgfsetbuttcap%
\pgfsetmiterjoin%
\definecolor{currentfill}{rgb}{0.580392,0.403922,0.741176}%
\pgfsetfillcolor{currentfill}%
\pgfsetlinewidth{0.000000pt}%
\definecolor{currentstroke}{rgb}{0.000000,0.000000,0.000000}%
\pgfsetstrokecolor{currentstroke}%
\pgfsetstrokeopacity{0.000000}%
\pgfsetdash{}{0pt}%
\pgfpathmoveto{\pgfqpoint{4.172025in}{0.370555in}}%
\pgfpathlineto{\pgfqpoint{4.364699in}{0.370555in}}%
\pgfpathlineto{\pgfqpoint{4.364699in}{0.370555in}}%
\pgfpathlineto{\pgfqpoint{4.172025in}{0.370555in}}%
\pgfpathclose%
\pgfusepath{fill}%
\end{pgfscope}%
\begin{pgfscope}%
\pgfpathrectangle{\pgfqpoint{0.703890in}{0.370555in}}{\pgfqpoint{4.624180in}{2.280445in}}%
\pgfusepath{clip}%
\pgfsetbuttcap%
\pgfsetmiterjoin%
\definecolor{currentfill}{rgb}{1.000000,0.498039,0.054902}%
\pgfsetfillcolor{currentfill}%
\pgfsetlinewidth{0.000000pt}%
\definecolor{currentstroke}{rgb}{0.000000,0.000000,0.000000}%
\pgfsetstrokecolor{currentstroke}%
\pgfsetstrokeopacity{0.000000}%
\pgfsetdash{}{0pt}%
\pgfpathmoveto{\pgfqpoint{1.281912in}{0.370555in}}%
\pgfpathlineto{\pgfqpoint{1.474586in}{0.370555in}}%
\pgfpathlineto{\pgfqpoint{1.474586in}{0.852396in}}%
\pgfpathlineto{\pgfqpoint{1.281912in}{0.852396in}}%
\pgfpathclose%
\pgfusepath{fill}%
\end{pgfscope}%
\begin{pgfscope}%
\pgfpathrectangle{\pgfqpoint{0.703890in}{0.370555in}}{\pgfqpoint{4.624180in}{2.280445in}}%
\pgfusepath{clip}%
\pgfsetbuttcap%
\pgfsetmiterjoin%
\definecolor{currentfill}{rgb}{1.000000,0.498039,0.054902}%
\pgfsetfillcolor{currentfill}%
\pgfsetlinewidth{0.000000pt}%
\definecolor{currentstroke}{rgb}{0.000000,0.000000,0.000000}%
\pgfsetstrokecolor{currentstroke}%
\pgfsetstrokeopacity{0.000000}%
\pgfsetdash{}{0pt}%
\pgfpathmoveto{\pgfqpoint{2.823306in}{0.370555in}}%
\pgfpathlineto{\pgfqpoint{3.015980in}{0.370555in}}%
\pgfpathlineto{\pgfqpoint{3.015980in}{0.479785in}}%
\pgfpathlineto{\pgfqpoint{2.823306in}{0.479785in}}%
\pgfpathclose%
\pgfusepath{fill}%
\end{pgfscope}%
\begin{pgfscope}%
\pgfpathrectangle{\pgfqpoint{0.703890in}{0.370555in}}{\pgfqpoint{4.624180in}{2.280445in}}%
\pgfusepath{clip}%
\pgfsetbuttcap%
\pgfsetmiterjoin%
\definecolor{currentfill}{rgb}{1.000000,0.498039,0.054902}%
\pgfsetfillcolor{currentfill}%
\pgfsetlinewidth{0.000000pt}%
\definecolor{currentstroke}{rgb}{0.000000,0.000000,0.000000}%
\pgfsetstrokecolor{currentstroke}%
\pgfsetstrokeopacity{0.000000}%
\pgfsetdash{}{0pt}%
\pgfpathmoveto{\pgfqpoint{4.364699in}{0.370555in}}%
\pgfpathlineto{\pgfqpoint{4.557373in}{0.370555in}}%
\pgfpathlineto{\pgfqpoint{4.557373in}{0.498201in}}%
\pgfpathlineto{\pgfqpoint{4.364699in}{0.498201in}}%
\pgfpathclose%
\pgfusepath{fill}%
\end{pgfscope}%
\begin{pgfscope}%
\pgfpathrectangle{\pgfqpoint{0.703890in}{0.370555in}}{\pgfqpoint{4.624180in}{2.280445in}}%
\pgfusepath{clip}%
\pgfsetbuttcap%
\pgfsetmiterjoin%
\definecolor{currentfill}{rgb}{0.890196,0.466667,0.760784}%
\pgfsetfillcolor{currentfill}%
\pgfsetlinewidth{0.000000pt}%
\definecolor{currentstroke}{rgb}{0.000000,0.000000,0.000000}%
\pgfsetstrokecolor{currentstroke}%
\pgfsetstrokeopacity{0.000000}%
\pgfsetdash{}{0pt}%
\pgfpathmoveto{\pgfqpoint{1.474586in}{0.370555in}}%
\pgfpathlineto{\pgfqpoint{1.667261in}{0.370555in}}%
\pgfpathlineto{\pgfqpoint{1.667261in}{0.370555in}}%
\pgfpathlineto{\pgfqpoint{1.474586in}{0.370555in}}%
\pgfpathclose%
\pgfusepath{fill}%
\end{pgfscope}%
\begin{pgfscope}%
\pgfpathrectangle{\pgfqpoint{0.703890in}{0.370555in}}{\pgfqpoint{4.624180in}{2.280445in}}%
\pgfusepath{clip}%
\pgfsetbuttcap%
\pgfsetmiterjoin%
\definecolor{currentfill}{rgb}{0.890196,0.466667,0.760784}%
\pgfsetfillcolor{currentfill}%
\pgfsetlinewidth{0.000000pt}%
\definecolor{currentstroke}{rgb}{0.000000,0.000000,0.000000}%
\pgfsetstrokecolor{currentstroke}%
\pgfsetstrokeopacity{0.000000}%
\pgfsetdash{}{0pt}%
\pgfpathmoveto{\pgfqpoint{3.015980in}{0.370555in}}%
\pgfpathlineto{\pgfqpoint{3.208654in}{0.370555in}}%
\pgfpathlineto{\pgfqpoint{3.208654in}{0.424653in}}%
\pgfpathlineto{\pgfqpoint{3.015980in}{0.424653in}}%
\pgfpathclose%
\pgfusepath{fill}%
\end{pgfscope}%
\begin{pgfscope}%
\pgfpathrectangle{\pgfqpoint{0.703890in}{0.370555in}}{\pgfqpoint{4.624180in}{2.280445in}}%
\pgfusepath{clip}%
\pgfsetbuttcap%
\pgfsetmiterjoin%
\definecolor{currentfill}{rgb}{0.890196,0.466667,0.760784}%
\pgfsetfillcolor{currentfill}%
\pgfsetlinewidth{0.000000pt}%
\definecolor{currentstroke}{rgb}{0.000000,0.000000,0.000000}%
\pgfsetstrokecolor{currentstroke}%
\pgfsetstrokeopacity{0.000000}%
\pgfsetdash{}{0pt}%
\pgfpathmoveto{\pgfqpoint{4.557373in}{0.370555in}}%
\pgfpathlineto{\pgfqpoint{4.750047in}{0.370555in}}%
\pgfpathlineto{\pgfqpoint{4.750047in}{0.487946in}}%
\pgfpathlineto{\pgfqpoint{4.557373in}{0.487946in}}%
\pgfpathclose%
\pgfusepath{fill}%
\end{pgfscope}%
\begin{pgfscope}%
\pgfpathrectangle{\pgfqpoint{0.703890in}{0.370555in}}{\pgfqpoint{4.624180in}{2.280445in}}%
\pgfusepath{clip}%
\pgfsetbuttcap%
\pgfsetmiterjoin%
\definecolor{currentfill}{rgb}{0.121569,0.466667,0.705882}%
\pgfsetfillcolor{currentfill}%
\pgfsetlinewidth{0.000000pt}%
\definecolor{currentstroke}{rgb}{0.000000,0.000000,0.000000}%
\pgfsetstrokecolor{currentstroke}%
\pgfsetstrokeopacity{0.000000}%
\pgfsetdash{}{0pt}%
\pgfpathmoveto{\pgfqpoint{1.667261in}{0.370555in}}%
\pgfpathlineto{\pgfqpoint{1.859935in}{0.370555in}}%
\pgfpathlineto{\pgfqpoint{1.859935in}{0.632990in}}%
\pgfpathlineto{\pgfqpoint{1.667261in}{0.632990in}}%
\pgfpathclose%
\pgfusepath{fill}%
\end{pgfscope}%
\begin{pgfscope}%
\pgfpathrectangle{\pgfqpoint{0.703890in}{0.370555in}}{\pgfqpoint{4.624180in}{2.280445in}}%
\pgfusepath{clip}%
\pgfsetbuttcap%
\pgfsetmiterjoin%
\definecolor{currentfill}{rgb}{0.121569,0.466667,0.705882}%
\pgfsetfillcolor{currentfill}%
\pgfsetlinewidth{0.000000pt}%
\definecolor{currentstroke}{rgb}{0.000000,0.000000,0.000000}%
\pgfsetstrokecolor{currentstroke}%
\pgfsetstrokeopacity{0.000000}%
\pgfsetdash{}{0pt}%
\pgfpathmoveto{\pgfqpoint{3.208654in}{0.370555in}}%
\pgfpathlineto{\pgfqpoint{3.401328in}{0.370555in}}%
\pgfpathlineto{\pgfqpoint{3.401328in}{0.412753in}}%
\pgfpathlineto{\pgfqpoint{3.208654in}{0.412753in}}%
\pgfpathclose%
\pgfusepath{fill}%
\end{pgfscope}%
\begin{pgfscope}%
\pgfpathrectangle{\pgfqpoint{0.703890in}{0.370555in}}{\pgfqpoint{4.624180in}{2.280445in}}%
\pgfusepath{clip}%
\pgfsetbuttcap%
\pgfsetmiterjoin%
\definecolor{currentfill}{rgb}{0.121569,0.466667,0.705882}%
\pgfsetfillcolor{currentfill}%
\pgfsetlinewidth{0.000000pt}%
\definecolor{currentstroke}{rgb}{0.000000,0.000000,0.000000}%
\pgfsetstrokecolor{currentstroke}%
\pgfsetstrokeopacity{0.000000}%
\pgfsetdash{}{0pt}%
\pgfpathmoveto{\pgfqpoint{4.750047in}{0.370555in}}%
\pgfpathlineto{\pgfqpoint{4.942722in}{0.370555in}}%
\pgfpathlineto{\pgfqpoint{4.942722in}{0.397320in}}%
\pgfpathlineto{\pgfqpoint{4.750047in}{0.397320in}}%
\pgfpathclose%
\pgfusepath{fill}%
\end{pgfscope}%
\begin{pgfscope}%
\pgfsetbuttcap%
\pgfsetroundjoin%
\definecolor{currentfill}{rgb}{0.000000,0.000000,0.000000}%
\pgfsetfillcolor{currentfill}%
\pgfsetlinewidth{0.803000pt}%
\definecolor{currentstroke}{rgb}{0.000000,0.000000,0.000000}%
\pgfsetstrokecolor{currentstroke}%
\pgfsetdash{}{0pt}%
\pgfsys@defobject{currentmarker}{\pgfqpoint{0.000000in}{-0.048611in}}{\pgfqpoint{0.000000in}{0.000000in}}{%
\pgfpathmoveto{\pgfqpoint{0.000000in}{0.000000in}}%
\pgfpathlineto{\pgfqpoint{0.000000in}{-0.048611in}}%
\pgfusepath{stroke,fill}%
}%
\begin{pgfscope}%
\pgfsys@transformshift{1.474586in}{0.370555in}%
\pgfsys@useobject{currentmarker}{}%
\end{pgfscope}%
\end{pgfscope}%
\begin{pgfscope}%
\definecolor{textcolor}{rgb}{0.000000,0.000000,0.000000}%
\pgfsetstrokecolor{textcolor}%
\pgfsetfillcolor{textcolor}%
\pgftext[x=1.474586in,y=0.273333in,,top]{\color{textcolor}\rmfamily\fontsize{10.000000}{12.000000}\selectfont 1 thread}%
\end{pgfscope}%
\begin{pgfscope}%
\pgfsetbuttcap%
\pgfsetroundjoin%
\definecolor{currentfill}{rgb}{0.000000,0.000000,0.000000}%
\pgfsetfillcolor{currentfill}%
\pgfsetlinewidth{0.803000pt}%
\definecolor{currentstroke}{rgb}{0.000000,0.000000,0.000000}%
\pgfsetstrokecolor{currentstroke}%
\pgfsetdash{}{0pt}%
\pgfsys@defobject{currentmarker}{\pgfqpoint{0.000000in}{-0.048611in}}{\pgfqpoint{0.000000in}{0.000000in}}{%
\pgfpathmoveto{\pgfqpoint{0.000000in}{0.000000in}}%
\pgfpathlineto{\pgfqpoint{0.000000in}{-0.048611in}}%
\pgfusepath{stroke,fill}%
}%
\begin{pgfscope}%
\pgfsys@transformshift{3.015980in}{0.370555in}%
\pgfsys@useobject{currentmarker}{}%
\end{pgfscope}%
\end{pgfscope}%
\begin{pgfscope}%
\definecolor{textcolor}{rgb}{0.000000,0.000000,0.000000}%
\pgfsetstrokecolor{textcolor}%
\pgfsetfillcolor{textcolor}%
\pgftext[x=3.015980in,y=0.273333in,,top]{\color{textcolor}\rmfamily\fontsize{10.000000}{12.000000}\selectfont 24 threads}%
\end{pgfscope}%
\begin{pgfscope}%
\pgfsetbuttcap%
\pgfsetroundjoin%
\definecolor{currentfill}{rgb}{0.000000,0.000000,0.000000}%
\pgfsetfillcolor{currentfill}%
\pgfsetlinewidth{0.803000pt}%
\definecolor{currentstroke}{rgb}{0.000000,0.000000,0.000000}%
\pgfsetstrokecolor{currentstroke}%
\pgfsetdash{}{0pt}%
\pgfsys@defobject{currentmarker}{\pgfqpoint{0.000000in}{-0.048611in}}{\pgfqpoint{0.000000in}{0.000000in}}{%
\pgfpathmoveto{\pgfqpoint{0.000000in}{0.000000in}}%
\pgfpathlineto{\pgfqpoint{0.000000in}{-0.048611in}}%
\pgfusepath{stroke,fill}%
}%
\begin{pgfscope}%
\pgfsys@transformshift{4.557373in}{0.370555in}%
\pgfsys@useobject{currentmarker}{}%
\end{pgfscope}%
\end{pgfscope}%
\begin{pgfscope}%
\definecolor{textcolor}{rgb}{0.000000,0.000000,0.000000}%
\pgfsetstrokecolor{textcolor}%
\pgfsetfillcolor{textcolor}%
\pgftext[x=4.557373in,y=0.273333in,,top]{\color{textcolor}\rmfamily\fontsize{10.000000}{12.000000}\selectfont CUDA}%
\end{pgfscope}%
\begin{pgfscope}%
\pgfsetbuttcap%
\pgfsetroundjoin%
\definecolor{currentfill}{rgb}{0.000000,0.000000,0.000000}%
\pgfsetfillcolor{currentfill}%
\pgfsetlinewidth{0.803000pt}%
\definecolor{currentstroke}{rgb}{0.000000,0.000000,0.000000}%
\pgfsetstrokecolor{currentstroke}%
\pgfsetdash{}{0pt}%
\pgfsys@defobject{currentmarker}{\pgfqpoint{-0.048611in}{0.000000in}}{\pgfqpoint{-0.000000in}{0.000000in}}{%
\pgfpathmoveto{\pgfqpoint{-0.000000in}{0.000000in}}%
\pgfpathlineto{\pgfqpoint{-0.048611in}{0.000000in}}%
\pgfusepath{stroke,fill}%
}%
\begin{pgfscope}%
\pgfsys@transformshift{0.703890in}{0.370555in}%
\pgfsys@useobject{currentmarker}{}%
\end{pgfscope}%
\end{pgfscope}%
\begin{pgfscope}%
\definecolor{textcolor}{rgb}{0.000000,0.000000,0.000000}%
\pgfsetstrokecolor{textcolor}%
\pgfsetfillcolor{textcolor}%
\pgftext[x=0.537223in, y=0.322361in, left, base]{\color{textcolor}\rmfamily\fontsize{10.000000}{12.000000}\selectfont \(\displaystyle {0}\)}%
\end{pgfscope}%
\begin{pgfscope}%
\pgfsetbuttcap%
\pgfsetroundjoin%
\definecolor{currentfill}{rgb}{0.000000,0.000000,0.000000}%
\pgfsetfillcolor{currentfill}%
\pgfsetlinewidth{0.803000pt}%
\definecolor{currentstroke}{rgb}{0.000000,0.000000,0.000000}%
\pgfsetstrokecolor{currentstroke}%
\pgfsetdash{}{0pt}%
\pgfsys@defobject{currentmarker}{\pgfqpoint{-0.048611in}{0.000000in}}{\pgfqpoint{-0.000000in}{0.000000in}}{%
\pgfpathmoveto{\pgfqpoint{-0.000000in}{0.000000in}}%
\pgfpathlineto{\pgfqpoint{-0.048611in}{0.000000in}}%
\pgfusepath{stroke,fill}%
}%
\begin{pgfscope}%
\pgfsys@transformshift{0.703890in}{0.919858in}%
\pgfsys@useobject{currentmarker}{}%
\end{pgfscope}%
\end{pgfscope}%
\begin{pgfscope}%
\definecolor{textcolor}{rgb}{0.000000,0.000000,0.000000}%
\pgfsetstrokecolor{textcolor}%
\pgfsetfillcolor{textcolor}%
\pgftext[x=0.398333in, y=0.871664in, left, base]{\color{textcolor}\rmfamily\fontsize{10.000000}{12.000000}\selectfont \(\displaystyle {500}\)}%
\end{pgfscope}%
\begin{pgfscope}%
\pgfsetbuttcap%
\pgfsetroundjoin%
\definecolor{currentfill}{rgb}{0.000000,0.000000,0.000000}%
\pgfsetfillcolor{currentfill}%
\pgfsetlinewidth{0.803000pt}%
\definecolor{currentstroke}{rgb}{0.000000,0.000000,0.000000}%
\pgfsetstrokecolor{currentstroke}%
\pgfsetdash{}{0pt}%
\pgfsys@defobject{currentmarker}{\pgfqpoint{-0.048611in}{0.000000in}}{\pgfqpoint{-0.000000in}{0.000000in}}{%
\pgfpathmoveto{\pgfqpoint{-0.000000in}{0.000000in}}%
\pgfpathlineto{\pgfqpoint{-0.048611in}{0.000000in}}%
\pgfusepath{stroke,fill}%
}%
\begin{pgfscope}%
\pgfsys@transformshift{0.703890in}{1.469161in}%
\pgfsys@useobject{currentmarker}{}%
\end{pgfscope}%
\end{pgfscope}%
\begin{pgfscope}%
\definecolor{textcolor}{rgb}{0.000000,0.000000,0.000000}%
\pgfsetstrokecolor{textcolor}%
\pgfsetfillcolor{textcolor}%
\pgftext[x=0.328889in, y=1.420966in, left, base]{\color{textcolor}\rmfamily\fontsize{10.000000}{12.000000}\selectfont \(\displaystyle {1000}\)}%
\end{pgfscope}%
\begin{pgfscope}%
\pgfsetbuttcap%
\pgfsetroundjoin%
\definecolor{currentfill}{rgb}{0.000000,0.000000,0.000000}%
\pgfsetfillcolor{currentfill}%
\pgfsetlinewidth{0.803000pt}%
\definecolor{currentstroke}{rgb}{0.000000,0.000000,0.000000}%
\pgfsetstrokecolor{currentstroke}%
\pgfsetdash{}{0pt}%
\pgfsys@defobject{currentmarker}{\pgfqpoint{-0.048611in}{0.000000in}}{\pgfqpoint{-0.000000in}{0.000000in}}{%
\pgfpathmoveto{\pgfqpoint{-0.000000in}{0.000000in}}%
\pgfpathlineto{\pgfqpoint{-0.048611in}{0.000000in}}%
\pgfusepath{stroke,fill}%
}%
\begin{pgfscope}%
\pgfsys@transformshift{0.703890in}{2.018463in}%
\pgfsys@useobject{currentmarker}{}%
\end{pgfscope}%
\end{pgfscope}%
\begin{pgfscope}%
\definecolor{textcolor}{rgb}{0.000000,0.000000,0.000000}%
\pgfsetstrokecolor{textcolor}%
\pgfsetfillcolor{textcolor}%
\pgftext[x=0.328889in, y=1.970269in, left, base]{\color{textcolor}\rmfamily\fontsize{10.000000}{12.000000}\selectfont \(\displaystyle {1500}\)}%
\end{pgfscope}%
\begin{pgfscope}%
\pgfsetbuttcap%
\pgfsetroundjoin%
\definecolor{currentfill}{rgb}{0.000000,0.000000,0.000000}%
\pgfsetfillcolor{currentfill}%
\pgfsetlinewidth{0.803000pt}%
\definecolor{currentstroke}{rgb}{0.000000,0.000000,0.000000}%
\pgfsetstrokecolor{currentstroke}%
\pgfsetdash{}{0pt}%
\pgfsys@defobject{currentmarker}{\pgfqpoint{-0.048611in}{0.000000in}}{\pgfqpoint{-0.000000in}{0.000000in}}{%
\pgfpathmoveto{\pgfqpoint{-0.000000in}{0.000000in}}%
\pgfpathlineto{\pgfqpoint{-0.048611in}{0.000000in}}%
\pgfusepath{stroke,fill}%
}%
\begin{pgfscope}%
\pgfsys@transformshift{0.703890in}{2.567766in}%
\pgfsys@useobject{currentmarker}{}%
\end{pgfscope}%
\end{pgfscope}%
\begin{pgfscope}%
\definecolor{textcolor}{rgb}{0.000000,0.000000,0.000000}%
\pgfsetstrokecolor{textcolor}%
\pgfsetfillcolor{textcolor}%
\pgftext[x=0.328889in, y=2.519572in, left, base]{\color{textcolor}\rmfamily\fontsize{10.000000}{12.000000}\selectfont \(\displaystyle {2000}\)}%
\end{pgfscope}%
\begin{pgfscope}%
\definecolor{textcolor}{rgb}{0.000000,0.000000,0.000000}%
\pgfsetstrokecolor{textcolor}%
\pgfsetfillcolor{textcolor}%
\pgftext[x=0.273333in,y=1.510778in,,bottom,rotate=90.000000]{\color{textcolor}\rmfamily\fontsize{10.000000}{12.000000}\selectfont Time in seconds}%
\end{pgfscope}%
\begin{pgfscope}%
\pgfsetrectcap%
\pgfsetmiterjoin%
\pgfsetlinewidth{0.803000pt}%
\definecolor{currentstroke}{rgb}{0.000000,0.000000,0.000000}%
\pgfsetstrokecolor{currentstroke}%
\pgfsetdash{}{0pt}%
\pgfpathmoveto{\pgfqpoint{0.703890in}{0.370555in}}%
\pgfpathlineto{\pgfqpoint{0.703890in}{2.651000in}}%
\pgfusepath{stroke}%
\end{pgfscope}%
\begin{pgfscope}%
\pgfsetrectcap%
\pgfsetmiterjoin%
\pgfsetlinewidth{0.803000pt}%
\definecolor{currentstroke}{rgb}{0.000000,0.000000,0.000000}%
\pgfsetstrokecolor{currentstroke}%
\pgfsetdash{}{0pt}%
\pgfpathmoveto{\pgfqpoint{5.328070in}{0.370555in}}%
\pgfpathlineto{\pgfqpoint{5.328070in}{2.651000in}}%
\pgfusepath{stroke}%
\end{pgfscope}%
\begin{pgfscope}%
\pgfsetrectcap%
\pgfsetmiterjoin%
\pgfsetlinewidth{0.803000pt}%
\definecolor{currentstroke}{rgb}{0.000000,0.000000,0.000000}%
\pgfsetstrokecolor{currentstroke}%
\pgfsetdash{}{0pt}%
\pgfpathmoveto{\pgfqpoint{0.703890in}{0.370555in}}%
\pgfpathlineto{\pgfqpoint{5.328070in}{0.370555in}}%
\pgfusepath{stroke}%
\end{pgfscope}%
\begin{pgfscope}%
\pgfsetrectcap%
\pgfsetmiterjoin%
\pgfsetlinewidth{0.803000pt}%
\definecolor{currentstroke}{rgb}{0.000000,0.000000,0.000000}%
\pgfsetstrokecolor{currentstroke}%
\pgfsetdash{}{0pt}%
\pgfpathmoveto{\pgfqpoint{0.703890in}{2.651000in}}%
\pgfpathlineto{\pgfqpoint{5.328070in}{2.651000in}}%
\pgfusepath{stroke}%
\end{pgfscope}%
\begin{pgfscope}%
\definecolor{textcolor}{rgb}{0.000000,0.000000,0.000000}%
\pgfsetstrokecolor{textcolor}%
\pgfsetfillcolor{textcolor}%
\pgftext[x=1.185575in,y=2.453118in,,bottom]{\color{textcolor}\rmfamily\fontsize{10.000000}{12.000000}\selectfont 1883}%
\end{pgfscope}%
\begin{pgfscope}%
\definecolor{textcolor}{rgb}{0.000000,0.000000,0.000000}%
\pgfsetstrokecolor{textcolor}%
\pgfsetfillcolor{textcolor}%
\pgftext[x=2.726969in,y=2.349849in,,bottom]{\color{textcolor}\rmfamily\fontsize{10.000000}{12.000000}\selectfont 1789}%
\end{pgfscope}%
\begin{pgfscope}%
\definecolor{textcolor}{rgb}{0.000000,0.000000,0.000000}%
\pgfsetstrokecolor{textcolor}%
\pgfsetfillcolor{textcolor}%
\pgftext[x=1.378249in,y=0.866732in,,bottom]{\color{textcolor}\rmfamily\fontsize{10.000000}{12.000000}\selectfont 439}%
\end{pgfscope}%
\begin{pgfscope}%
\definecolor{textcolor}{rgb}{0.000000,0.000000,0.000000}%
\pgfsetstrokecolor{textcolor}%
\pgfsetfillcolor{textcolor}%
\pgftext[x=2.919643in,y=0.493206in,,bottom]{\color{textcolor}\rmfamily\fontsize{10.000000}{12.000000}\selectfont 99}%
\end{pgfscope}%
\begin{pgfscope}%
\definecolor{textcolor}{rgb}{0.000000,0.000000,0.000000}%
\pgfsetstrokecolor{textcolor}%
\pgfsetfillcolor{textcolor}%
\pgftext[x=4.461036in,y=0.511883in,,bottom]{\color{textcolor}\rmfamily\fontsize{10.000000}{12.000000}\selectfont 116}%
\end{pgfscope}%
\begin{pgfscope}%
\definecolor{textcolor}{rgb}{0.000000,0.000000,0.000000}%
\pgfsetstrokecolor{textcolor}%
\pgfsetfillcolor{textcolor}%
\pgftext[x=3.112317in,y=0.438276in,,bottom]{\color{textcolor}\rmfamily\fontsize{10.000000}{12.000000}\selectfont 49}%
\end{pgfscope}%
\begin{pgfscope}%
\definecolor{textcolor}{rgb}{0.000000,0.000000,0.000000}%
\pgfsetstrokecolor{textcolor}%
\pgfsetfillcolor{textcolor}%
\pgftext[x=4.653710in,y=0.501995in,,bottom]{\color{textcolor}\rmfamily\fontsize{10.000000}{12.000000}\selectfont 107}%
\end{pgfscope}%
\begin{pgfscope}%
\definecolor{textcolor}{rgb}{0.000000,0.000000,0.000000}%
\pgfsetstrokecolor{textcolor}%
\pgfsetfillcolor{textcolor}%
\pgftext[x=1.763598in,y=0.647011in,,bottom]{\color{textcolor}\rmfamily\fontsize{10.000000}{12.000000}\selectfont 239}%
\end{pgfscope}%
\begin{pgfscope}%
\definecolor{textcolor}{rgb}{0.000000,0.000000,0.000000}%
\pgfsetstrokecolor{textcolor}%
\pgfsetfillcolor{textcolor}%
\pgftext[x=3.304991in,y=0.426191in,,bottom]{\color{textcolor}\rmfamily\fontsize{10.000000}{12.000000}\selectfont 38}%
\end{pgfscope}%
\begin{pgfscope}%
\definecolor{textcolor}{rgb}{0.000000,0.000000,0.000000}%
\pgfsetstrokecolor{textcolor}%
\pgfsetfillcolor{textcolor}%
\pgftext[x=4.846385in,y=0.410811in,,bottom]{\color{textcolor}\rmfamily\fontsize{10.000000}{12.000000}\selectfont 24}%
\end{pgfscope}%
\begin{pgfscope}%
\definecolor{textcolor}{rgb}{0.000000,0.000000,0.000000}%
\pgfsetstrokecolor{textcolor}%
\pgfsetfillcolor{textcolor}%
\pgftext[x=3.015980in,y=2.734333in,,base]{\color{textcolor}\rmfamily\fontsize{12.000000}{14.400000}\selectfont 405\(\displaystyle \times\)136\(\displaystyle \times\)19}%
\end{pgfscope}%
\begin{pgfscope}%
\pgfsetbuttcap%
\pgfsetmiterjoin%
\definecolor{currentfill}{rgb}{1.000000,1.000000,1.000000}%
\pgfsetfillcolor{currentfill}%
\pgfsetfillopacity{0.800000}%
\pgfsetlinewidth{1.003750pt}%
\definecolor{currentstroke}{rgb}{0.800000,0.800000,0.800000}%
\pgfsetstrokecolor{currentstroke}%
\pgfsetstrokeopacity{0.800000}%
\pgfsetdash{}{0pt}%
\pgfpathmoveto{\pgfqpoint{4.141959in}{1.765445in}}%
\pgfpathlineto{\pgfqpoint{5.230848in}{1.765445in}}%
\pgfpathquadraticcurveto{\pgfqpoint{5.258626in}{1.765445in}}{\pgfqpoint{5.258626in}{1.793223in}}%
\pgfpathlineto{\pgfqpoint{5.258626in}{2.553778in}}%
\pgfpathquadraticcurveto{\pgfqpoint{5.258626in}{2.581556in}}{\pgfqpoint{5.230848in}{2.581556in}}%
\pgfpathlineto{\pgfqpoint{4.141959in}{2.581556in}}%
\pgfpathquadraticcurveto{\pgfqpoint{4.114181in}{2.581556in}}{\pgfqpoint{4.114181in}{2.553778in}}%
\pgfpathlineto{\pgfqpoint{4.114181in}{1.793223in}}%
\pgfpathquadraticcurveto{\pgfqpoint{4.114181in}{1.765445in}}{\pgfqpoint{4.141959in}{1.765445in}}%
\pgfpathclose%
\pgfusepath{stroke,fill}%
\end{pgfscope}%
\begin{pgfscope}%
\pgfsetbuttcap%
\pgfsetmiterjoin%
\definecolor{currentfill}{rgb}{0.580392,0.403922,0.741176}%
\pgfsetfillcolor{currentfill}%
\pgfsetlinewidth{0.000000pt}%
\definecolor{currentstroke}{rgb}{0.000000,0.000000,0.000000}%
\pgfsetstrokecolor{currentstroke}%
\pgfsetstrokeopacity{0.000000}%
\pgfsetdash{}{0pt}%
\pgfpathmoveto{\pgfqpoint{4.169737in}{2.428778in}}%
\pgfpathlineto{\pgfqpoint{4.447514in}{2.428778in}}%
\pgfpathlineto{\pgfqpoint{4.447514in}{2.526000in}}%
\pgfpathlineto{\pgfqpoint{4.169737in}{2.526000in}}%
\pgfpathclose%
\pgfusepath{fill}%
\end{pgfscope}%
\begin{pgfscope}%
\definecolor{textcolor}{rgb}{0.000000,0.000000,0.000000}%
\pgfsetstrokecolor{textcolor}%
\pgfsetfillcolor{textcolor}%
\pgftext[x=4.558626in,y=2.428778in,left,base]{\color{textcolor}\rmfamily\fontsize{10.000000}{12.000000}\selectfont TTB}%
\end{pgfscope}%
\begin{pgfscope}%
\pgfsetbuttcap%
\pgfsetmiterjoin%
\definecolor{currentfill}{rgb}{1.000000,0.498039,0.054902}%
\pgfsetfillcolor{currentfill}%
\pgfsetlinewidth{0.000000pt}%
\definecolor{currentstroke}{rgb}{0.000000,0.000000,0.000000}%
\pgfsetstrokecolor{currentstroke}%
\pgfsetstrokeopacity{0.000000}%
\pgfsetdash{}{0pt}%
\pgfpathmoveto{\pgfqpoint{4.169737in}{2.235167in}}%
\pgfpathlineto{\pgfqpoint{4.447514in}{2.235167in}}%
\pgfpathlineto{\pgfqpoint{4.447514in}{2.332389in}}%
\pgfpathlineto{\pgfqpoint{4.169737in}{2.332389in}}%
\pgfpathclose%
\pgfusepath{fill}%
\end{pgfscope}%
\begin{pgfscope}%
\definecolor{textcolor}{rgb}{0.000000,0.000000,0.000000}%
\pgfsetstrokecolor{textcolor}%
\pgfsetfillcolor{textcolor}%
\pgftext[x=4.558626in,y=2.235167in,left,base]{\color{textcolor}\rmfamily\fontsize{10.000000}{12.000000}\selectfont CP-ALS}%
\end{pgfscope}%
\begin{pgfscope}%
\pgfsetbuttcap%
\pgfsetmiterjoin%
\definecolor{currentfill}{rgb}{0.890196,0.466667,0.760784}%
\pgfsetfillcolor{currentfill}%
\pgfsetlinewidth{0.000000pt}%
\definecolor{currentstroke}{rgb}{0.000000,0.000000,0.000000}%
\pgfsetstrokecolor{currentstroke}%
\pgfsetstrokeopacity{0.000000}%
\pgfsetdash{}{0pt}%
\pgfpathmoveto{\pgfqpoint{4.169737in}{2.041556in}}%
\pgfpathlineto{\pgfqpoint{4.447514in}{2.041556in}}%
\pgfpathlineto{\pgfqpoint{4.447514in}{2.138778in}}%
\pgfpathlineto{\pgfqpoint{4.169737in}{2.138778in}}%
\pgfpathclose%
\pgfusepath{fill}%
\end{pgfscope}%
\begin{pgfscope}%
\definecolor{textcolor}{rgb}{0.000000,0.000000,0.000000}%
\pgfsetstrokecolor{textcolor}%
\pgfsetfillcolor{textcolor}%
\pgftext[x=4.558626in,y=2.041556in,left,base]{\color{textcolor}\rmfamily\fontsize{10.000000}{12.000000}\selectfont OMP ALS}%
\end{pgfscope}%
\begin{pgfscope}%
\pgfsetbuttcap%
\pgfsetmiterjoin%
\definecolor{currentfill}{rgb}{0.121569,0.466667,0.705882}%
\pgfsetfillcolor{currentfill}%
\pgfsetlinewidth{0.000000pt}%
\definecolor{currentstroke}{rgb}{0.000000,0.000000,0.000000}%
\pgfsetstrokecolor{currentstroke}%
\pgfsetstrokeopacity{0.000000}%
\pgfsetdash{}{0pt}%
\pgfpathmoveto{\pgfqpoint{4.169737in}{1.847945in}}%
\pgfpathlineto{\pgfqpoint{4.447514in}{1.847945in}}%
\pgfpathlineto{\pgfqpoint{4.447514in}{1.945167in}}%
\pgfpathlineto{\pgfqpoint{4.169737in}{1.945167in}}%
\pgfpathclose%
\pgfusepath{fill}%
\end{pgfscope}%
\begin{pgfscope}%
\definecolor{textcolor}{rgb}{0.000000,0.000000,0.000000}%
\pgfsetstrokecolor{textcolor}%
\pgfsetfillcolor{textcolor}%
\pgftext[x=4.558626in,y=1.847945in,left,base]{\color{textcolor}\rmfamily\fontsize{10.000000}{12.000000}\selectfont CALS}%
\end{pgfscope}%
\end{pgfpicture}%
\makeatother%
\endgroup%

%% file: data/ALS_v_CALS_MKL_real_255-281-25.pgf
\begingroup%
\makeatletter%
\begin{pgfpicture}%
\pgfpathrectangle{\pgfpointorigin}{\pgfqpoint{5.478070in}{3.000000in}}%
\pgfusepath{use as bounding box, clip}%
\begin{pgfscope}%
\pgfsetbuttcap%
\pgfsetmiterjoin%
\definecolor{currentfill}{rgb}{1.000000,1.000000,1.000000}%
\pgfsetfillcolor{currentfill}%
\pgfsetlinewidth{0.000000pt}%
\definecolor{currentstroke}{rgb}{1.000000,1.000000,1.000000}%
\pgfsetstrokecolor{currentstroke}%
\pgfsetdash{}{0pt}%
\pgfpathmoveto{\pgfqpoint{0.000000in}{0.000000in}}%
\pgfpathlineto{\pgfqpoint{5.478070in}{0.000000in}}%
\pgfpathlineto{\pgfqpoint{5.478070in}{3.000000in}}%
\pgfpathlineto{\pgfqpoint{0.000000in}{3.000000in}}%
\pgfpathclose%
\pgfusepath{fill}%
\end{pgfscope}%
\begin{pgfscope}%
\pgfsetbuttcap%
\pgfsetmiterjoin%
\definecolor{currentfill}{rgb}{1.000000,1.000000,1.000000}%
\pgfsetfillcolor{currentfill}%
\pgfsetlinewidth{0.000000pt}%
\definecolor{currentstroke}{rgb}{0.000000,0.000000,0.000000}%
\pgfsetstrokecolor{currentstroke}%
\pgfsetstrokeopacity{0.000000}%
\pgfsetdash{}{0pt}%
\pgfpathmoveto{\pgfqpoint{0.703890in}{0.370555in}}%
\pgfpathlineto{\pgfqpoint{5.328070in}{0.370555in}}%
\pgfpathlineto{\pgfqpoint{5.328070in}{2.651000in}}%
\pgfpathlineto{\pgfqpoint{0.703890in}{2.651000in}}%
\pgfpathclose%
\pgfusepath{fill}%
\end{pgfscope}%
\begin{pgfscope}%
\pgfpathrectangle{\pgfqpoint{0.703890in}{0.370555in}}{\pgfqpoint{4.624180in}{2.280445in}}%
\pgfusepath{clip}%
\pgfsetbuttcap%
\pgfsetmiterjoin%
\definecolor{currentfill}{rgb}{0.580392,0.403922,0.741176}%
\pgfsetfillcolor{currentfill}%
\pgfsetlinewidth{0.000000pt}%
\definecolor{currentstroke}{rgb}{0.000000,0.000000,0.000000}%
\pgfsetstrokecolor{currentstroke}%
\pgfsetstrokeopacity{0.000000}%
\pgfsetdash{}{0pt}%
\pgfpathmoveto{\pgfqpoint{1.089238in}{0.370555in}}%
\pgfpathlineto{\pgfqpoint{1.281912in}{0.370555in}}%
\pgfpathlineto{\pgfqpoint{1.281912in}{2.438986in}}%
\pgfpathlineto{\pgfqpoint{1.089238in}{2.438986in}}%
\pgfpathclose%
\pgfusepath{fill}%
\end{pgfscope}%
\begin{pgfscope}%
\pgfpathrectangle{\pgfqpoint{0.703890in}{0.370555in}}{\pgfqpoint{4.624180in}{2.280445in}}%
\pgfusepath{clip}%
\pgfsetbuttcap%
\pgfsetmiterjoin%
\definecolor{currentfill}{rgb}{0.580392,0.403922,0.741176}%
\pgfsetfillcolor{currentfill}%
\pgfsetlinewidth{0.000000pt}%
\definecolor{currentstroke}{rgb}{0.000000,0.000000,0.000000}%
\pgfsetstrokecolor{currentstroke}%
\pgfsetstrokeopacity{0.000000}%
\pgfsetdash{}{0pt}%
\pgfpathmoveto{\pgfqpoint{2.630631in}{0.370555in}}%
\pgfpathlineto{\pgfqpoint{2.823306in}{0.370555in}}%
\pgfpathlineto{\pgfqpoint{2.823306in}{1.520370in}}%
\pgfpathlineto{\pgfqpoint{2.630631in}{1.520370in}}%
\pgfpathclose%
\pgfusepath{fill}%
\end{pgfscope}%
\begin{pgfscope}%
\pgfpathrectangle{\pgfqpoint{0.703890in}{0.370555in}}{\pgfqpoint{4.624180in}{2.280445in}}%
\pgfusepath{clip}%
\pgfsetbuttcap%
\pgfsetmiterjoin%
\definecolor{currentfill}{rgb}{0.580392,0.403922,0.741176}%
\pgfsetfillcolor{currentfill}%
\pgfsetlinewidth{0.000000pt}%
\definecolor{currentstroke}{rgb}{0.000000,0.000000,0.000000}%
\pgfsetstrokecolor{currentstroke}%
\pgfsetstrokeopacity{0.000000}%
\pgfsetdash{}{0pt}%
\pgfpathmoveto{\pgfqpoint{4.172025in}{0.370555in}}%
\pgfpathlineto{\pgfqpoint{4.364699in}{0.370555in}}%
\pgfpathlineto{\pgfqpoint{4.364699in}{0.370555in}}%
\pgfpathlineto{\pgfqpoint{4.172025in}{0.370555in}}%
\pgfpathclose%
\pgfusepath{fill}%
\end{pgfscope}%
\begin{pgfscope}%
\pgfpathrectangle{\pgfqpoint{0.703890in}{0.370555in}}{\pgfqpoint{4.624180in}{2.280445in}}%
\pgfusepath{clip}%
\pgfsetbuttcap%
\pgfsetmiterjoin%
\definecolor{currentfill}{rgb}{1.000000,0.498039,0.054902}%
\pgfsetfillcolor{currentfill}%
\pgfsetlinewidth{0.000000pt}%
\definecolor{currentstroke}{rgb}{0.000000,0.000000,0.000000}%
\pgfsetstrokecolor{currentstroke}%
\pgfsetstrokeopacity{0.000000}%
\pgfsetdash{}{0pt}%
\pgfpathmoveto{\pgfqpoint{1.281912in}{0.370555in}}%
\pgfpathlineto{\pgfqpoint{1.474586in}{0.370555in}}%
\pgfpathlineto{\pgfqpoint{1.474586in}{0.863612in}}%
\pgfpathlineto{\pgfqpoint{1.281912in}{0.863612in}}%
\pgfpathclose%
\pgfusepath{fill}%
\end{pgfscope}%
\begin{pgfscope}%
\pgfpathrectangle{\pgfqpoint{0.703890in}{0.370555in}}{\pgfqpoint{4.624180in}{2.280445in}}%
\pgfusepath{clip}%
\pgfsetbuttcap%
\pgfsetmiterjoin%
\definecolor{currentfill}{rgb}{1.000000,0.498039,0.054902}%
\pgfsetfillcolor{currentfill}%
\pgfsetlinewidth{0.000000pt}%
\definecolor{currentstroke}{rgb}{0.000000,0.000000,0.000000}%
\pgfsetstrokecolor{currentstroke}%
\pgfsetstrokeopacity{0.000000}%
\pgfsetdash{}{0pt}%
\pgfpathmoveto{\pgfqpoint{2.823306in}{0.370555in}}%
\pgfpathlineto{\pgfqpoint{3.015980in}{0.370555in}}%
\pgfpathlineto{\pgfqpoint{3.015980in}{0.469646in}}%
\pgfpathlineto{\pgfqpoint{2.823306in}{0.469646in}}%
\pgfpathclose%
\pgfusepath{fill}%
\end{pgfscope}%
\begin{pgfscope}%
\pgfpathrectangle{\pgfqpoint{0.703890in}{0.370555in}}{\pgfqpoint{4.624180in}{2.280445in}}%
\pgfusepath{clip}%
\pgfsetbuttcap%
\pgfsetmiterjoin%
\definecolor{currentfill}{rgb}{1.000000,0.498039,0.054902}%
\pgfsetfillcolor{currentfill}%
\pgfsetlinewidth{0.000000pt}%
\definecolor{currentstroke}{rgb}{0.000000,0.000000,0.000000}%
\pgfsetstrokecolor{currentstroke}%
\pgfsetstrokeopacity{0.000000}%
\pgfsetdash{}{0pt}%
\pgfpathmoveto{\pgfqpoint{4.364699in}{0.370555in}}%
\pgfpathlineto{\pgfqpoint{4.557373in}{0.370555in}}%
\pgfpathlineto{\pgfqpoint{4.557373in}{0.440548in}}%
\pgfpathlineto{\pgfqpoint{4.364699in}{0.440548in}}%
\pgfpathclose%
\pgfusepath{fill}%
\end{pgfscope}%
\begin{pgfscope}%
\pgfpathrectangle{\pgfqpoint{0.703890in}{0.370555in}}{\pgfqpoint{4.624180in}{2.280445in}}%
\pgfusepath{clip}%
\pgfsetbuttcap%
\pgfsetmiterjoin%
\definecolor{currentfill}{rgb}{0.890196,0.466667,0.760784}%
\pgfsetfillcolor{currentfill}%
\pgfsetlinewidth{0.000000pt}%
\definecolor{currentstroke}{rgb}{0.000000,0.000000,0.000000}%
\pgfsetstrokecolor{currentstroke}%
\pgfsetstrokeopacity{0.000000}%
\pgfsetdash{}{0pt}%
\pgfpathmoveto{\pgfqpoint{1.474586in}{0.370555in}}%
\pgfpathlineto{\pgfqpoint{1.667261in}{0.370555in}}%
\pgfpathlineto{\pgfqpoint{1.667261in}{0.370555in}}%
\pgfpathlineto{\pgfqpoint{1.474586in}{0.370555in}}%
\pgfpathclose%
\pgfusepath{fill}%
\end{pgfscope}%
\begin{pgfscope}%
\pgfpathrectangle{\pgfqpoint{0.703890in}{0.370555in}}{\pgfqpoint{4.624180in}{2.280445in}}%
\pgfusepath{clip}%
\pgfsetbuttcap%
\pgfsetmiterjoin%
\definecolor{currentfill}{rgb}{0.890196,0.466667,0.760784}%
\pgfsetfillcolor{currentfill}%
\pgfsetlinewidth{0.000000pt}%
\definecolor{currentstroke}{rgb}{0.000000,0.000000,0.000000}%
\pgfsetstrokecolor{currentstroke}%
\pgfsetstrokeopacity{0.000000}%
\pgfsetdash{}{0pt}%
\pgfpathmoveto{\pgfqpoint{3.015980in}{0.370555in}}%
\pgfpathlineto{\pgfqpoint{3.208654in}{0.370555in}}%
\pgfpathlineto{\pgfqpoint{3.208654in}{0.420908in}}%
\pgfpathlineto{\pgfqpoint{3.015980in}{0.420908in}}%
\pgfpathclose%
\pgfusepath{fill}%
\end{pgfscope}%
\begin{pgfscope}%
\pgfpathrectangle{\pgfqpoint{0.703890in}{0.370555in}}{\pgfqpoint{4.624180in}{2.280445in}}%
\pgfusepath{clip}%
\pgfsetbuttcap%
\pgfsetmiterjoin%
\definecolor{currentfill}{rgb}{0.890196,0.466667,0.760784}%
\pgfsetfillcolor{currentfill}%
\pgfsetlinewidth{0.000000pt}%
\definecolor{currentstroke}{rgb}{0.000000,0.000000,0.000000}%
\pgfsetstrokecolor{currentstroke}%
\pgfsetstrokeopacity{0.000000}%
\pgfsetdash{}{0pt}%
\pgfpathmoveto{\pgfqpoint{4.557373in}{0.370555in}}%
\pgfpathlineto{\pgfqpoint{4.750047in}{0.370555in}}%
\pgfpathlineto{\pgfqpoint{4.750047in}{0.428602in}}%
\pgfpathlineto{\pgfqpoint{4.557373in}{0.428602in}}%
\pgfpathclose%
\pgfusepath{fill}%
\end{pgfscope}%
\begin{pgfscope}%
\pgfpathrectangle{\pgfqpoint{0.703890in}{0.370555in}}{\pgfqpoint{4.624180in}{2.280445in}}%
\pgfusepath{clip}%
\pgfsetbuttcap%
\pgfsetmiterjoin%
\definecolor{currentfill}{rgb}{0.121569,0.466667,0.705882}%
\pgfsetfillcolor{currentfill}%
\pgfsetlinewidth{0.000000pt}%
\definecolor{currentstroke}{rgb}{0.000000,0.000000,0.000000}%
\pgfsetstrokecolor{currentstroke}%
\pgfsetstrokeopacity{0.000000}%
\pgfsetdash{}{0pt}%
\pgfpathmoveto{\pgfqpoint{1.667261in}{0.370555in}}%
\pgfpathlineto{\pgfqpoint{1.859935in}{0.370555in}}%
\pgfpathlineto{\pgfqpoint{1.859935in}{0.633891in}}%
\pgfpathlineto{\pgfqpoint{1.667261in}{0.633891in}}%
\pgfpathclose%
\pgfusepath{fill}%
\end{pgfscope}%
\begin{pgfscope}%
\pgfpathrectangle{\pgfqpoint{0.703890in}{0.370555in}}{\pgfqpoint{4.624180in}{2.280445in}}%
\pgfusepath{clip}%
\pgfsetbuttcap%
\pgfsetmiterjoin%
\definecolor{currentfill}{rgb}{0.121569,0.466667,0.705882}%
\pgfsetfillcolor{currentfill}%
\pgfsetlinewidth{0.000000pt}%
\definecolor{currentstroke}{rgb}{0.000000,0.000000,0.000000}%
\pgfsetstrokecolor{currentstroke}%
\pgfsetstrokeopacity{0.000000}%
\pgfsetdash{}{0pt}%
\pgfpathmoveto{\pgfqpoint{3.208654in}{0.370555in}}%
\pgfpathlineto{\pgfqpoint{3.401328in}{0.370555in}}%
\pgfpathlineto{\pgfqpoint{3.401328in}{0.411503in}}%
\pgfpathlineto{\pgfqpoint{3.208654in}{0.411503in}}%
\pgfpathclose%
\pgfusepath{fill}%
\end{pgfscope}%
\begin{pgfscope}%
\pgfpathrectangle{\pgfqpoint{0.703890in}{0.370555in}}{\pgfqpoint{4.624180in}{2.280445in}}%
\pgfusepath{clip}%
\pgfsetbuttcap%
\pgfsetmiterjoin%
\definecolor{currentfill}{rgb}{0.121569,0.466667,0.705882}%
\pgfsetfillcolor{currentfill}%
\pgfsetlinewidth{0.000000pt}%
\definecolor{currentstroke}{rgb}{0.000000,0.000000,0.000000}%
\pgfsetstrokecolor{currentstroke}%
\pgfsetstrokeopacity{0.000000}%
\pgfsetdash{}{0pt}%
\pgfpathmoveto{\pgfqpoint{4.750047in}{0.370555in}}%
\pgfpathlineto{\pgfqpoint{4.942722in}{0.370555in}}%
\pgfpathlineto{\pgfqpoint{4.942722in}{0.384501in}}%
\pgfpathlineto{\pgfqpoint{4.750047in}{0.384501in}}%
\pgfpathclose%
\pgfusepath{fill}%
\end{pgfscope}%
\begin{pgfscope}%
\pgfsetbuttcap%
\pgfsetroundjoin%
\definecolor{currentfill}{rgb}{0.000000,0.000000,0.000000}%
\pgfsetfillcolor{currentfill}%
\pgfsetlinewidth{0.803000pt}%
\definecolor{currentstroke}{rgb}{0.000000,0.000000,0.000000}%
\pgfsetstrokecolor{currentstroke}%
\pgfsetdash{}{0pt}%
\pgfsys@defobject{currentmarker}{\pgfqpoint{0.000000in}{-0.048611in}}{\pgfqpoint{0.000000in}{0.000000in}}{%
\pgfpathmoveto{\pgfqpoint{0.000000in}{0.000000in}}%
\pgfpathlineto{\pgfqpoint{0.000000in}{-0.048611in}}%
\pgfusepath{stroke,fill}%
}%
\begin{pgfscope}%
\pgfsys@transformshift{1.474586in}{0.370555in}%
\pgfsys@useobject{currentmarker}{}%
\end{pgfscope}%
\end{pgfscope}%
\begin{pgfscope}%
\definecolor{textcolor}{rgb}{0.000000,0.000000,0.000000}%
\pgfsetstrokecolor{textcolor}%
\pgfsetfillcolor{textcolor}%
\pgftext[x=1.474586in,y=0.273333in,,top]{\color{textcolor}\rmfamily\fontsize{10.000000}{12.000000}\selectfont 1 thread}%
\end{pgfscope}%
\begin{pgfscope}%
\pgfsetbuttcap%
\pgfsetroundjoin%
\definecolor{currentfill}{rgb}{0.000000,0.000000,0.000000}%
\pgfsetfillcolor{currentfill}%
\pgfsetlinewidth{0.803000pt}%
\definecolor{currentstroke}{rgb}{0.000000,0.000000,0.000000}%
\pgfsetstrokecolor{currentstroke}%
\pgfsetdash{}{0pt}%
\pgfsys@defobject{currentmarker}{\pgfqpoint{0.000000in}{-0.048611in}}{\pgfqpoint{0.000000in}{0.000000in}}{%
\pgfpathmoveto{\pgfqpoint{0.000000in}{0.000000in}}%
\pgfpathlineto{\pgfqpoint{0.000000in}{-0.048611in}}%
\pgfusepath{stroke,fill}%
}%
\begin{pgfscope}%
\pgfsys@transformshift{3.015980in}{0.370555in}%
\pgfsys@useobject{currentmarker}{}%
\end{pgfscope}%
\end{pgfscope}%
\begin{pgfscope}%
\definecolor{textcolor}{rgb}{0.000000,0.000000,0.000000}%
\pgfsetstrokecolor{textcolor}%
\pgfsetfillcolor{textcolor}%
\pgftext[x=3.015980in,y=0.273333in,,top]{\color{textcolor}\rmfamily\fontsize{10.000000}{12.000000}\selectfont 24 threads}%
\end{pgfscope}%
\begin{pgfscope}%
\pgfsetbuttcap%
\pgfsetroundjoin%
\definecolor{currentfill}{rgb}{0.000000,0.000000,0.000000}%
\pgfsetfillcolor{currentfill}%
\pgfsetlinewidth{0.803000pt}%
\definecolor{currentstroke}{rgb}{0.000000,0.000000,0.000000}%
\pgfsetstrokecolor{currentstroke}%
\pgfsetdash{}{0pt}%
\pgfsys@defobject{currentmarker}{\pgfqpoint{0.000000in}{-0.048611in}}{\pgfqpoint{0.000000in}{0.000000in}}{%
\pgfpathmoveto{\pgfqpoint{0.000000in}{0.000000in}}%
\pgfpathlineto{\pgfqpoint{0.000000in}{-0.048611in}}%
\pgfusepath{stroke,fill}%
}%
\begin{pgfscope}%
\pgfsys@transformshift{4.557373in}{0.370555in}%
\pgfsys@useobject{currentmarker}{}%
\end{pgfscope}%
\end{pgfscope}%
\begin{pgfscope}%
\definecolor{textcolor}{rgb}{0.000000,0.000000,0.000000}%
\pgfsetstrokecolor{textcolor}%
\pgfsetfillcolor{textcolor}%
\pgftext[x=4.557373in,y=0.273333in,,top]{\color{textcolor}\rmfamily\fontsize{10.000000}{12.000000}\selectfont CUDA}%
\end{pgfscope}%
\begin{pgfscope}%
\pgfsetbuttcap%
\pgfsetroundjoin%
\definecolor{currentfill}{rgb}{0.000000,0.000000,0.000000}%
\pgfsetfillcolor{currentfill}%
\pgfsetlinewidth{0.803000pt}%
\definecolor{currentstroke}{rgb}{0.000000,0.000000,0.000000}%
\pgfsetstrokecolor{currentstroke}%
\pgfsetdash{}{0pt}%
\pgfsys@defobject{currentmarker}{\pgfqpoint{-0.048611in}{0.000000in}}{\pgfqpoint{-0.000000in}{0.000000in}}{%
\pgfpathmoveto{\pgfqpoint{-0.000000in}{0.000000in}}%
\pgfpathlineto{\pgfqpoint{-0.048611in}{0.000000in}}%
\pgfusepath{stroke,fill}%
}%
\begin{pgfscope}%
\pgfsys@transformshift{0.703890in}{0.370555in}%
\pgfsys@useobject{currentmarker}{}%
\end{pgfscope}%
\end{pgfscope}%
\begin{pgfscope}%
\definecolor{textcolor}{rgb}{0.000000,0.000000,0.000000}%
\pgfsetstrokecolor{textcolor}%
\pgfsetfillcolor{textcolor}%
\pgftext[x=0.537223in, y=0.322361in, left, base]{\color{textcolor}\rmfamily\fontsize{10.000000}{12.000000}\selectfont \(\displaystyle {0}\)}%
\end{pgfscope}%
\begin{pgfscope}%
\pgfsetbuttcap%
\pgfsetroundjoin%
\definecolor{currentfill}{rgb}{0.000000,0.000000,0.000000}%
\pgfsetfillcolor{currentfill}%
\pgfsetlinewidth{0.803000pt}%
\definecolor{currentstroke}{rgb}{0.000000,0.000000,0.000000}%
\pgfsetstrokecolor{currentstroke}%
\pgfsetdash{}{0pt}%
\pgfsys@defobject{currentmarker}{\pgfqpoint{-0.048611in}{0.000000in}}{\pgfqpoint{-0.000000in}{0.000000in}}{%
\pgfpathmoveto{\pgfqpoint{-0.000000in}{0.000000in}}%
\pgfpathlineto{\pgfqpoint{-0.048611in}{0.000000in}}%
\pgfusepath{stroke,fill}%
}%
\begin{pgfscope}%
\pgfsys@transformshift{0.703890in}{0.695026in}%
\pgfsys@useobject{currentmarker}{}%
\end{pgfscope}%
\end{pgfscope}%
\begin{pgfscope}%
\definecolor{textcolor}{rgb}{0.000000,0.000000,0.000000}%
\pgfsetstrokecolor{textcolor}%
\pgfsetfillcolor{textcolor}%
\pgftext[x=0.398333in, y=0.646831in, left, base]{\color{textcolor}\rmfamily\fontsize{10.000000}{12.000000}\selectfont \(\displaystyle {500}\)}%
\end{pgfscope}%
\begin{pgfscope}%
\pgfsetbuttcap%
\pgfsetroundjoin%
\definecolor{currentfill}{rgb}{0.000000,0.000000,0.000000}%
\pgfsetfillcolor{currentfill}%
\pgfsetlinewidth{0.803000pt}%
\definecolor{currentstroke}{rgb}{0.000000,0.000000,0.000000}%
\pgfsetstrokecolor{currentstroke}%
\pgfsetdash{}{0pt}%
\pgfsys@defobject{currentmarker}{\pgfqpoint{-0.048611in}{0.000000in}}{\pgfqpoint{-0.000000in}{0.000000in}}{%
\pgfpathmoveto{\pgfqpoint{-0.000000in}{0.000000in}}%
\pgfpathlineto{\pgfqpoint{-0.048611in}{0.000000in}}%
\pgfusepath{stroke,fill}%
}%
\begin{pgfscope}%
\pgfsys@transformshift{0.703890in}{1.019496in}%
\pgfsys@useobject{currentmarker}{}%
\end{pgfscope}%
\end{pgfscope}%
\begin{pgfscope}%
\definecolor{textcolor}{rgb}{0.000000,0.000000,0.000000}%
\pgfsetstrokecolor{textcolor}%
\pgfsetfillcolor{textcolor}%
\pgftext[x=0.328889in, y=0.971302in, left, base]{\color{textcolor}\rmfamily\fontsize{10.000000}{12.000000}\selectfont \(\displaystyle {1000}\)}%
\end{pgfscope}%
\begin{pgfscope}%
\pgfsetbuttcap%
\pgfsetroundjoin%
\definecolor{currentfill}{rgb}{0.000000,0.000000,0.000000}%
\pgfsetfillcolor{currentfill}%
\pgfsetlinewidth{0.803000pt}%
\definecolor{currentstroke}{rgb}{0.000000,0.000000,0.000000}%
\pgfsetstrokecolor{currentstroke}%
\pgfsetdash{}{0pt}%
\pgfsys@defobject{currentmarker}{\pgfqpoint{-0.048611in}{0.000000in}}{\pgfqpoint{-0.000000in}{0.000000in}}{%
\pgfpathmoveto{\pgfqpoint{-0.000000in}{0.000000in}}%
\pgfpathlineto{\pgfqpoint{-0.048611in}{0.000000in}}%
\pgfusepath{stroke,fill}%
}%
\begin{pgfscope}%
\pgfsys@transformshift{0.703890in}{1.343966in}%
\pgfsys@useobject{currentmarker}{}%
\end{pgfscope}%
\end{pgfscope}%
\begin{pgfscope}%
\definecolor{textcolor}{rgb}{0.000000,0.000000,0.000000}%
\pgfsetstrokecolor{textcolor}%
\pgfsetfillcolor{textcolor}%
\pgftext[x=0.328889in, y=1.295772in, left, base]{\color{textcolor}\rmfamily\fontsize{10.000000}{12.000000}\selectfont \(\displaystyle {1500}\)}%
\end{pgfscope}%
\begin{pgfscope}%
\pgfsetbuttcap%
\pgfsetroundjoin%
\definecolor{currentfill}{rgb}{0.000000,0.000000,0.000000}%
\pgfsetfillcolor{currentfill}%
\pgfsetlinewidth{0.803000pt}%
\definecolor{currentstroke}{rgb}{0.000000,0.000000,0.000000}%
\pgfsetstrokecolor{currentstroke}%
\pgfsetdash{}{0pt}%
\pgfsys@defobject{currentmarker}{\pgfqpoint{-0.048611in}{0.000000in}}{\pgfqpoint{-0.000000in}{0.000000in}}{%
\pgfpathmoveto{\pgfqpoint{-0.000000in}{0.000000in}}%
\pgfpathlineto{\pgfqpoint{-0.048611in}{0.000000in}}%
\pgfusepath{stroke,fill}%
}%
\begin{pgfscope}%
\pgfsys@transformshift{0.703890in}{1.668437in}%
\pgfsys@useobject{currentmarker}{}%
\end{pgfscope}%
\end{pgfscope}%
\begin{pgfscope}%
\definecolor{textcolor}{rgb}{0.000000,0.000000,0.000000}%
\pgfsetstrokecolor{textcolor}%
\pgfsetfillcolor{textcolor}%
\pgftext[x=0.328889in, y=1.620242in, left, base]{\color{textcolor}\rmfamily\fontsize{10.000000}{12.000000}\selectfont \(\displaystyle {2000}\)}%
\end{pgfscope}%
\begin{pgfscope}%
\pgfsetbuttcap%
\pgfsetroundjoin%
\definecolor{currentfill}{rgb}{0.000000,0.000000,0.000000}%
\pgfsetfillcolor{currentfill}%
\pgfsetlinewidth{0.803000pt}%
\definecolor{currentstroke}{rgb}{0.000000,0.000000,0.000000}%
\pgfsetstrokecolor{currentstroke}%
\pgfsetdash{}{0pt}%
\pgfsys@defobject{currentmarker}{\pgfqpoint{-0.048611in}{0.000000in}}{\pgfqpoint{-0.000000in}{0.000000in}}{%
\pgfpathmoveto{\pgfqpoint{-0.000000in}{0.000000in}}%
\pgfpathlineto{\pgfqpoint{-0.048611in}{0.000000in}}%
\pgfusepath{stroke,fill}%
}%
\begin{pgfscope}%
\pgfsys@transformshift{0.703890in}{1.992907in}%
\pgfsys@useobject{currentmarker}{}%
\end{pgfscope}%
\end{pgfscope}%
\begin{pgfscope}%
\definecolor{textcolor}{rgb}{0.000000,0.000000,0.000000}%
\pgfsetstrokecolor{textcolor}%
\pgfsetfillcolor{textcolor}%
\pgftext[x=0.328889in, y=1.944713in, left, base]{\color{textcolor}\rmfamily\fontsize{10.000000}{12.000000}\selectfont \(\displaystyle {2500}\)}%
\end{pgfscope}%
\begin{pgfscope}%
\pgfsetbuttcap%
\pgfsetroundjoin%
\definecolor{currentfill}{rgb}{0.000000,0.000000,0.000000}%
\pgfsetfillcolor{currentfill}%
\pgfsetlinewidth{0.803000pt}%
\definecolor{currentstroke}{rgb}{0.000000,0.000000,0.000000}%
\pgfsetstrokecolor{currentstroke}%
\pgfsetdash{}{0pt}%
\pgfsys@defobject{currentmarker}{\pgfqpoint{-0.048611in}{0.000000in}}{\pgfqpoint{-0.000000in}{0.000000in}}{%
\pgfpathmoveto{\pgfqpoint{-0.000000in}{0.000000in}}%
\pgfpathlineto{\pgfqpoint{-0.048611in}{0.000000in}}%
\pgfusepath{stroke,fill}%
}%
\begin{pgfscope}%
\pgfsys@transformshift{0.703890in}{2.317377in}%
\pgfsys@useobject{currentmarker}{}%
\end{pgfscope}%
\end{pgfscope}%
\begin{pgfscope}%
\definecolor{textcolor}{rgb}{0.000000,0.000000,0.000000}%
\pgfsetstrokecolor{textcolor}%
\pgfsetfillcolor{textcolor}%
\pgftext[x=0.328889in, y=2.269183in, left, base]{\color{textcolor}\rmfamily\fontsize{10.000000}{12.000000}\selectfont \(\displaystyle {3000}\)}%
\end{pgfscope}%
\begin{pgfscope}%
\pgfsetbuttcap%
\pgfsetroundjoin%
\definecolor{currentfill}{rgb}{0.000000,0.000000,0.000000}%
\pgfsetfillcolor{currentfill}%
\pgfsetlinewidth{0.803000pt}%
\definecolor{currentstroke}{rgb}{0.000000,0.000000,0.000000}%
\pgfsetstrokecolor{currentstroke}%
\pgfsetdash{}{0pt}%
\pgfsys@defobject{currentmarker}{\pgfqpoint{-0.048611in}{0.000000in}}{\pgfqpoint{-0.000000in}{0.000000in}}{%
\pgfpathmoveto{\pgfqpoint{-0.000000in}{0.000000in}}%
\pgfpathlineto{\pgfqpoint{-0.048611in}{0.000000in}}%
\pgfusepath{stroke,fill}%
}%
\begin{pgfscope}%
\pgfsys@transformshift{0.703890in}{2.641848in}%
\pgfsys@useobject{currentmarker}{}%
\end{pgfscope}%
\end{pgfscope}%
\begin{pgfscope}%
\definecolor{textcolor}{rgb}{0.000000,0.000000,0.000000}%
\pgfsetstrokecolor{textcolor}%
\pgfsetfillcolor{textcolor}%
\pgftext[x=0.328889in, y=2.593653in, left, base]{\color{textcolor}\rmfamily\fontsize{10.000000}{12.000000}\selectfont \(\displaystyle {3500}\)}%
\end{pgfscope}%
\begin{pgfscope}%
\definecolor{textcolor}{rgb}{0.000000,0.000000,0.000000}%
\pgfsetstrokecolor{textcolor}%
\pgfsetfillcolor{textcolor}%
\pgftext[x=0.273333in,y=1.510778in,,bottom,rotate=90.000000]{\color{textcolor}\rmfamily\fontsize{10.000000}{12.000000}\selectfont Time in seconds}%
\end{pgfscope}%
\begin{pgfscope}%
\pgfsetrectcap%
\pgfsetmiterjoin%
\pgfsetlinewidth{0.803000pt}%
\definecolor{currentstroke}{rgb}{0.000000,0.000000,0.000000}%
\pgfsetstrokecolor{currentstroke}%
\pgfsetdash{}{0pt}%
\pgfpathmoveto{\pgfqpoint{0.703890in}{0.370555in}}%
\pgfpathlineto{\pgfqpoint{0.703890in}{2.651000in}}%
\pgfusepath{stroke}%
\end{pgfscope}%
\begin{pgfscope}%
\pgfsetrectcap%
\pgfsetmiterjoin%
\pgfsetlinewidth{0.803000pt}%
\definecolor{currentstroke}{rgb}{0.000000,0.000000,0.000000}%
\pgfsetstrokecolor{currentstroke}%
\pgfsetdash{}{0pt}%
\pgfpathmoveto{\pgfqpoint{5.328070in}{0.370555in}}%
\pgfpathlineto{\pgfqpoint{5.328070in}{2.651000in}}%
\pgfusepath{stroke}%
\end{pgfscope}%
\begin{pgfscope}%
\pgfsetrectcap%
\pgfsetmiterjoin%
\pgfsetlinewidth{0.803000pt}%
\definecolor{currentstroke}{rgb}{0.000000,0.000000,0.000000}%
\pgfsetstrokecolor{currentstroke}%
\pgfsetdash{}{0pt}%
\pgfpathmoveto{\pgfqpoint{0.703890in}{0.370555in}}%
\pgfpathlineto{\pgfqpoint{5.328070in}{0.370555in}}%
\pgfusepath{stroke}%
\end{pgfscope}%
\begin{pgfscope}%
\pgfsetrectcap%
\pgfsetmiterjoin%
\pgfsetlinewidth{0.803000pt}%
\definecolor{currentstroke}{rgb}{0.000000,0.000000,0.000000}%
\pgfsetstrokecolor{currentstroke}%
\pgfsetdash{}{0pt}%
\pgfpathmoveto{\pgfqpoint{0.703890in}{2.651000in}}%
\pgfpathlineto{\pgfqpoint{5.328070in}{2.651000in}}%
\pgfusepath{stroke}%
\end{pgfscope}%
\begin{pgfscope}%
\definecolor{textcolor}{rgb}{0.000000,0.000000,0.000000}%
\pgfsetstrokecolor{textcolor}%
\pgfsetfillcolor{textcolor}%
\pgftext[x=1.185575in,y=2.452618in,,bottom]{\color{textcolor}\rmfamily\fontsize{10.000000}{12.000000}\selectfont 3187}%
\end{pgfscope}%
\begin{pgfscope}%
\definecolor{textcolor}{rgb}{0.000000,0.000000,0.000000}%
\pgfsetstrokecolor{textcolor}%
\pgfsetfillcolor{textcolor}%
\pgftext[x=2.726969in,y=1.534367in,,bottom]{\color{textcolor}\rmfamily\fontsize{10.000000}{12.000000}\selectfont 1772}%
\end{pgfscope}%
\begin{pgfscope}%
\definecolor{textcolor}{rgb}{0.000000,0.000000,0.000000}%
\pgfsetstrokecolor{textcolor}%
\pgfsetfillcolor{textcolor}%
\pgftext[x=1.378249in,y=0.877639in,,bottom]{\color{textcolor}\rmfamily\fontsize{10.000000}{12.000000}\selectfont 760}%
\end{pgfscope}%
\begin{pgfscope}%
\definecolor{textcolor}{rgb}{0.000000,0.000000,0.000000}%
\pgfsetstrokecolor{textcolor}%
\pgfsetfillcolor{textcolor}%
\pgftext[x=2.919643in,y=0.483732in,,bottom]{\color{textcolor}\rmfamily\fontsize{10.000000}{12.000000}\selectfont 153}%
\end{pgfscope}%
\begin{pgfscope}%
\definecolor{textcolor}{rgb}{0.000000,0.000000,0.000000}%
\pgfsetstrokecolor{textcolor}%
\pgfsetfillcolor{textcolor}%
\pgftext[x=4.461036in,y=0.454530in,,bottom]{\color{textcolor}\rmfamily\fontsize{10.000000}{12.000000}\selectfont 108}%
\end{pgfscope}%
\begin{pgfscope}%
\definecolor{textcolor}{rgb}{0.000000,0.000000,0.000000}%
\pgfsetstrokecolor{textcolor}%
\pgfsetfillcolor{textcolor}%
\pgftext[x=3.112317in,y=0.435062in,,bottom]{\color{textcolor}\rmfamily\fontsize{10.000000}{12.000000}\selectfont 78}%
\end{pgfscope}%
\begin{pgfscope}%
\definecolor{textcolor}{rgb}{0.000000,0.000000,0.000000}%
\pgfsetstrokecolor{textcolor}%
\pgfsetfillcolor{textcolor}%
\pgftext[x=4.653710in,y=0.442200in,,bottom]{\color{textcolor}\rmfamily\fontsize{10.000000}{12.000000}\selectfont 89}%
\end{pgfscope}%
\begin{pgfscope}%
\definecolor{textcolor}{rgb}{0.000000,0.000000,0.000000}%
\pgfsetstrokecolor{textcolor}%
\pgfsetfillcolor{textcolor}%
\pgftext[x=1.763598in,y=0.647914in,,bottom]{\color{textcolor}\rmfamily\fontsize{10.000000}{12.000000}\selectfont 406}%
\end{pgfscope}%
\begin{pgfscope}%
\definecolor{textcolor}{rgb}{0.000000,0.000000,0.000000}%
\pgfsetstrokecolor{textcolor}%
\pgfsetfillcolor{textcolor}%
\pgftext[x=3.304991in,y=0.425328in,,bottom]{\color{textcolor}\rmfamily\fontsize{10.000000}{12.000000}\selectfont 63}%
\end{pgfscope}%
\begin{pgfscope}%
\definecolor{textcolor}{rgb}{0.000000,0.000000,0.000000}%
\pgfsetstrokecolor{textcolor}%
\pgfsetfillcolor{textcolor}%
\pgftext[x=4.846385in,y=0.398072in,,bottom]{\color{textcolor}\rmfamily\fontsize{10.000000}{12.000000}\selectfont 21}%
\end{pgfscope}%
\begin{pgfscope}%
\definecolor{textcolor}{rgb}{0.000000,0.000000,0.000000}%
\pgfsetstrokecolor{textcolor}%
\pgfsetfillcolor{textcolor}%
\pgftext[x=3.015980in,y=2.734333in,,base]{\color{textcolor}\rmfamily\fontsize{12.000000}{14.400000}\selectfont 255\(\displaystyle \times\)281\(\displaystyle \times\)25}%
\end{pgfscope}%
\begin{pgfscope}%
\pgfsetbuttcap%
\pgfsetmiterjoin%
\definecolor{currentfill}{rgb}{1.000000,1.000000,1.000000}%
\pgfsetfillcolor{currentfill}%
\pgfsetfillopacity{0.800000}%
\pgfsetlinewidth{1.003750pt}%
\definecolor{currentstroke}{rgb}{0.800000,0.800000,0.800000}%
\pgfsetstrokecolor{currentstroke}%
\pgfsetstrokeopacity{0.800000}%
\pgfsetdash{}{0pt}%
\pgfpathmoveto{\pgfqpoint{4.141959in}{1.765445in}}%
\pgfpathlineto{\pgfqpoint{5.230848in}{1.765445in}}%
\pgfpathquadraticcurveto{\pgfqpoint{5.258626in}{1.765445in}}{\pgfqpoint{5.258626in}{1.793223in}}%
\pgfpathlineto{\pgfqpoint{5.258626in}{2.553778in}}%
\pgfpathquadraticcurveto{\pgfqpoint{5.258626in}{2.581556in}}{\pgfqpoint{5.230848in}{2.581556in}}%
\pgfpathlineto{\pgfqpoint{4.141959in}{2.581556in}}%
\pgfpathquadraticcurveto{\pgfqpoint{4.114181in}{2.581556in}}{\pgfqpoint{4.114181in}{2.553778in}}%
\pgfpathlineto{\pgfqpoint{4.114181in}{1.793223in}}%
\pgfpathquadraticcurveto{\pgfqpoint{4.114181in}{1.765445in}}{\pgfqpoint{4.141959in}{1.765445in}}%
\pgfpathclose%
\pgfusepath{stroke,fill}%
\end{pgfscope}%
\begin{pgfscope}%
\pgfsetbuttcap%
\pgfsetmiterjoin%
\definecolor{currentfill}{rgb}{0.580392,0.403922,0.741176}%
\pgfsetfillcolor{currentfill}%
\pgfsetlinewidth{0.000000pt}%
\definecolor{currentstroke}{rgb}{0.000000,0.000000,0.000000}%
\pgfsetstrokecolor{currentstroke}%
\pgfsetstrokeopacity{0.000000}%
\pgfsetdash{}{0pt}%
\pgfpathmoveto{\pgfqpoint{4.169737in}{2.428778in}}%
\pgfpathlineto{\pgfqpoint{4.447514in}{2.428778in}}%
\pgfpathlineto{\pgfqpoint{4.447514in}{2.526000in}}%
\pgfpathlineto{\pgfqpoint{4.169737in}{2.526000in}}%
\pgfpathclose%
\pgfusepath{fill}%
\end{pgfscope}%
\begin{pgfscope}%
\definecolor{textcolor}{rgb}{0.000000,0.000000,0.000000}%
\pgfsetstrokecolor{textcolor}%
\pgfsetfillcolor{textcolor}%
\pgftext[x=4.558626in,y=2.428778in,left,base]{\color{textcolor}\rmfamily\fontsize{10.000000}{12.000000}\selectfont TTB}%
\end{pgfscope}%
\begin{pgfscope}%
\pgfsetbuttcap%
\pgfsetmiterjoin%
\definecolor{currentfill}{rgb}{1.000000,0.498039,0.054902}%
\pgfsetfillcolor{currentfill}%
\pgfsetlinewidth{0.000000pt}%
\definecolor{currentstroke}{rgb}{0.000000,0.000000,0.000000}%
\pgfsetstrokecolor{currentstroke}%
\pgfsetstrokeopacity{0.000000}%
\pgfsetdash{}{0pt}%
\pgfpathmoveto{\pgfqpoint{4.169737in}{2.235167in}}%
\pgfpathlineto{\pgfqpoint{4.447514in}{2.235167in}}%
\pgfpathlineto{\pgfqpoint{4.447514in}{2.332389in}}%
\pgfpathlineto{\pgfqpoint{4.169737in}{2.332389in}}%
\pgfpathclose%
\pgfusepath{fill}%
\end{pgfscope}%
\begin{pgfscope}%
\definecolor{textcolor}{rgb}{0.000000,0.000000,0.000000}%
\pgfsetstrokecolor{textcolor}%
\pgfsetfillcolor{textcolor}%
\pgftext[x=4.558626in,y=2.235167in,left,base]{\color{textcolor}\rmfamily\fontsize{10.000000}{12.000000}\selectfont CP-ALS}%
\end{pgfscope}%
\begin{pgfscope}%
\pgfsetbuttcap%
\pgfsetmiterjoin%
\definecolor{currentfill}{rgb}{0.890196,0.466667,0.760784}%
\pgfsetfillcolor{currentfill}%
\pgfsetlinewidth{0.000000pt}%
\definecolor{currentstroke}{rgb}{0.000000,0.000000,0.000000}%
\pgfsetstrokecolor{currentstroke}%
\pgfsetstrokeopacity{0.000000}%
\pgfsetdash{}{0pt}%
\pgfpathmoveto{\pgfqpoint{4.169737in}{2.041556in}}%
\pgfpathlineto{\pgfqpoint{4.447514in}{2.041556in}}%
\pgfpathlineto{\pgfqpoint{4.447514in}{2.138778in}}%
\pgfpathlineto{\pgfqpoint{4.169737in}{2.138778in}}%
\pgfpathclose%
\pgfusepath{fill}%
\end{pgfscope}%
\begin{pgfscope}%
\definecolor{textcolor}{rgb}{0.000000,0.000000,0.000000}%
\pgfsetstrokecolor{textcolor}%
\pgfsetfillcolor{textcolor}%
\pgftext[x=4.558626in,y=2.041556in,left,base]{\color{textcolor}\rmfamily\fontsize{10.000000}{12.000000}\selectfont OMP ALS}%
\end{pgfscope}%
\begin{pgfscope}%
\pgfsetbuttcap%
\pgfsetmiterjoin%
\definecolor{currentfill}{rgb}{0.121569,0.466667,0.705882}%
\pgfsetfillcolor{currentfill}%
\pgfsetlinewidth{0.000000pt}%
\definecolor{currentstroke}{rgb}{0.000000,0.000000,0.000000}%
\pgfsetstrokecolor{currentstroke}%
\pgfsetstrokeopacity{0.000000}%
\pgfsetdash{}{0pt}%
\pgfpathmoveto{\pgfqpoint{4.169737in}{1.847945in}}%
\pgfpathlineto{\pgfqpoint{4.447514in}{1.847945in}}%
\pgfpathlineto{\pgfqpoint{4.447514in}{1.945167in}}%
\pgfpathlineto{\pgfqpoint{4.169737in}{1.945167in}}%
\pgfpathclose%
\pgfusepath{fill}%
\end{pgfscope}%
\begin{pgfscope}%
\definecolor{textcolor}{rgb}{0.000000,0.000000,0.000000}%
\pgfsetstrokecolor{textcolor}%
\pgfsetfillcolor{textcolor}%
\pgftext[x=4.558626in,y=1.847945in,left,base]{\color{textcolor}\rmfamily\fontsize{10.000000}{12.000000}\selectfont CALS}%
\end{pgfscope}%
\end{pgfpicture}%
\makeatother%
\endgroup%

%% file: data/ALS_v_CALS_MKL_real_299-301-41.pgf
\begingroup%
\makeatletter%
\begin{pgfpicture}%
\pgfpathrectangle{\pgfpointorigin}{\pgfqpoint{5.478070in}{3.000000in}}%
\pgfusepath{use as bounding box, clip}%
\begin{pgfscope}%
\pgfsetbuttcap%
\pgfsetmiterjoin%
\definecolor{currentfill}{rgb}{1.000000,1.000000,1.000000}%
\pgfsetfillcolor{currentfill}%
\pgfsetlinewidth{0.000000pt}%
\definecolor{currentstroke}{rgb}{1.000000,1.000000,1.000000}%
\pgfsetstrokecolor{currentstroke}%
\pgfsetdash{}{0pt}%
\pgfpathmoveto{\pgfqpoint{0.000000in}{0.000000in}}%
\pgfpathlineto{\pgfqpoint{5.478070in}{0.000000in}}%
\pgfpathlineto{\pgfqpoint{5.478070in}{3.000000in}}%
\pgfpathlineto{\pgfqpoint{0.000000in}{3.000000in}}%
\pgfpathclose%
\pgfusepath{fill}%
\end{pgfscope}%
\begin{pgfscope}%
\pgfsetbuttcap%
\pgfsetmiterjoin%
\definecolor{currentfill}{rgb}{1.000000,1.000000,1.000000}%
\pgfsetfillcolor{currentfill}%
\pgfsetlinewidth{0.000000pt}%
\definecolor{currentstroke}{rgb}{0.000000,0.000000,0.000000}%
\pgfsetstrokecolor{currentstroke}%
\pgfsetstrokeopacity{0.000000}%
\pgfsetdash{}{0pt}%
\pgfpathmoveto{\pgfqpoint{0.703890in}{0.370555in}}%
\pgfpathlineto{\pgfqpoint{5.328070in}{0.370555in}}%
\pgfpathlineto{\pgfqpoint{5.328070in}{2.651000in}}%
\pgfpathlineto{\pgfqpoint{0.703890in}{2.651000in}}%
\pgfpathclose%
\pgfusepath{fill}%
\end{pgfscope}%
\begin{pgfscope}%
\pgfpathrectangle{\pgfqpoint{0.703890in}{0.370555in}}{\pgfqpoint{4.624180in}{2.280445in}}%
\pgfusepath{clip}%
\pgfsetbuttcap%
\pgfsetmiterjoin%
\definecolor{currentfill}{rgb}{0.580392,0.403922,0.741176}%
\pgfsetfillcolor{currentfill}%
\pgfsetlinewidth{0.000000pt}%
\definecolor{currentstroke}{rgb}{0.000000,0.000000,0.000000}%
\pgfsetstrokecolor{currentstroke}%
\pgfsetstrokeopacity{0.000000}%
\pgfsetdash{}{0pt}%
\pgfpathmoveto{\pgfqpoint{1.089238in}{0.370555in}}%
\pgfpathlineto{\pgfqpoint{1.281912in}{0.370555in}}%
\pgfpathlineto{\pgfqpoint{1.281912in}{2.438986in}}%
\pgfpathlineto{\pgfqpoint{1.089238in}{2.438986in}}%
\pgfpathclose%
\pgfusepath{fill}%
\end{pgfscope}%
\begin{pgfscope}%
\pgfpathrectangle{\pgfqpoint{0.703890in}{0.370555in}}{\pgfqpoint{4.624180in}{2.280445in}}%
\pgfusepath{clip}%
\pgfsetbuttcap%
\pgfsetmiterjoin%
\definecolor{currentfill}{rgb}{0.580392,0.403922,0.741176}%
\pgfsetfillcolor{currentfill}%
\pgfsetlinewidth{0.000000pt}%
\definecolor{currentstroke}{rgb}{0.000000,0.000000,0.000000}%
\pgfsetstrokecolor{currentstroke}%
\pgfsetstrokeopacity{0.000000}%
\pgfsetdash{}{0pt}%
\pgfpathmoveto{\pgfqpoint{2.630631in}{0.370555in}}%
\pgfpathlineto{\pgfqpoint{2.823306in}{0.370555in}}%
\pgfpathlineto{\pgfqpoint{2.823306in}{1.075197in}}%
\pgfpathlineto{\pgfqpoint{2.630631in}{1.075197in}}%
\pgfpathclose%
\pgfusepath{fill}%
\end{pgfscope}%
\begin{pgfscope}%
\pgfpathrectangle{\pgfqpoint{0.703890in}{0.370555in}}{\pgfqpoint{4.624180in}{2.280445in}}%
\pgfusepath{clip}%
\pgfsetbuttcap%
\pgfsetmiterjoin%
\definecolor{currentfill}{rgb}{0.580392,0.403922,0.741176}%
\pgfsetfillcolor{currentfill}%
\pgfsetlinewidth{0.000000pt}%
\definecolor{currentstroke}{rgb}{0.000000,0.000000,0.000000}%
\pgfsetstrokecolor{currentstroke}%
\pgfsetstrokeopacity{0.000000}%
\pgfsetdash{}{0pt}%
\pgfpathmoveto{\pgfqpoint{4.172025in}{0.370555in}}%
\pgfpathlineto{\pgfqpoint{4.364699in}{0.370555in}}%
\pgfpathlineto{\pgfqpoint{4.364699in}{0.370555in}}%
\pgfpathlineto{\pgfqpoint{4.172025in}{0.370555in}}%
\pgfpathclose%
\pgfusepath{fill}%
\end{pgfscope}%
\begin{pgfscope}%
\pgfpathrectangle{\pgfqpoint{0.703890in}{0.370555in}}{\pgfqpoint{4.624180in}{2.280445in}}%
\pgfusepath{clip}%
\pgfsetbuttcap%
\pgfsetmiterjoin%
\definecolor{currentfill}{rgb}{1.000000,0.498039,0.054902}%
\pgfsetfillcolor{currentfill}%
\pgfsetlinewidth{0.000000pt}%
\definecolor{currentstroke}{rgb}{0.000000,0.000000,0.000000}%
\pgfsetstrokecolor{currentstroke}%
\pgfsetstrokeopacity{0.000000}%
\pgfsetdash{}{0pt}%
\pgfpathmoveto{\pgfqpoint{1.281912in}{0.370555in}}%
\pgfpathlineto{\pgfqpoint{1.474586in}{0.370555in}}%
\pgfpathlineto{\pgfqpoint{1.474586in}{1.130249in}}%
\pgfpathlineto{\pgfqpoint{1.281912in}{1.130249in}}%
\pgfpathclose%
\pgfusepath{fill}%
\end{pgfscope}%
\begin{pgfscope}%
\pgfpathrectangle{\pgfqpoint{0.703890in}{0.370555in}}{\pgfqpoint{4.624180in}{2.280445in}}%
\pgfusepath{clip}%
\pgfsetbuttcap%
\pgfsetmiterjoin%
\definecolor{currentfill}{rgb}{1.000000,0.498039,0.054902}%
\pgfsetfillcolor{currentfill}%
\pgfsetlinewidth{0.000000pt}%
\definecolor{currentstroke}{rgb}{0.000000,0.000000,0.000000}%
\pgfsetstrokecolor{currentstroke}%
\pgfsetstrokeopacity{0.000000}%
\pgfsetdash{}{0pt}%
\pgfpathmoveto{\pgfqpoint{2.823306in}{0.370555in}}%
\pgfpathlineto{\pgfqpoint{3.015980in}{0.370555in}}%
\pgfpathlineto{\pgfqpoint{3.015980in}{0.485065in}}%
\pgfpathlineto{\pgfqpoint{2.823306in}{0.485065in}}%
\pgfpathclose%
\pgfusepath{fill}%
\end{pgfscope}%
\begin{pgfscope}%
\pgfpathrectangle{\pgfqpoint{0.703890in}{0.370555in}}{\pgfqpoint{4.624180in}{2.280445in}}%
\pgfusepath{clip}%
\pgfsetbuttcap%
\pgfsetmiterjoin%
\definecolor{currentfill}{rgb}{1.000000,0.498039,0.054902}%
\pgfsetfillcolor{currentfill}%
\pgfsetlinewidth{0.000000pt}%
\definecolor{currentstroke}{rgb}{0.000000,0.000000,0.000000}%
\pgfsetstrokecolor{currentstroke}%
\pgfsetstrokeopacity{0.000000}%
\pgfsetdash{}{0pt}%
\pgfpathmoveto{\pgfqpoint{4.364699in}{0.370555in}}%
\pgfpathlineto{\pgfqpoint{4.557373in}{0.370555in}}%
\pgfpathlineto{\pgfqpoint{4.557373in}{0.416533in}}%
\pgfpathlineto{\pgfqpoint{4.364699in}{0.416533in}}%
\pgfpathclose%
\pgfusepath{fill}%
\end{pgfscope}%
\begin{pgfscope}%
\pgfpathrectangle{\pgfqpoint{0.703890in}{0.370555in}}{\pgfqpoint{4.624180in}{2.280445in}}%
\pgfusepath{clip}%
\pgfsetbuttcap%
\pgfsetmiterjoin%
\definecolor{currentfill}{rgb}{0.890196,0.466667,0.760784}%
\pgfsetfillcolor{currentfill}%
\pgfsetlinewidth{0.000000pt}%
\definecolor{currentstroke}{rgb}{0.000000,0.000000,0.000000}%
\pgfsetstrokecolor{currentstroke}%
\pgfsetstrokeopacity{0.000000}%
\pgfsetdash{}{0pt}%
\pgfpathmoveto{\pgfqpoint{1.474586in}{0.370555in}}%
\pgfpathlineto{\pgfqpoint{1.667261in}{0.370555in}}%
\pgfpathlineto{\pgfqpoint{1.667261in}{0.370555in}}%
\pgfpathlineto{\pgfqpoint{1.474586in}{0.370555in}}%
\pgfpathclose%
\pgfusepath{fill}%
\end{pgfscope}%
\begin{pgfscope}%
\pgfpathrectangle{\pgfqpoint{0.703890in}{0.370555in}}{\pgfqpoint{4.624180in}{2.280445in}}%
\pgfusepath{clip}%
\pgfsetbuttcap%
\pgfsetmiterjoin%
\definecolor{currentfill}{rgb}{0.890196,0.466667,0.760784}%
\pgfsetfillcolor{currentfill}%
\pgfsetlinewidth{0.000000pt}%
\definecolor{currentstroke}{rgb}{0.000000,0.000000,0.000000}%
\pgfsetstrokecolor{currentstroke}%
\pgfsetstrokeopacity{0.000000}%
\pgfsetdash{}{0pt}%
\pgfpathmoveto{\pgfqpoint{3.015980in}{0.370555in}}%
\pgfpathlineto{\pgfqpoint{3.208654in}{0.370555in}}%
\pgfpathlineto{\pgfqpoint{3.208654in}{0.439656in}}%
\pgfpathlineto{\pgfqpoint{3.015980in}{0.439656in}}%
\pgfpathclose%
\pgfusepath{fill}%
\end{pgfscope}%
\begin{pgfscope}%
\pgfpathrectangle{\pgfqpoint{0.703890in}{0.370555in}}{\pgfqpoint{4.624180in}{2.280445in}}%
\pgfusepath{clip}%
\pgfsetbuttcap%
\pgfsetmiterjoin%
\definecolor{currentfill}{rgb}{0.890196,0.466667,0.760784}%
\pgfsetfillcolor{currentfill}%
\pgfsetlinewidth{0.000000pt}%
\definecolor{currentstroke}{rgb}{0.000000,0.000000,0.000000}%
\pgfsetstrokecolor{currentstroke}%
\pgfsetstrokeopacity{0.000000}%
\pgfsetdash{}{0pt}%
\pgfpathmoveto{\pgfqpoint{4.557373in}{0.370555in}}%
\pgfpathlineto{\pgfqpoint{4.750047in}{0.370555in}}%
\pgfpathlineto{\pgfqpoint{4.750047in}{0.407567in}}%
\pgfpathlineto{\pgfqpoint{4.557373in}{0.407567in}}%
\pgfpathclose%
\pgfusepath{fill}%
\end{pgfscope}%
\begin{pgfscope}%
\pgfpathrectangle{\pgfqpoint{0.703890in}{0.370555in}}{\pgfqpoint{4.624180in}{2.280445in}}%
\pgfusepath{clip}%
\pgfsetbuttcap%
\pgfsetmiterjoin%
\definecolor{currentfill}{rgb}{0.121569,0.466667,0.705882}%
\pgfsetfillcolor{currentfill}%
\pgfsetlinewidth{0.000000pt}%
\definecolor{currentstroke}{rgb}{0.000000,0.000000,0.000000}%
\pgfsetstrokecolor{currentstroke}%
\pgfsetstrokeopacity{0.000000}%
\pgfsetdash{}{0pt}%
\pgfpathmoveto{\pgfqpoint{1.667261in}{0.370555in}}%
\pgfpathlineto{\pgfqpoint{1.859935in}{0.370555in}}%
\pgfpathlineto{\pgfqpoint{1.859935in}{0.624596in}}%
\pgfpathlineto{\pgfqpoint{1.667261in}{0.624596in}}%
\pgfpathclose%
\pgfusepath{fill}%
\end{pgfscope}%
\begin{pgfscope}%
\pgfpathrectangle{\pgfqpoint{0.703890in}{0.370555in}}{\pgfqpoint{4.624180in}{2.280445in}}%
\pgfusepath{clip}%
\pgfsetbuttcap%
\pgfsetmiterjoin%
\definecolor{currentfill}{rgb}{0.121569,0.466667,0.705882}%
\pgfsetfillcolor{currentfill}%
\pgfsetlinewidth{0.000000pt}%
\definecolor{currentstroke}{rgb}{0.000000,0.000000,0.000000}%
\pgfsetstrokecolor{currentstroke}%
\pgfsetstrokeopacity{0.000000}%
\pgfsetdash{}{0pt}%
\pgfpathmoveto{\pgfqpoint{3.208654in}{0.370555in}}%
\pgfpathlineto{\pgfqpoint{3.401328in}{0.370555in}}%
\pgfpathlineto{\pgfqpoint{3.401328in}{0.400120in}}%
\pgfpathlineto{\pgfqpoint{3.208654in}{0.400120in}}%
\pgfpathclose%
\pgfusepath{fill}%
\end{pgfscope}%
\begin{pgfscope}%
\pgfpathrectangle{\pgfqpoint{0.703890in}{0.370555in}}{\pgfqpoint{4.624180in}{2.280445in}}%
\pgfusepath{clip}%
\pgfsetbuttcap%
\pgfsetmiterjoin%
\definecolor{currentfill}{rgb}{0.121569,0.466667,0.705882}%
\pgfsetfillcolor{currentfill}%
\pgfsetlinewidth{0.000000pt}%
\definecolor{currentstroke}{rgb}{0.000000,0.000000,0.000000}%
\pgfsetstrokecolor{currentstroke}%
\pgfsetstrokeopacity{0.000000}%
\pgfsetdash{}{0pt}%
\pgfpathmoveto{\pgfqpoint{4.750047in}{0.370555in}}%
\pgfpathlineto{\pgfqpoint{4.942722in}{0.370555in}}%
\pgfpathlineto{\pgfqpoint{4.942722in}{0.380391in}}%
\pgfpathlineto{\pgfqpoint{4.750047in}{0.380391in}}%
\pgfpathclose%
\pgfusepath{fill}%
\end{pgfscope}%
\begin{pgfscope}%
\pgfsetbuttcap%
\pgfsetroundjoin%
\definecolor{currentfill}{rgb}{0.000000,0.000000,0.000000}%
\pgfsetfillcolor{currentfill}%
\pgfsetlinewidth{0.803000pt}%
\definecolor{currentstroke}{rgb}{0.000000,0.000000,0.000000}%
\pgfsetstrokecolor{currentstroke}%
\pgfsetdash{}{0pt}%
\pgfsys@defobject{currentmarker}{\pgfqpoint{0.000000in}{-0.048611in}}{\pgfqpoint{0.000000in}{0.000000in}}{%
\pgfpathmoveto{\pgfqpoint{0.000000in}{0.000000in}}%
\pgfpathlineto{\pgfqpoint{0.000000in}{-0.048611in}}%
\pgfusepath{stroke,fill}%
}%
\begin{pgfscope}%
\pgfsys@transformshift{1.474586in}{0.370555in}%
\pgfsys@useobject{currentmarker}{}%
\end{pgfscope}%
\end{pgfscope}%
\begin{pgfscope}%
\definecolor{textcolor}{rgb}{0.000000,0.000000,0.000000}%
\pgfsetstrokecolor{textcolor}%
\pgfsetfillcolor{textcolor}%
\pgftext[x=1.474586in,y=0.273333in,,top]{\color{textcolor}\rmfamily\fontsize{10.000000}{12.000000}\selectfont 1 thread}%
\end{pgfscope}%
\begin{pgfscope}%
\pgfsetbuttcap%
\pgfsetroundjoin%
\definecolor{currentfill}{rgb}{0.000000,0.000000,0.000000}%
\pgfsetfillcolor{currentfill}%
\pgfsetlinewidth{0.803000pt}%
\definecolor{currentstroke}{rgb}{0.000000,0.000000,0.000000}%
\pgfsetstrokecolor{currentstroke}%
\pgfsetdash{}{0pt}%
\pgfsys@defobject{currentmarker}{\pgfqpoint{0.000000in}{-0.048611in}}{\pgfqpoint{0.000000in}{0.000000in}}{%
\pgfpathmoveto{\pgfqpoint{0.000000in}{0.000000in}}%
\pgfpathlineto{\pgfqpoint{0.000000in}{-0.048611in}}%
\pgfusepath{stroke,fill}%
}%
\begin{pgfscope}%
\pgfsys@transformshift{3.015980in}{0.370555in}%
\pgfsys@useobject{currentmarker}{}%
\end{pgfscope}%
\end{pgfscope}%
\begin{pgfscope}%
\definecolor{textcolor}{rgb}{0.000000,0.000000,0.000000}%
\pgfsetstrokecolor{textcolor}%
\pgfsetfillcolor{textcolor}%
\pgftext[x=3.015980in,y=0.273333in,,top]{\color{textcolor}\rmfamily\fontsize{10.000000}{12.000000}\selectfont 24 threads}%
\end{pgfscope}%
\begin{pgfscope}%
\pgfsetbuttcap%
\pgfsetroundjoin%
\definecolor{currentfill}{rgb}{0.000000,0.000000,0.000000}%
\pgfsetfillcolor{currentfill}%
\pgfsetlinewidth{0.803000pt}%
\definecolor{currentstroke}{rgb}{0.000000,0.000000,0.000000}%
\pgfsetstrokecolor{currentstroke}%
\pgfsetdash{}{0pt}%
\pgfsys@defobject{currentmarker}{\pgfqpoint{0.000000in}{-0.048611in}}{\pgfqpoint{0.000000in}{0.000000in}}{%
\pgfpathmoveto{\pgfqpoint{0.000000in}{0.000000in}}%
\pgfpathlineto{\pgfqpoint{0.000000in}{-0.048611in}}%
\pgfusepath{stroke,fill}%
}%
\begin{pgfscope}%
\pgfsys@transformshift{4.557373in}{0.370555in}%
\pgfsys@useobject{currentmarker}{}%
\end{pgfscope}%
\end{pgfscope}%
\begin{pgfscope}%
\definecolor{textcolor}{rgb}{0.000000,0.000000,0.000000}%
\pgfsetstrokecolor{textcolor}%
\pgfsetfillcolor{textcolor}%
\pgftext[x=4.557373in,y=0.273333in,,top]{\color{textcolor}\rmfamily\fontsize{10.000000}{12.000000}\selectfont CUDA}%
\end{pgfscope}%
\begin{pgfscope}%
\pgfsetbuttcap%
\pgfsetroundjoin%
\definecolor{currentfill}{rgb}{0.000000,0.000000,0.000000}%
\pgfsetfillcolor{currentfill}%
\pgfsetlinewidth{0.803000pt}%
\definecolor{currentstroke}{rgb}{0.000000,0.000000,0.000000}%
\pgfsetstrokecolor{currentstroke}%
\pgfsetdash{}{0pt}%
\pgfsys@defobject{currentmarker}{\pgfqpoint{-0.048611in}{0.000000in}}{\pgfqpoint{-0.000000in}{0.000000in}}{%
\pgfpathmoveto{\pgfqpoint{-0.000000in}{0.000000in}}%
\pgfpathlineto{\pgfqpoint{-0.048611in}{0.000000in}}%
\pgfusepath{stroke,fill}%
}%
\begin{pgfscope}%
\pgfsys@transformshift{0.703890in}{0.370555in}%
\pgfsys@useobject{currentmarker}{}%
\end{pgfscope}%
\end{pgfscope}%
\begin{pgfscope}%
\definecolor{textcolor}{rgb}{0.000000,0.000000,0.000000}%
\pgfsetstrokecolor{textcolor}%
\pgfsetfillcolor{textcolor}%
\pgftext[x=0.537223in, y=0.322361in, left, base]{\color{textcolor}\rmfamily\fontsize{10.000000}{12.000000}\selectfont \(\displaystyle {0}\)}%
\end{pgfscope}%
\begin{pgfscope}%
\pgfsetbuttcap%
\pgfsetroundjoin%
\definecolor{currentfill}{rgb}{0.000000,0.000000,0.000000}%
\pgfsetfillcolor{currentfill}%
\pgfsetlinewidth{0.803000pt}%
\definecolor{currentstroke}{rgb}{0.000000,0.000000,0.000000}%
\pgfsetstrokecolor{currentstroke}%
\pgfsetdash{}{0pt}%
\pgfsys@defobject{currentmarker}{\pgfqpoint{-0.048611in}{0.000000in}}{\pgfqpoint{-0.000000in}{0.000000in}}{%
\pgfpathmoveto{\pgfqpoint{-0.000000in}{0.000000in}}%
\pgfpathlineto{\pgfqpoint{-0.048611in}{0.000000in}}%
\pgfusepath{stroke,fill}%
}%
\begin{pgfscope}%
\pgfsys@transformshift{0.703890in}{0.701699in}%
\pgfsys@useobject{currentmarker}{}%
\end{pgfscope}%
\end{pgfscope}%
\begin{pgfscope}%
\definecolor{textcolor}{rgb}{0.000000,0.000000,0.000000}%
\pgfsetstrokecolor{textcolor}%
\pgfsetfillcolor{textcolor}%
\pgftext[x=0.328889in, y=0.653504in, left, base]{\color{textcolor}\rmfamily\fontsize{10.000000}{12.000000}\selectfont \(\displaystyle {1000}\)}%
\end{pgfscope}%
\begin{pgfscope}%
\pgfsetbuttcap%
\pgfsetroundjoin%
\definecolor{currentfill}{rgb}{0.000000,0.000000,0.000000}%
\pgfsetfillcolor{currentfill}%
\pgfsetlinewidth{0.803000pt}%
\definecolor{currentstroke}{rgb}{0.000000,0.000000,0.000000}%
\pgfsetstrokecolor{currentstroke}%
\pgfsetdash{}{0pt}%
\pgfsys@defobject{currentmarker}{\pgfqpoint{-0.048611in}{0.000000in}}{\pgfqpoint{-0.000000in}{0.000000in}}{%
\pgfpathmoveto{\pgfqpoint{-0.000000in}{0.000000in}}%
\pgfpathlineto{\pgfqpoint{-0.048611in}{0.000000in}}%
\pgfusepath{stroke,fill}%
}%
\begin{pgfscope}%
\pgfsys@transformshift{0.703890in}{1.032842in}%
\pgfsys@useobject{currentmarker}{}%
\end{pgfscope}%
\end{pgfscope}%
\begin{pgfscope}%
\definecolor{textcolor}{rgb}{0.000000,0.000000,0.000000}%
\pgfsetstrokecolor{textcolor}%
\pgfsetfillcolor{textcolor}%
\pgftext[x=0.328889in, y=0.984648in, left, base]{\color{textcolor}\rmfamily\fontsize{10.000000}{12.000000}\selectfont \(\displaystyle {2000}\)}%
\end{pgfscope}%
\begin{pgfscope}%
\pgfsetbuttcap%
\pgfsetroundjoin%
\definecolor{currentfill}{rgb}{0.000000,0.000000,0.000000}%
\pgfsetfillcolor{currentfill}%
\pgfsetlinewidth{0.803000pt}%
\definecolor{currentstroke}{rgb}{0.000000,0.000000,0.000000}%
\pgfsetstrokecolor{currentstroke}%
\pgfsetdash{}{0pt}%
\pgfsys@defobject{currentmarker}{\pgfqpoint{-0.048611in}{0.000000in}}{\pgfqpoint{-0.000000in}{0.000000in}}{%
\pgfpathmoveto{\pgfqpoint{-0.000000in}{0.000000in}}%
\pgfpathlineto{\pgfqpoint{-0.048611in}{0.000000in}}%
\pgfusepath{stroke,fill}%
}%
\begin{pgfscope}%
\pgfsys@transformshift{0.703890in}{1.363986in}%
\pgfsys@useobject{currentmarker}{}%
\end{pgfscope}%
\end{pgfscope}%
\begin{pgfscope}%
\definecolor{textcolor}{rgb}{0.000000,0.000000,0.000000}%
\pgfsetstrokecolor{textcolor}%
\pgfsetfillcolor{textcolor}%
\pgftext[x=0.328889in, y=1.315791in, left, base]{\color{textcolor}\rmfamily\fontsize{10.000000}{12.000000}\selectfont \(\displaystyle {3000}\)}%
\end{pgfscope}%
\begin{pgfscope}%
\pgfsetbuttcap%
\pgfsetroundjoin%
\definecolor{currentfill}{rgb}{0.000000,0.000000,0.000000}%
\pgfsetfillcolor{currentfill}%
\pgfsetlinewidth{0.803000pt}%
\definecolor{currentstroke}{rgb}{0.000000,0.000000,0.000000}%
\pgfsetstrokecolor{currentstroke}%
\pgfsetdash{}{0pt}%
\pgfsys@defobject{currentmarker}{\pgfqpoint{-0.048611in}{0.000000in}}{\pgfqpoint{-0.000000in}{0.000000in}}{%
\pgfpathmoveto{\pgfqpoint{-0.000000in}{0.000000in}}%
\pgfpathlineto{\pgfqpoint{-0.048611in}{0.000000in}}%
\pgfusepath{stroke,fill}%
}%
\begin{pgfscope}%
\pgfsys@transformshift{0.703890in}{1.695129in}%
\pgfsys@useobject{currentmarker}{}%
\end{pgfscope}%
\end{pgfscope}%
\begin{pgfscope}%
\definecolor{textcolor}{rgb}{0.000000,0.000000,0.000000}%
\pgfsetstrokecolor{textcolor}%
\pgfsetfillcolor{textcolor}%
\pgftext[x=0.328889in, y=1.646935in, left, base]{\color{textcolor}\rmfamily\fontsize{10.000000}{12.000000}\selectfont \(\displaystyle {4000}\)}%
\end{pgfscope}%
\begin{pgfscope}%
\pgfsetbuttcap%
\pgfsetroundjoin%
\definecolor{currentfill}{rgb}{0.000000,0.000000,0.000000}%
\pgfsetfillcolor{currentfill}%
\pgfsetlinewidth{0.803000pt}%
\definecolor{currentstroke}{rgb}{0.000000,0.000000,0.000000}%
\pgfsetstrokecolor{currentstroke}%
\pgfsetdash{}{0pt}%
\pgfsys@defobject{currentmarker}{\pgfqpoint{-0.048611in}{0.000000in}}{\pgfqpoint{-0.000000in}{0.000000in}}{%
\pgfpathmoveto{\pgfqpoint{-0.000000in}{0.000000in}}%
\pgfpathlineto{\pgfqpoint{-0.048611in}{0.000000in}}%
\pgfusepath{stroke,fill}%
}%
\begin{pgfscope}%
\pgfsys@transformshift{0.703890in}{2.026273in}%
\pgfsys@useobject{currentmarker}{}%
\end{pgfscope}%
\end{pgfscope}%
\begin{pgfscope}%
\definecolor{textcolor}{rgb}{0.000000,0.000000,0.000000}%
\pgfsetstrokecolor{textcolor}%
\pgfsetfillcolor{textcolor}%
\pgftext[x=0.328889in, y=1.978078in, left, base]{\color{textcolor}\rmfamily\fontsize{10.000000}{12.000000}\selectfont \(\displaystyle {5000}\)}%
\end{pgfscope}%
\begin{pgfscope}%
\pgfsetbuttcap%
\pgfsetroundjoin%
\definecolor{currentfill}{rgb}{0.000000,0.000000,0.000000}%
\pgfsetfillcolor{currentfill}%
\pgfsetlinewidth{0.803000pt}%
\definecolor{currentstroke}{rgb}{0.000000,0.000000,0.000000}%
\pgfsetstrokecolor{currentstroke}%
\pgfsetdash{}{0pt}%
\pgfsys@defobject{currentmarker}{\pgfqpoint{-0.048611in}{0.000000in}}{\pgfqpoint{-0.000000in}{0.000000in}}{%
\pgfpathmoveto{\pgfqpoint{-0.000000in}{0.000000in}}%
\pgfpathlineto{\pgfqpoint{-0.048611in}{0.000000in}}%
\pgfusepath{stroke,fill}%
}%
\begin{pgfscope}%
\pgfsys@transformshift{0.703890in}{2.357416in}%
\pgfsys@useobject{currentmarker}{}%
\end{pgfscope}%
\end{pgfscope}%
\begin{pgfscope}%
\definecolor{textcolor}{rgb}{0.000000,0.000000,0.000000}%
\pgfsetstrokecolor{textcolor}%
\pgfsetfillcolor{textcolor}%
\pgftext[x=0.328889in, y=2.309221in, left, base]{\color{textcolor}\rmfamily\fontsize{10.000000}{12.000000}\selectfont \(\displaystyle {6000}\)}%
\end{pgfscope}%
\begin{pgfscope}%
\definecolor{textcolor}{rgb}{0.000000,0.000000,0.000000}%
\pgfsetstrokecolor{textcolor}%
\pgfsetfillcolor{textcolor}%
\pgftext[x=0.273333in,y=1.510778in,,bottom,rotate=90.000000]{\color{textcolor}\rmfamily\fontsize{10.000000}{12.000000}\selectfont Time in seconds}%
\end{pgfscope}%
\begin{pgfscope}%
\pgfsetrectcap%
\pgfsetmiterjoin%
\pgfsetlinewidth{0.803000pt}%
\definecolor{currentstroke}{rgb}{0.000000,0.000000,0.000000}%
\pgfsetstrokecolor{currentstroke}%
\pgfsetdash{}{0pt}%
\pgfpathmoveto{\pgfqpoint{0.703890in}{0.370555in}}%
\pgfpathlineto{\pgfqpoint{0.703890in}{2.651000in}}%
\pgfusepath{stroke}%
\end{pgfscope}%
\begin{pgfscope}%
\pgfsetrectcap%
\pgfsetmiterjoin%
\pgfsetlinewidth{0.803000pt}%
\definecolor{currentstroke}{rgb}{0.000000,0.000000,0.000000}%
\pgfsetstrokecolor{currentstroke}%
\pgfsetdash{}{0pt}%
\pgfpathmoveto{\pgfqpoint{5.328070in}{0.370555in}}%
\pgfpathlineto{\pgfqpoint{5.328070in}{2.651000in}}%
\pgfusepath{stroke}%
\end{pgfscope}%
\begin{pgfscope}%
\pgfsetrectcap%
\pgfsetmiterjoin%
\pgfsetlinewidth{0.803000pt}%
\definecolor{currentstroke}{rgb}{0.000000,0.000000,0.000000}%
\pgfsetstrokecolor{currentstroke}%
\pgfsetdash{}{0pt}%
\pgfpathmoveto{\pgfqpoint{0.703890in}{0.370555in}}%
\pgfpathlineto{\pgfqpoint{5.328070in}{0.370555in}}%
\pgfusepath{stroke}%
\end{pgfscope}%
\begin{pgfscope}%
\pgfsetrectcap%
\pgfsetmiterjoin%
\pgfsetlinewidth{0.803000pt}%
\definecolor{currentstroke}{rgb}{0.000000,0.000000,0.000000}%
\pgfsetstrokecolor{currentstroke}%
\pgfsetdash{}{0pt}%
\pgfpathmoveto{\pgfqpoint{0.703890in}{2.651000in}}%
\pgfpathlineto{\pgfqpoint{5.328070in}{2.651000in}}%
\pgfusepath{stroke}%
\end{pgfscope}%
\begin{pgfscope}%
\definecolor{textcolor}{rgb}{0.000000,0.000000,0.000000}%
\pgfsetstrokecolor{textcolor}%
\pgfsetfillcolor{textcolor}%
\pgftext[x=1.185575in,y=2.452766in,,bottom]{\color{textcolor}\rmfamily\fontsize{10.000000}{12.000000}\selectfont 6246}%
\end{pgfscope}%
\begin{pgfscope}%
\definecolor{textcolor}{rgb}{0.000000,0.000000,0.000000}%
\pgfsetstrokecolor{textcolor}%
\pgfsetfillcolor{textcolor}%
\pgftext[x=2.726969in,y=1.089118in,,bottom]{\color{textcolor}\rmfamily\fontsize{10.000000}{12.000000}\selectfont 2128}%
\end{pgfscope}%
\begin{pgfscope}%
\definecolor{textcolor}{rgb}{0.000000,0.000000,0.000000}%
\pgfsetstrokecolor{textcolor}%
\pgfsetfillcolor{textcolor}%
\pgftext[x=1.378249in,y=1.144087in,,bottom]{\color{textcolor}\rmfamily\fontsize{10.000000}{12.000000}\selectfont 2294}%
\end{pgfscope}%
\begin{pgfscope}%
\definecolor{textcolor}{rgb}{0.000000,0.000000,0.000000}%
\pgfsetstrokecolor{textcolor}%
\pgfsetfillcolor{textcolor}%
\pgftext[x=2.919643in,y=0.499020in,,bottom]{\color{textcolor}\rmfamily\fontsize{10.000000}{12.000000}\selectfont 346}%
\end{pgfscope}%
\begin{pgfscope}%
\definecolor{textcolor}{rgb}{0.000000,0.000000,0.000000}%
\pgfsetstrokecolor{textcolor}%
\pgfsetfillcolor{textcolor}%
\pgftext[x=4.461036in,y=0.430473in,,bottom]{\color{textcolor}\rmfamily\fontsize{10.000000}{12.000000}\selectfont 139}%
\end{pgfscope}%
\begin{pgfscope}%
\definecolor{textcolor}{rgb}{0.000000,0.000000,0.000000}%
\pgfsetstrokecolor{textcolor}%
\pgfsetfillcolor{textcolor}%
\pgftext[x=3.112317in,y=0.453653in,,bottom]{\color{textcolor}\rmfamily\fontsize{10.000000}{12.000000}\selectfont 209}%
\end{pgfscope}%
\begin{pgfscope}%
\definecolor{textcolor}{rgb}{0.000000,0.000000,0.000000}%
\pgfsetstrokecolor{textcolor}%
\pgfsetfillcolor{textcolor}%
\pgftext[x=4.653710in,y=0.421532in,,bottom]{\color{textcolor}\rmfamily\fontsize{10.000000}{12.000000}\selectfont 112}%
\end{pgfscope}%
\begin{pgfscope}%
\definecolor{textcolor}{rgb}{0.000000,0.000000,0.000000}%
\pgfsetstrokecolor{textcolor}%
\pgfsetfillcolor{textcolor}%
\pgftext[x=1.763598in,y=0.638431in,,bottom]{\color{textcolor}\rmfamily\fontsize{10.000000}{12.000000}\selectfont 767}%
\end{pgfscope}%
\begin{pgfscope}%
\definecolor{textcolor}{rgb}{0.000000,0.000000,0.000000}%
\pgfsetstrokecolor{textcolor}%
\pgfsetfillcolor{textcolor}%
\pgftext[x=3.304991in,y=0.413916in,,bottom]{\color{textcolor}\rmfamily\fontsize{10.000000}{12.000000}\selectfont 89}%
\end{pgfscope}%
\begin{pgfscope}%
\definecolor{textcolor}{rgb}{0.000000,0.000000,0.000000}%
\pgfsetstrokecolor{textcolor}%
\pgfsetfillcolor{textcolor}%
\pgftext[x=4.846385in,y=0.394379in,,bottom]{\color{textcolor}\rmfamily\fontsize{10.000000}{12.000000}\selectfont 30}%
\end{pgfscope}%
\begin{pgfscope}%
\definecolor{textcolor}{rgb}{0.000000,0.000000,0.000000}%
\pgfsetstrokecolor{textcolor}%
\pgfsetfillcolor{textcolor}%
\pgftext[x=3.015980in,y=2.734333in,,base]{\color{textcolor}\rmfamily\fontsize{12.000000}{14.400000}\selectfont 299\(\displaystyle \times\)301\(\displaystyle \times\)41}%
\end{pgfscope}%
\begin{pgfscope}%
\pgfsetbuttcap%
\pgfsetmiterjoin%
\definecolor{currentfill}{rgb}{1.000000,1.000000,1.000000}%
\pgfsetfillcolor{currentfill}%
\pgfsetfillopacity{0.800000}%
\pgfsetlinewidth{1.003750pt}%
\definecolor{currentstroke}{rgb}{0.800000,0.800000,0.800000}%
\pgfsetstrokecolor{currentstroke}%
\pgfsetstrokeopacity{0.800000}%
\pgfsetdash{}{0pt}%
\pgfpathmoveto{\pgfqpoint{4.141959in}{1.765445in}}%
\pgfpathlineto{\pgfqpoint{5.230848in}{1.765445in}}%
\pgfpathquadraticcurveto{\pgfqpoint{5.258626in}{1.765445in}}{\pgfqpoint{5.258626in}{1.793223in}}%
\pgfpathlineto{\pgfqpoint{5.258626in}{2.553778in}}%
\pgfpathquadraticcurveto{\pgfqpoint{5.258626in}{2.581556in}}{\pgfqpoint{5.230848in}{2.581556in}}%
\pgfpathlineto{\pgfqpoint{4.141959in}{2.581556in}}%
\pgfpathquadraticcurveto{\pgfqpoint{4.114181in}{2.581556in}}{\pgfqpoint{4.114181in}{2.553778in}}%
\pgfpathlineto{\pgfqpoint{4.114181in}{1.793223in}}%
\pgfpathquadraticcurveto{\pgfqpoint{4.114181in}{1.765445in}}{\pgfqpoint{4.141959in}{1.765445in}}%
\pgfpathclose%
\pgfusepath{stroke,fill}%
\end{pgfscope}%
\begin{pgfscope}%
\pgfsetbuttcap%
\pgfsetmiterjoin%
\definecolor{currentfill}{rgb}{0.580392,0.403922,0.741176}%
\pgfsetfillcolor{currentfill}%
\pgfsetlinewidth{0.000000pt}%
\definecolor{currentstroke}{rgb}{0.000000,0.000000,0.000000}%
\pgfsetstrokecolor{currentstroke}%
\pgfsetstrokeopacity{0.000000}%
\pgfsetdash{}{0pt}%
\pgfpathmoveto{\pgfqpoint{4.169737in}{2.428778in}}%
\pgfpathlineto{\pgfqpoint{4.447514in}{2.428778in}}%
\pgfpathlineto{\pgfqpoint{4.447514in}{2.526000in}}%
\pgfpathlineto{\pgfqpoint{4.169737in}{2.526000in}}%
\pgfpathclose%
\pgfusepath{fill}%
\end{pgfscope}%
\begin{pgfscope}%
\definecolor{textcolor}{rgb}{0.000000,0.000000,0.000000}%
\pgfsetstrokecolor{textcolor}%
\pgfsetfillcolor{textcolor}%
\pgftext[x=4.558626in,y=2.428778in,left,base]{\color{textcolor}\rmfamily\fontsize{10.000000}{12.000000}\selectfont TTB}%
\end{pgfscope}%
\begin{pgfscope}%
\pgfsetbuttcap%
\pgfsetmiterjoin%
\definecolor{currentfill}{rgb}{1.000000,0.498039,0.054902}%
\pgfsetfillcolor{currentfill}%
\pgfsetlinewidth{0.000000pt}%
\definecolor{currentstroke}{rgb}{0.000000,0.000000,0.000000}%
\pgfsetstrokecolor{currentstroke}%
\pgfsetstrokeopacity{0.000000}%
\pgfsetdash{}{0pt}%
\pgfpathmoveto{\pgfqpoint{4.169737in}{2.235167in}}%
\pgfpathlineto{\pgfqpoint{4.447514in}{2.235167in}}%
\pgfpathlineto{\pgfqpoint{4.447514in}{2.332389in}}%
\pgfpathlineto{\pgfqpoint{4.169737in}{2.332389in}}%
\pgfpathclose%
\pgfusepath{fill}%
\end{pgfscope}%
\begin{pgfscope}%
\definecolor{textcolor}{rgb}{0.000000,0.000000,0.000000}%
\pgfsetstrokecolor{textcolor}%
\pgfsetfillcolor{textcolor}%
\pgftext[x=4.558626in,y=2.235167in,left,base]{\color{textcolor}\rmfamily\fontsize{10.000000}{12.000000}\selectfont CP-ALS}%
\end{pgfscope}%
\begin{pgfscope}%
\pgfsetbuttcap%
\pgfsetmiterjoin%
\definecolor{currentfill}{rgb}{0.890196,0.466667,0.760784}%
\pgfsetfillcolor{currentfill}%
\pgfsetlinewidth{0.000000pt}%
\definecolor{currentstroke}{rgb}{0.000000,0.000000,0.000000}%
\pgfsetstrokecolor{currentstroke}%
\pgfsetstrokeopacity{0.000000}%
\pgfsetdash{}{0pt}%
\pgfpathmoveto{\pgfqpoint{4.169737in}{2.041556in}}%
\pgfpathlineto{\pgfqpoint{4.447514in}{2.041556in}}%
\pgfpathlineto{\pgfqpoint{4.447514in}{2.138778in}}%
\pgfpathlineto{\pgfqpoint{4.169737in}{2.138778in}}%
\pgfpathclose%
\pgfusepath{fill}%
\end{pgfscope}%
\begin{pgfscope}%
\definecolor{textcolor}{rgb}{0.000000,0.000000,0.000000}%
\pgfsetstrokecolor{textcolor}%
\pgfsetfillcolor{textcolor}%
\pgftext[x=4.558626in,y=2.041556in,left,base]{\color{textcolor}\rmfamily\fontsize{10.000000}{12.000000}\selectfont OMP ALS}%
\end{pgfscope}%
\begin{pgfscope}%
\pgfsetbuttcap%
\pgfsetmiterjoin%
\definecolor{currentfill}{rgb}{0.121569,0.466667,0.705882}%
\pgfsetfillcolor{currentfill}%
\pgfsetlinewidth{0.000000pt}%
\definecolor{currentstroke}{rgb}{0.000000,0.000000,0.000000}%
\pgfsetstrokecolor{currentstroke}%
\pgfsetstrokeopacity{0.000000}%
\pgfsetdash{}{0pt}%
\pgfpathmoveto{\pgfqpoint{4.169737in}{1.847945in}}%
\pgfpathlineto{\pgfqpoint{4.447514in}{1.847945in}}%
\pgfpathlineto{\pgfqpoint{4.447514in}{1.945167in}}%
\pgfpathlineto{\pgfqpoint{4.169737in}{1.945167in}}%
\pgfpathclose%
\pgfusepath{fill}%
\end{pgfscope}%
\begin{pgfscope}%
\definecolor{textcolor}{rgb}{0.000000,0.000000,0.000000}%
\pgfsetstrokecolor{textcolor}%
\pgfsetfillcolor{textcolor}%
\pgftext[x=4.558626in,y=1.847945in,left,base]{\color{textcolor}\rmfamily\fontsize{10.000000}{12.000000}\selectfont CALS}%
\end{pgfscope}%
\end{pgfpicture}%
\makeatother%
\endgroup%

%% file: data/CUDA_v_CALS_MKL.pgf
\begingroup%
\makeatletter%
\begin{pgfpicture}%
\pgfpathrectangle{\pgfpointorigin}{\pgfqpoint{5.478070in}{3.000000in}}%
\pgfusepath{use as bounding box, clip}%
\begin{pgfscope}%
\pgfsetbuttcap%
\pgfsetmiterjoin%
\definecolor{currentfill}{rgb}{1.000000,1.000000,1.000000}%
\pgfsetfillcolor{currentfill}%
\pgfsetlinewidth{0.000000pt}%
\definecolor{currentstroke}{rgb}{1.000000,1.000000,1.000000}%
\pgfsetstrokecolor{currentstroke}%
\pgfsetdash{}{0pt}%
\pgfpathmoveto{\pgfqpoint{0.000000in}{0.000000in}}%
\pgfpathlineto{\pgfqpoint{5.478070in}{0.000000in}}%
\pgfpathlineto{\pgfqpoint{5.478070in}{3.000000in}}%
\pgfpathlineto{\pgfqpoint{0.000000in}{3.000000in}}%
\pgfpathclose%
\pgfusepath{fill}%
\end{pgfscope}%
\begin{pgfscope}%
\pgfsetbuttcap%
\pgfsetmiterjoin%
\definecolor{currentfill}{rgb}{1.000000,1.000000,1.000000}%
\pgfsetfillcolor{currentfill}%
\pgfsetlinewidth{0.000000pt}%
\definecolor{currentstroke}{rgb}{0.000000,0.000000,0.000000}%
\pgfsetstrokecolor{currentstroke}%
\pgfsetstrokeopacity{0.000000}%
\pgfsetdash{}{0pt}%
\pgfpathmoveto{\pgfqpoint{0.634445in}{0.370555in}}%
\pgfpathlineto{\pgfqpoint{5.328070in}{0.370555in}}%
\pgfpathlineto{\pgfqpoint{5.328070in}{2.850000in}}%
\pgfpathlineto{\pgfqpoint{0.634445in}{2.850000in}}%
\pgfpathclose%
\pgfusepath{fill}%
\end{pgfscope}%
\begin{pgfscope}%
\pgfpathrectangle{\pgfqpoint{0.634445in}{0.370555in}}{\pgfqpoint{4.693625in}{2.479445in}}%
\pgfusepath{clip}%
\pgfsetbuttcap%
\pgfsetmiterjoin%
\definecolor{currentfill}{rgb}{0.372549,0.827451,0.372549}%
\pgfsetfillcolor{currentfill}%
\pgfsetlinewidth{0.000000pt}%
\definecolor{currentstroke}{rgb}{0.000000,0.000000,0.000000}%
\pgfsetstrokecolor{currentstroke}%
\pgfsetstrokeopacity{0.000000}%
\pgfsetdash{}{0pt}%
\pgfpathmoveto{\pgfqpoint{1.025580in}{0.370555in}}%
\pgfpathlineto{\pgfqpoint{1.286337in}{0.370555in}}%
\pgfpathlineto{\pgfqpoint{1.286337in}{0.611746in}}%
\pgfpathlineto{\pgfqpoint{1.025580in}{0.611746in}}%
\pgfpathclose%
\pgfusepath{fill}%
\end{pgfscope}%
\begin{pgfscope}%
\pgfpathrectangle{\pgfqpoint{0.634445in}{0.370555in}}{\pgfqpoint{4.693625in}{2.479445in}}%
\pgfusepath{clip}%
\pgfsetbuttcap%
\pgfsetmiterjoin%
\definecolor{currentfill}{rgb}{0.372549,0.827451,0.372549}%
\pgfsetfillcolor{currentfill}%
\pgfsetlinewidth{0.000000pt}%
\definecolor{currentstroke}{rgb}{0.000000,0.000000,0.000000}%
\pgfsetstrokecolor{currentstroke}%
\pgfsetstrokeopacity{0.000000}%
\pgfsetdash{}{0pt}%
\pgfpathmoveto{\pgfqpoint{2.590122in}{0.370555in}}%
\pgfpathlineto{\pgfqpoint{2.850879in}{0.370555in}}%
\pgfpathlineto{\pgfqpoint{2.850879in}{0.805994in}}%
\pgfpathlineto{\pgfqpoint{2.590122in}{0.805994in}}%
\pgfpathclose%
\pgfusepath{fill}%
\end{pgfscope}%
\begin{pgfscope}%
\pgfpathrectangle{\pgfqpoint{0.634445in}{0.370555in}}{\pgfqpoint{4.693625in}{2.479445in}}%
\pgfusepath{clip}%
\pgfsetbuttcap%
\pgfsetmiterjoin%
\definecolor{currentfill}{rgb}{0.372549,0.827451,0.372549}%
\pgfsetfillcolor{currentfill}%
\pgfsetlinewidth{0.000000pt}%
\definecolor{currentstroke}{rgb}{0.000000,0.000000,0.000000}%
\pgfsetstrokecolor{currentstroke}%
\pgfsetstrokeopacity{0.000000}%
\pgfsetdash{}{0pt}%
\pgfpathmoveto{\pgfqpoint{4.154664in}{0.370555in}}%
\pgfpathlineto{\pgfqpoint{4.415421in}{0.370555in}}%
\pgfpathlineto{\pgfqpoint{4.415421in}{2.619485in}}%
\pgfpathlineto{\pgfqpoint{4.154664in}{2.619485in}}%
\pgfpathclose%
\pgfusepath{fill}%
\end{pgfscope}%
\begin{pgfscope}%
\pgfpathrectangle{\pgfqpoint{0.634445in}{0.370555in}}{\pgfqpoint{4.693625in}{2.479445in}}%
\pgfusepath{clip}%
\pgfsetbuttcap%
\pgfsetmiterjoin%
\definecolor{currentfill}{rgb}{0.172549,0.627451,0.172549}%
\pgfsetfillcolor{currentfill}%
\pgfsetlinewidth{0.000000pt}%
\definecolor{currentstroke}{rgb}{0.000000,0.000000,0.000000}%
\pgfsetstrokecolor{currentstroke}%
\pgfsetstrokeopacity{0.000000}%
\pgfsetdash{}{0pt}%
\pgfpathmoveto{\pgfqpoint{1.286337in}{0.370555in}}%
\pgfpathlineto{\pgfqpoint{1.547094in}{0.370555in}}%
\pgfpathlineto{\pgfqpoint{1.547094in}{0.559797in}}%
\pgfpathlineto{\pgfqpoint{1.286337in}{0.559797in}}%
\pgfpathclose%
\pgfusepath{fill}%
\end{pgfscope}%
\begin{pgfscope}%
\pgfpathrectangle{\pgfqpoint{0.634445in}{0.370555in}}{\pgfqpoint{4.693625in}{2.479445in}}%
\pgfusepath{clip}%
\pgfsetbuttcap%
\pgfsetmiterjoin%
\definecolor{currentfill}{rgb}{0.172549,0.627451,0.172549}%
\pgfsetfillcolor{currentfill}%
\pgfsetlinewidth{0.000000pt}%
\definecolor{currentstroke}{rgb}{0.000000,0.000000,0.000000}%
\pgfsetstrokecolor{currentstroke}%
\pgfsetstrokeopacity{0.000000}%
\pgfsetdash{}{0pt}%
\pgfpathmoveto{\pgfqpoint{2.850879in}{0.370555in}}%
\pgfpathlineto{\pgfqpoint{3.111636in}{0.370555in}}%
\pgfpathlineto{\pgfqpoint{3.111636in}{0.719632in}}%
\pgfpathlineto{\pgfqpoint{2.850879in}{0.719632in}}%
\pgfpathclose%
\pgfusepath{fill}%
\end{pgfscope}%
\begin{pgfscope}%
\pgfpathrectangle{\pgfqpoint{0.634445in}{0.370555in}}{\pgfqpoint{4.693625in}{2.479445in}}%
\pgfusepath{clip}%
\pgfsetbuttcap%
\pgfsetmiterjoin%
\definecolor{currentfill}{rgb}{0.172549,0.627451,0.172549}%
\pgfsetfillcolor{currentfill}%
\pgfsetlinewidth{0.000000pt}%
\definecolor{currentstroke}{rgb}{0.000000,0.000000,0.000000}%
\pgfsetstrokecolor{currentstroke}%
\pgfsetstrokeopacity{0.000000}%
\pgfsetdash{}{0pt}%
\pgfpathmoveto{\pgfqpoint{4.415421in}{0.370555in}}%
\pgfpathlineto{\pgfqpoint{4.676178in}{0.370555in}}%
\pgfpathlineto{\pgfqpoint{4.676178in}{1.045434in}}%
\pgfpathlineto{\pgfqpoint{4.415421in}{1.045434in}}%
\pgfpathclose%
\pgfusepath{fill}%
\end{pgfscope}%
\begin{pgfscope}%
\pgfpathrectangle{\pgfqpoint{0.634445in}{0.370555in}}{\pgfqpoint{4.693625in}{2.479445in}}%
\pgfusepath{clip}%
\pgfsetbuttcap%
\pgfsetmiterjoin%
\definecolor{currentfill}{rgb}{0.086275,0.313725,0.086275}%
\pgfsetfillcolor{currentfill}%
\pgfsetlinewidth{0.000000pt}%
\definecolor{currentstroke}{rgb}{0.000000,0.000000,0.000000}%
\pgfsetstrokecolor{currentstroke}%
\pgfsetstrokeopacity{0.000000}%
\pgfsetdash{}{0pt}%
\pgfpathmoveto{\pgfqpoint{1.547094in}{0.370555in}}%
\pgfpathlineto{\pgfqpoint{1.807851in}{0.370555in}}%
\pgfpathlineto{\pgfqpoint{1.807851in}{0.441742in}}%
\pgfpathlineto{\pgfqpoint{1.547094in}{0.441742in}}%
\pgfpathclose%
\pgfusepath{fill}%
\end{pgfscope}%
\begin{pgfscope}%
\pgfpathrectangle{\pgfqpoint{0.634445in}{0.370555in}}{\pgfqpoint{4.693625in}{2.479445in}}%
\pgfusepath{clip}%
\pgfsetbuttcap%
\pgfsetmiterjoin%
\definecolor{currentfill}{rgb}{0.086275,0.313725,0.086275}%
\pgfsetfillcolor{currentfill}%
\pgfsetlinewidth{0.000000pt}%
\definecolor{currentstroke}{rgb}{0.000000,0.000000,0.000000}%
\pgfsetstrokecolor{currentstroke}%
\pgfsetstrokeopacity{0.000000}%
\pgfsetdash{}{0pt}%
\pgfpathmoveto{\pgfqpoint{3.111636in}{0.370555in}}%
\pgfpathlineto{\pgfqpoint{3.372393in}{0.370555in}}%
\pgfpathlineto{\pgfqpoint{3.372393in}{0.503656in}}%
\pgfpathlineto{\pgfqpoint{3.111636in}{0.503656in}}%
\pgfpathclose%
\pgfusepath{fill}%
\end{pgfscope}%
\begin{pgfscope}%
\pgfpathrectangle{\pgfqpoint{0.634445in}{0.370555in}}{\pgfqpoint{4.693625in}{2.479445in}}%
\pgfusepath{clip}%
\pgfsetbuttcap%
\pgfsetmiterjoin%
\definecolor{currentfill}{rgb}{0.086275,0.313725,0.086275}%
\pgfsetfillcolor{currentfill}%
\pgfsetlinewidth{0.000000pt}%
\definecolor{currentstroke}{rgb}{0.000000,0.000000,0.000000}%
\pgfsetstrokecolor{currentstroke}%
\pgfsetstrokeopacity{0.000000}%
\pgfsetdash{}{0pt}%
\pgfpathmoveto{\pgfqpoint{4.676178in}{0.370555in}}%
\pgfpathlineto{\pgfqpoint{4.936935in}{0.370555in}}%
\pgfpathlineto{\pgfqpoint{4.936935in}{0.595652in}}%
\pgfpathlineto{\pgfqpoint{4.676178in}{0.595652in}}%
\pgfpathclose%
\pgfusepath{fill}%
\end{pgfscope}%
\begin{pgfscope}%
\pgfsetbuttcap%
\pgfsetroundjoin%
\definecolor{currentfill}{rgb}{0.000000,0.000000,0.000000}%
\pgfsetfillcolor{currentfill}%
\pgfsetlinewidth{0.803000pt}%
\definecolor{currentstroke}{rgb}{0.000000,0.000000,0.000000}%
\pgfsetstrokecolor{currentstroke}%
\pgfsetdash{}{0pt}%
\pgfsys@defobject{currentmarker}{\pgfqpoint{0.000000in}{-0.048611in}}{\pgfqpoint{0.000000in}{0.000000in}}{%
\pgfpathmoveto{\pgfqpoint{0.000000in}{0.000000in}}%
\pgfpathlineto{\pgfqpoint{0.000000in}{-0.048611in}}%
\pgfusepath{stroke,fill}%
}%
\begin{pgfscope}%
\pgfsys@transformshift{1.416716in}{0.370555in}%
\pgfsys@useobject{currentmarker}{}%
\end{pgfscope}%
\end{pgfscope}%
\begin{pgfscope}%
\definecolor{textcolor}{rgb}{0.000000,0.000000,0.000000}%
\pgfsetstrokecolor{textcolor}%
\pgfsetfillcolor{textcolor}%
\pgftext[x=1.416716in,y=0.273333in,,top]{\color{textcolor}\rmfamily\fontsize{10.000000}{12.000000}\selectfont 100\(\displaystyle \times\)100\(\displaystyle \times\)100}%
\end{pgfscope}%
\begin{pgfscope}%
\pgfsetbuttcap%
\pgfsetroundjoin%
\definecolor{currentfill}{rgb}{0.000000,0.000000,0.000000}%
\pgfsetfillcolor{currentfill}%
\pgfsetlinewidth{0.803000pt}%
\definecolor{currentstroke}{rgb}{0.000000,0.000000,0.000000}%
\pgfsetstrokecolor{currentstroke}%
\pgfsetdash{}{0pt}%
\pgfsys@defobject{currentmarker}{\pgfqpoint{0.000000in}{-0.048611in}}{\pgfqpoint{0.000000in}{0.000000in}}{%
\pgfpathmoveto{\pgfqpoint{0.000000in}{0.000000in}}%
\pgfpathlineto{\pgfqpoint{0.000000in}{-0.048611in}}%
\pgfusepath{stroke,fill}%
}%
\begin{pgfscope}%
\pgfsys@transformshift{2.981257in}{0.370555in}%
\pgfsys@useobject{currentmarker}{}%
\end{pgfscope}%
\end{pgfscope}%
\begin{pgfscope}%
\definecolor{textcolor}{rgb}{0.000000,0.000000,0.000000}%
\pgfsetstrokecolor{textcolor}%
\pgfsetfillcolor{textcolor}%
\pgftext[x=2.981257in,y=0.273333in,,top]{\color{textcolor}\rmfamily\fontsize{10.000000}{12.000000}\selectfont 200\(\displaystyle \times\)200\(\displaystyle \times\)200}%
\end{pgfscope}%
\begin{pgfscope}%
\pgfsetbuttcap%
\pgfsetroundjoin%
\definecolor{currentfill}{rgb}{0.000000,0.000000,0.000000}%
\pgfsetfillcolor{currentfill}%
\pgfsetlinewidth{0.803000pt}%
\definecolor{currentstroke}{rgb}{0.000000,0.000000,0.000000}%
\pgfsetstrokecolor{currentstroke}%
\pgfsetdash{}{0pt}%
\pgfsys@defobject{currentmarker}{\pgfqpoint{0.000000in}{-0.048611in}}{\pgfqpoint{0.000000in}{0.000000in}}{%
\pgfpathmoveto{\pgfqpoint{0.000000in}{0.000000in}}%
\pgfpathlineto{\pgfqpoint{0.000000in}{-0.048611in}}%
\pgfusepath{stroke,fill}%
}%
\begin{pgfscope}%
\pgfsys@transformshift{4.545799in}{0.370555in}%
\pgfsys@useobject{currentmarker}{}%
\end{pgfscope}%
\end{pgfscope}%
\begin{pgfscope}%
\definecolor{textcolor}{rgb}{0.000000,0.000000,0.000000}%
\pgfsetstrokecolor{textcolor}%
\pgfsetfillcolor{textcolor}%
\pgftext[x=4.545799in,y=0.273333in,,top]{\color{textcolor}\rmfamily\fontsize{10.000000}{12.000000}\selectfont 300\(\displaystyle \times\)300\(\displaystyle \times\)300}%
\end{pgfscope}%
\begin{pgfscope}%
\pgfsetbuttcap%
\pgfsetroundjoin%
\definecolor{currentfill}{rgb}{0.000000,0.000000,0.000000}%
\pgfsetfillcolor{currentfill}%
\pgfsetlinewidth{0.803000pt}%
\definecolor{currentstroke}{rgb}{0.000000,0.000000,0.000000}%
\pgfsetstrokecolor{currentstroke}%
\pgfsetdash{}{0pt}%
\pgfsys@defobject{currentmarker}{\pgfqpoint{-0.048611in}{0.000000in}}{\pgfqpoint{-0.000000in}{0.000000in}}{%
\pgfpathmoveto{\pgfqpoint{-0.000000in}{0.000000in}}%
\pgfpathlineto{\pgfqpoint{-0.048611in}{0.000000in}}%
\pgfusepath{stroke,fill}%
}%
\begin{pgfscope}%
\pgfsys@transformshift{0.634445in}{0.370555in}%
\pgfsys@useobject{currentmarker}{}%
\end{pgfscope}%
\end{pgfscope}%
\begin{pgfscope}%
\definecolor{textcolor}{rgb}{0.000000,0.000000,0.000000}%
\pgfsetstrokecolor{textcolor}%
\pgfsetfillcolor{textcolor}%
\pgftext[x=0.467778in, y=0.322361in, left, base]{\color{textcolor}\rmfamily\fontsize{10.000000}{12.000000}\selectfont \(\displaystyle {0}\)}%
\end{pgfscope}%
\begin{pgfscope}%
\pgfsetbuttcap%
\pgfsetroundjoin%
\definecolor{currentfill}{rgb}{0.000000,0.000000,0.000000}%
\pgfsetfillcolor{currentfill}%
\pgfsetlinewidth{0.803000pt}%
\definecolor{currentstroke}{rgb}{0.000000,0.000000,0.000000}%
\pgfsetstrokecolor{currentstroke}%
\pgfsetdash{}{0pt}%
\pgfsys@defobject{currentmarker}{\pgfqpoint{-0.048611in}{0.000000in}}{\pgfqpoint{-0.000000in}{0.000000in}}{%
\pgfpathmoveto{\pgfqpoint{-0.000000in}{0.000000in}}%
\pgfpathlineto{\pgfqpoint{-0.048611in}{0.000000in}}%
\pgfusepath{stroke,fill}%
}%
\begin{pgfscope}%
\pgfsys@transformshift{0.634445in}{0.754036in}%
\pgfsys@useobject{currentmarker}{}%
\end{pgfscope}%
\end{pgfscope}%
\begin{pgfscope}%
\definecolor{textcolor}{rgb}{0.000000,0.000000,0.000000}%
\pgfsetstrokecolor{textcolor}%
\pgfsetfillcolor{textcolor}%
\pgftext[x=0.398333in, y=0.705842in, left, base]{\color{textcolor}\rmfamily\fontsize{10.000000}{12.000000}\selectfont \(\displaystyle {20}\)}%
\end{pgfscope}%
\begin{pgfscope}%
\pgfsetbuttcap%
\pgfsetroundjoin%
\definecolor{currentfill}{rgb}{0.000000,0.000000,0.000000}%
\pgfsetfillcolor{currentfill}%
\pgfsetlinewidth{0.803000pt}%
\definecolor{currentstroke}{rgb}{0.000000,0.000000,0.000000}%
\pgfsetstrokecolor{currentstroke}%
\pgfsetdash{}{0pt}%
\pgfsys@defobject{currentmarker}{\pgfqpoint{-0.048611in}{0.000000in}}{\pgfqpoint{-0.000000in}{0.000000in}}{%
\pgfpathmoveto{\pgfqpoint{-0.000000in}{0.000000in}}%
\pgfpathlineto{\pgfqpoint{-0.048611in}{0.000000in}}%
\pgfusepath{stroke,fill}%
}%
\begin{pgfscope}%
\pgfsys@transformshift{0.634445in}{1.137517in}%
\pgfsys@useobject{currentmarker}{}%
\end{pgfscope}%
\end{pgfscope}%
\begin{pgfscope}%
\definecolor{textcolor}{rgb}{0.000000,0.000000,0.000000}%
\pgfsetstrokecolor{textcolor}%
\pgfsetfillcolor{textcolor}%
\pgftext[x=0.398333in, y=1.089323in, left, base]{\color{textcolor}\rmfamily\fontsize{10.000000}{12.000000}\selectfont \(\displaystyle {40}\)}%
\end{pgfscope}%
\begin{pgfscope}%
\pgfsetbuttcap%
\pgfsetroundjoin%
\definecolor{currentfill}{rgb}{0.000000,0.000000,0.000000}%
\pgfsetfillcolor{currentfill}%
\pgfsetlinewidth{0.803000pt}%
\definecolor{currentstroke}{rgb}{0.000000,0.000000,0.000000}%
\pgfsetstrokecolor{currentstroke}%
\pgfsetdash{}{0pt}%
\pgfsys@defobject{currentmarker}{\pgfqpoint{-0.048611in}{0.000000in}}{\pgfqpoint{-0.000000in}{0.000000in}}{%
\pgfpathmoveto{\pgfqpoint{-0.000000in}{0.000000in}}%
\pgfpathlineto{\pgfqpoint{-0.048611in}{0.000000in}}%
\pgfusepath{stroke,fill}%
}%
\begin{pgfscope}%
\pgfsys@transformshift{0.634445in}{1.520998in}%
\pgfsys@useobject{currentmarker}{}%
\end{pgfscope}%
\end{pgfscope}%
\begin{pgfscope}%
\definecolor{textcolor}{rgb}{0.000000,0.000000,0.000000}%
\pgfsetstrokecolor{textcolor}%
\pgfsetfillcolor{textcolor}%
\pgftext[x=0.398333in, y=1.472803in, left, base]{\color{textcolor}\rmfamily\fontsize{10.000000}{12.000000}\selectfont \(\displaystyle {60}\)}%
\end{pgfscope}%
\begin{pgfscope}%
\pgfsetbuttcap%
\pgfsetroundjoin%
\definecolor{currentfill}{rgb}{0.000000,0.000000,0.000000}%
\pgfsetfillcolor{currentfill}%
\pgfsetlinewidth{0.803000pt}%
\definecolor{currentstroke}{rgb}{0.000000,0.000000,0.000000}%
\pgfsetstrokecolor{currentstroke}%
\pgfsetdash{}{0pt}%
\pgfsys@defobject{currentmarker}{\pgfqpoint{-0.048611in}{0.000000in}}{\pgfqpoint{-0.000000in}{0.000000in}}{%
\pgfpathmoveto{\pgfqpoint{-0.000000in}{0.000000in}}%
\pgfpathlineto{\pgfqpoint{-0.048611in}{0.000000in}}%
\pgfusepath{stroke,fill}%
}%
\begin{pgfscope}%
\pgfsys@transformshift{0.634445in}{1.904479in}%
\pgfsys@useobject{currentmarker}{}%
\end{pgfscope}%
\end{pgfscope}%
\begin{pgfscope}%
\definecolor{textcolor}{rgb}{0.000000,0.000000,0.000000}%
\pgfsetstrokecolor{textcolor}%
\pgfsetfillcolor{textcolor}%
\pgftext[x=0.398333in, y=1.856284in, left, base]{\color{textcolor}\rmfamily\fontsize{10.000000}{12.000000}\selectfont \(\displaystyle {80}\)}%
\end{pgfscope}%
\begin{pgfscope}%
\pgfsetbuttcap%
\pgfsetroundjoin%
\definecolor{currentfill}{rgb}{0.000000,0.000000,0.000000}%
\pgfsetfillcolor{currentfill}%
\pgfsetlinewidth{0.803000pt}%
\definecolor{currentstroke}{rgb}{0.000000,0.000000,0.000000}%
\pgfsetstrokecolor{currentstroke}%
\pgfsetdash{}{0pt}%
\pgfsys@defobject{currentmarker}{\pgfqpoint{-0.048611in}{0.000000in}}{\pgfqpoint{-0.000000in}{0.000000in}}{%
\pgfpathmoveto{\pgfqpoint{-0.000000in}{0.000000in}}%
\pgfpathlineto{\pgfqpoint{-0.048611in}{0.000000in}}%
\pgfusepath{stroke,fill}%
}%
\begin{pgfscope}%
\pgfsys@transformshift{0.634445in}{2.287959in}%
\pgfsys@useobject{currentmarker}{}%
\end{pgfscope}%
\end{pgfscope}%
\begin{pgfscope}%
\definecolor{textcolor}{rgb}{0.000000,0.000000,0.000000}%
\pgfsetstrokecolor{textcolor}%
\pgfsetfillcolor{textcolor}%
\pgftext[x=0.328889in, y=2.239765in, left, base]{\color{textcolor}\rmfamily\fontsize{10.000000}{12.000000}\selectfont \(\displaystyle {100}\)}%
\end{pgfscope}%
\begin{pgfscope}%
\pgfsetbuttcap%
\pgfsetroundjoin%
\definecolor{currentfill}{rgb}{0.000000,0.000000,0.000000}%
\pgfsetfillcolor{currentfill}%
\pgfsetlinewidth{0.803000pt}%
\definecolor{currentstroke}{rgb}{0.000000,0.000000,0.000000}%
\pgfsetstrokecolor{currentstroke}%
\pgfsetdash{}{0pt}%
\pgfsys@defobject{currentmarker}{\pgfqpoint{-0.048611in}{0.000000in}}{\pgfqpoint{-0.000000in}{0.000000in}}{%
\pgfpathmoveto{\pgfqpoint{-0.000000in}{0.000000in}}%
\pgfpathlineto{\pgfqpoint{-0.048611in}{0.000000in}}%
\pgfusepath{stroke,fill}%
}%
\begin{pgfscope}%
\pgfsys@transformshift{0.634445in}{2.671440in}%
\pgfsys@useobject{currentmarker}{}%
\end{pgfscope}%
\end{pgfscope}%
\begin{pgfscope}%
\definecolor{textcolor}{rgb}{0.000000,0.000000,0.000000}%
\pgfsetstrokecolor{textcolor}%
\pgfsetfillcolor{textcolor}%
\pgftext[x=0.328889in, y=2.623246in, left, base]{\color{textcolor}\rmfamily\fontsize{10.000000}{12.000000}\selectfont \(\displaystyle {120}\)}%
\end{pgfscope}%
\begin{pgfscope}%
\definecolor{textcolor}{rgb}{0.000000,0.000000,0.000000}%
\pgfsetstrokecolor{textcolor}%
\pgfsetfillcolor{textcolor}%
\pgftext[x=0.273333in,y=1.610278in,,bottom,rotate=90.000000]{\color{textcolor}\rmfamily\fontsize{10.000000}{12.000000}\selectfont Time in seconds}%
\end{pgfscope}%
\begin{pgfscope}%
\pgfsetrectcap%
\pgfsetmiterjoin%
\pgfsetlinewidth{0.803000pt}%
\definecolor{currentstroke}{rgb}{0.000000,0.000000,0.000000}%
\pgfsetstrokecolor{currentstroke}%
\pgfsetdash{}{0pt}%
\pgfpathmoveto{\pgfqpoint{0.634445in}{0.370555in}}%
\pgfpathlineto{\pgfqpoint{0.634445in}{2.850000in}}%
\pgfusepath{stroke}%
\end{pgfscope}%
\begin{pgfscope}%
\pgfsetrectcap%
\pgfsetmiterjoin%
\pgfsetlinewidth{0.803000pt}%
\definecolor{currentstroke}{rgb}{0.000000,0.000000,0.000000}%
\pgfsetstrokecolor{currentstroke}%
\pgfsetdash{}{0pt}%
\pgfpathmoveto{\pgfqpoint{5.328070in}{0.370555in}}%
\pgfpathlineto{\pgfqpoint{5.328070in}{2.850000in}}%
\pgfusepath{stroke}%
\end{pgfscope}%
\begin{pgfscope}%
\pgfsetrectcap%
\pgfsetmiterjoin%
\pgfsetlinewidth{0.803000pt}%
\definecolor{currentstroke}{rgb}{0.000000,0.000000,0.000000}%
\pgfsetstrokecolor{currentstroke}%
\pgfsetdash{}{0pt}%
\pgfpathmoveto{\pgfqpoint{0.634445in}{0.370555in}}%
\pgfpathlineto{\pgfqpoint{5.328070in}{0.370555in}}%
\pgfusepath{stroke}%
\end{pgfscope}%
\begin{pgfscope}%
\pgfsetrectcap%
\pgfsetmiterjoin%
\pgfsetlinewidth{0.803000pt}%
\definecolor{currentstroke}{rgb}{0.000000,0.000000,0.000000}%
\pgfsetstrokecolor{currentstroke}%
\pgfsetdash{}{0pt}%
\pgfpathmoveto{\pgfqpoint{0.634445in}{2.850000in}}%
\pgfpathlineto{\pgfqpoint{5.328070in}{2.850000in}}%
\pgfusepath{stroke}%
\end{pgfscope}%
\begin{pgfscope}%
\definecolor{textcolor}{rgb}{0.000000,0.000000,0.000000}%
\pgfsetstrokecolor{textcolor}%
\pgfsetfillcolor{textcolor}%
\pgftext[x=1.155959in,y=0.626037in,,bottom]{\color{textcolor}\rmfamily\fontsize{10.000000}{12.000000}\selectfont 12.6}%
\end{pgfscope}%
\begin{pgfscope}%
\definecolor{textcolor}{rgb}{0.000000,0.000000,0.000000}%
\pgfsetstrokecolor{textcolor}%
\pgfsetfillcolor{textcolor}%
\pgftext[x=2.720501in,y=0.819695in,,bottom]{\color{textcolor}\rmfamily\fontsize{10.000000}{12.000000}\selectfont 22.7}%
\end{pgfscope}%
\begin{pgfscope}%
\definecolor{textcolor}{rgb}{0.000000,0.000000,0.000000}%
\pgfsetstrokecolor{textcolor}%
\pgfsetfillcolor{textcolor}%
\pgftext[x=4.285042in,y=2.633559in,,bottom]{\color{textcolor}\rmfamily\fontsize{10.000000}{12.000000}\selectfont 117.3}%
\end{pgfscope}%
\begin{pgfscope}%
\definecolor{textcolor}{rgb}{0.000000,0.000000,0.000000}%
\pgfsetstrokecolor{textcolor}%
\pgfsetfillcolor{textcolor}%
\pgftext[x=1.416716in,y=0.574267in,,bottom]{\color{textcolor}\rmfamily\fontsize{10.000000}{12.000000}\selectfont 9.9}%
\end{pgfscope}%
\begin{pgfscope}%
\definecolor{textcolor}{rgb}{0.000000,0.000000,0.000000}%
\pgfsetstrokecolor{textcolor}%
\pgfsetfillcolor{textcolor}%
\pgftext[x=2.981257in,y=0.733412in,,bottom]{\color{textcolor}\rmfamily\fontsize{10.000000}{12.000000}\selectfont 18.2}%
\end{pgfscope}%
\begin{pgfscope}%
\definecolor{textcolor}{rgb}{0.000000,0.000000,0.000000}%
\pgfsetstrokecolor{textcolor}%
\pgfsetfillcolor{textcolor}%
\pgftext[x=4.545799in,y=1.059370in,,bottom]{\color{textcolor}\rmfamily\fontsize{10.000000}{12.000000}\selectfont 35.2}%
\end{pgfscope}%
\begin{pgfscope}%
\definecolor{textcolor}{rgb}{0.000000,0.000000,0.000000}%
\pgfsetstrokecolor{textcolor}%
\pgfsetfillcolor{textcolor}%
\pgftext[x=1.677473in,y=0.455388in,,bottom]{\color{textcolor}\rmfamily\fontsize{10.000000}{12.000000}\selectfont 3.7}%
\end{pgfscope}%
\begin{pgfscope}%
\definecolor{textcolor}{rgb}{0.000000,0.000000,0.000000}%
\pgfsetstrokecolor{textcolor}%
\pgfsetfillcolor{textcolor}%
\pgftext[x=3.242014in,y=0.516745in,,bottom]{\color{textcolor}\rmfamily\fontsize{10.000000}{12.000000}\selectfont 6.9}%
\end{pgfscope}%
\begin{pgfscope}%
\definecolor{textcolor}{rgb}{0.000000,0.000000,0.000000}%
\pgfsetstrokecolor{textcolor}%
\pgfsetfillcolor{textcolor}%
\pgftext[x=4.806556in,y=0.608781in,,bottom]{\color{textcolor}\rmfamily\fontsize{10.000000}{12.000000}\selectfont 11.7}%
\end{pgfscope}%
\begin{pgfscope}%
\pgfsetbuttcap%
\pgfsetmiterjoin%
\definecolor{currentfill}{rgb}{1.000000,1.000000,1.000000}%
\pgfsetfillcolor{currentfill}%
\pgfsetfillopacity{0.800000}%
\pgfsetlinewidth{1.003750pt}%
\definecolor{currentstroke}{rgb}{0.800000,0.800000,0.800000}%
\pgfsetstrokecolor{currentstroke}%
\pgfsetstrokeopacity{0.800000}%
\pgfsetdash{}{0pt}%
\pgfpathmoveto{\pgfqpoint{0.731667in}{2.158056in}}%
\pgfpathlineto{\pgfqpoint{2.277639in}{2.158056in}}%
\pgfpathquadraticcurveto{\pgfqpoint{2.305417in}{2.158056in}}{\pgfqpoint{2.305417in}{2.185834in}}%
\pgfpathlineto{\pgfqpoint{2.305417in}{2.752778in}}%
\pgfpathquadraticcurveto{\pgfqpoint{2.305417in}{2.780556in}}{\pgfqpoint{2.277639in}{2.780556in}}%
\pgfpathlineto{\pgfqpoint{0.731667in}{2.780556in}}%
\pgfpathquadraticcurveto{\pgfqpoint{0.703889in}{2.780556in}}{\pgfqpoint{0.703889in}{2.752778in}}%
\pgfpathlineto{\pgfqpoint{0.703889in}{2.185834in}}%
\pgfpathquadraticcurveto{\pgfqpoint{0.703889in}{2.158056in}}{\pgfqpoint{0.731667in}{2.158056in}}%
\pgfpathclose%
\pgfusepath{stroke,fill}%
\end{pgfscope}%
\begin{pgfscope}%
\pgfsetbuttcap%
\pgfsetmiterjoin%
\definecolor{currentfill}{rgb}{0.372549,0.827451,0.372549}%
\pgfsetfillcolor{currentfill}%
\pgfsetlinewidth{0.000000pt}%
\definecolor{currentstroke}{rgb}{0.000000,0.000000,0.000000}%
\pgfsetstrokecolor{currentstroke}%
\pgfsetstrokeopacity{0.000000}%
\pgfsetdash{}{0pt}%
\pgfpathmoveto{\pgfqpoint{0.759445in}{2.627778in}}%
\pgfpathlineto{\pgfqpoint{1.037223in}{2.627778in}}%
\pgfpathlineto{\pgfqpoint{1.037223in}{2.725000in}}%
\pgfpathlineto{\pgfqpoint{0.759445in}{2.725000in}}%
\pgfpathclose%
\pgfusepath{fill}%
\end{pgfscope}%
\begin{pgfscope}%
\definecolor{textcolor}{rgb}{0.000000,0.000000,0.000000}%
\pgfsetstrokecolor{textcolor}%
\pgfsetfillcolor{textcolor}%
\pgftext[x=1.148334in,y=2.627778in,left,base]{\color{textcolor}\rmfamily\fontsize{10.000000}{12.000000}\selectfont ALS CUDA}%
\end{pgfscope}%
\begin{pgfscope}%
\pgfsetbuttcap%
\pgfsetmiterjoin%
\definecolor{currentfill}{rgb}{0.172549,0.627451,0.172549}%
\pgfsetfillcolor{currentfill}%
\pgfsetlinewidth{0.000000pt}%
\definecolor{currentstroke}{rgb}{0.000000,0.000000,0.000000}%
\pgfsetstrokecolor{currentstroke}%
\pgfsetstrokeopacity{0.000000}%
\pgfsetdash{}{0pt}%
\pgfpathmoveto{\pgfqpoint{0.759445in}{2.434167in}}%
\pgfpathlineto{\pgfqpoint{1.037223in}{2.434167in}}%
\pgfpathlineto{\pgfqpoint{1.037223in}{2.531389in}}%
\pgfpathlineto{\pgfqpoint{0.759445in}{2.531389in}}%
\pgfpathclose%
\pgfusepath{fill}%
\end{pgfscope}%
\begin{pgfscope}%
\definecolor{textcolor}{rgb}{0.000000,0.000000,0.000000}%
\pgfsetstrokecolor{textcolor}%
\pgfsetfillcolor{textcolor}%
\pgftext[x=1.148334in,y=2.434167in,left,base]{\color{textcolor}\rmfamily\fontsize{10.000000}{12.000000}\selectfont OMP ALS CUDA}%
\end{pgfscope}%
\begin{pgfscope}%
\pgfsetbuttcap%
\pgfsetmiterjoin%
\definecolor{currentfill}{rgb}{0.086275,0.313725,0.086275}%
\pgfsetfillcolor{currentfill}%
\pgfsetlinewidth{0.000000pt}%
\definecolor{currentstroke}{rgb}{0.000000,0.000000,0.000000}%
\pgfsetstrokecolor{currentstroke}%
\pgfsetstrokeopacity{0.000000}%
\pgfsetdash{}{0pt}%
\pgfpathmoveto{\pgfqpoint{0.759445in}{2.240556in}}%
\pgfpathlineto{\pgfqpoint{1.037223in}{2.240556in}}%
\pgfpathlineto{\pgfqpoint{1.037223in}{2.337778in}}%
\pgfpathlineto{\pgfqpoint{0.759445in}{2.337778in}}%
\pgfpathclose%
\pgfusepath{fill}%
\end{pgfscope}%
\begin{pgfscope}%
\definecolor{textcolor}{rgb}{0.000000,0.000000,0.000000}%
\pgfsetstrokecolor{textcolor}%
\pgfsetfillcolor{textcolor}%
\pgftext[x=1.148334in,y=2.240556in,left,base]{\color{textcolor}\rmfamily\fontsize{10.000000}{12.000000}\selectfont CALS CUDA}%
\end{pgfscope}%
\end{pgfpicture}%
\makeatother%
\endgroup%

%% file: sections/07-conclusion.tex

\section{Conclusion}
\label{sec:conclusion}

In this paper we present Concurrent ALS (CALS), an algorithm and library, which offers an interface to MATLAB, for computing multiple, concurrent Alternating Least Squares algorithms for the Canonical Polyadic Decomposition.
We show that CALS is able to accommodate applications that fit multiple models of different \changed{number of components} and starting points, by achieving better efficiency for the same computation.
Furthermore, we demonstrate how higher efficiency favors the, otherwise impractical under \decompname{}-ALS, offloading of the computation to GPUs to further speed up computation.
Finally, we showcase the effectiveness of CALS over our own optimized version of \decompname{}-ALS as well as Tensor Toolbox's implementation \texttt{cp\_als} on both artificial and real datasets.